\documentclass[lettersize,journal]{IEEEtran}
\usepackage{amsmath,amsfonts}
\usepackage{amssymb}
\usepackage{algorithmic}
\usepackage{algorithm}
\usepackage{array}
\usepackage{subfig}
\usepackage{textcomp}
\usepackage{stfloats}
\usepackage{url}
\usepackage{verbatim}
\usepackage{graphicx}
\usepackage{cite}
\usepackage{bm}
\usepackage{booktabs}  
\usepackage{multirow}
\usepackage{enumerate}
\usepackage{color}
\newcommand\figref{Fig.~\ref}

\newcommand\tabref{Tab.~\ref}
\newcommand\secref{Section~\ref}

\def\bee{\begin{equation}}
	\def\ene{\end{equation}}

\def\beq{\begin{eqnarray}}
	\def\enq{\end{eqnarray}}

\begin{document}

\title{A Robust Super-resolution Gridless Imaging Framework for UAV-borne SAR Tomography}

\author{Silin Gao, Wenlong Wang, Muhan Wang, Zhe Zhang,~\IEEEmembership{Member,~IEEE}, Zai Yang,~\IEEEmembership{Senior Member,~IEEE}, Xiaolan Qiu,~\IEEEmembership{Senior Member,~IEEE}, Bingchen Zhang and Yirong Wu
	\thanks{Corresponding author: Zhe Zhang (zhangzhe01@aircas.ac.cn)}
	\thanks{
		S. Gao, M. Wang, B. Zhang and Y. Wu are with Key Laboratory of Technology in Geo-spatial Information Processing and Application System, Chinese Academy of Sciences, Beijing, China, and School of Electronic, Electrical and Communication Engineering, University of Chinese Academy of Sciences, Beijing, China.}
	\thanks{
		S. Gao, M. Wang, Z. Zhang, X. Qiu, B. Zhang and Y. Wu are also with Aerospace Information Research Institute, Chinese Academy of Sciences, Beijing 100094, China.}
	\thanks{W. Wang and Z. Yang are with School of Mathematics \& Statistics, Xi'an Jiaotong University, Xi'an 710049, China.}
	\thanks{
		M. Wang and Z. Zhang are with Suzhou Key Laboratory of Microwave Imaging, Processing and Application Technology, and Suzhou Aerospace Information Research Institute, Suzhou 215123, China.}
	\thanks{
		X. Qiu is with China National Key Laboratory of Microwave Imaging Technology, Aerospace Information Research Institute, Chinese Academy of Sciences, Beijing 100190, China.}
}

\maketitle

\begin{abstract}
	Synthetic aperture radar (SAR) tomography (TomoSAR) retrieves three-dimensional (3-D) information from multiple SAR images, effectively addresses the layover problem, and has become pivotal in urban mapping.
	Unmanned aerial vehicle (UAV) has gained popularity as a TomoSAR platform, offering distinct advantages such as the ability to achieve 3-D imaging in a single flight, cost-effectiveness, rapid deployment, and flexible trajectory planning.
	The evolution of compressed sensing (CS) has led to the widespread adoption of sparse reconstruction techniques in TomoSAR signal processing, with a focus on $\ell _1$ norm regularization and other grid-based CS methods. However, the discretization of illuminated scene along elevation introduces modeling errors, resulting in reduced reconstruction accuracy, known as the ``off-grid" effect. Recent advancements have introduced gridless CS algorithms to mitigate this issue.
	This paper presents an innovative gridless 3-D imaging framework tailored for UAV-borne TomoSAR. Capitalizing on the pulse repetition frequency (PRF) redundancy inherent in slow UAV platforms, a multiple measurement vectors (MMV) model is constructed to enhance noise immunity without compromising azimuth-range resolution. Given the sparsely placed array elements due to mounting platform constraints, an atomic norm soft thresholding algorithm is proposed for partially observed MMV, offering gridless reconstruction capability and super-resolution. An efficient alternative optimization algorithm is also employed to enhance computational efficiency.
	Validation of the proposed framework is achieved through computer simulations and flight experiments, affirming its efficacy in UAV-borne TomoSAR applications.
\end{abstract}

\begin{IEEEkeywords}
	Synthetic aperture radar tomography (TomoSAR), unmanned aerial vehicle (UAV), gridless compressed sensing, atomic norm, multiple measurement vectors (MMV)
\end{IEEEkeywords}

\section{Introduction}
Synthetic aperture radar (SAR) stands out for its weather-independent, high-resolution imaging capabilities, serving as a vital tool in diverse applications across civilian and military domains~\cite{SAR1,SAR2,SAR3}. SAR tomography (TomoSAR) emerges as a technique dedicated to retrieving three-dimensional (3-D) information from multiple two-dimensional (2-D) SAR images acquired from distinct view angles, often referred to as single-look complex (SLC) images~\cite{Nonlocal}. As an extension of interferometric SAR (InSAR) technology, TomoSAR addresses the layover problem, offering an efficient solution for 3-D imaging, particularly beneficial in urban mapping~\cite{UAV}.

The present SAR tomography systems comprise the repeated multiple-pass multi-baseline SAR and array InSAR~\cite{3D}. The former imposes no specific requirements on the SAR system itself but requires stringent control over the orbit or flight trajectory to ensure data coherence, which introduces challenges such as extended data acquisition cycles and operational complexities. In contrast, array InSAR, involving three or more antenna elements, facilitates the acquisition of multiple spatially coherent observations in a single flight, enabling efficient 3-D imaging in a single mission~\cite{MV3DSAR}. Currently, array InSAR has become a significant system for SAR tomography, garnering attention globally.
However, practical array InSAR usually suffers from incomplete observations and limited resolution due to mounting constraints~\cite{TomoANM1}.
The resolution along elevation direction is much lower than range and azimuth, making it difficult to achieve high-resolution 3-D imaging~\cite{3D}.

In the evolution of TomoSAR technology, airborne platforms initially validate the feasibility of 3-D imaging through multipass TomoSAR~\cite{TomoSAR}. Subsequently, spaceborne systems (e.g. ERS, TerraSAR-X, and COSMO-SkyMed) provide crucial datasets, advancing research in processing frameworks and algorithms~\cite{TomoSAR1,TomoSAR3,TomoSAR5}.
In recent years, spaceborne TomoSAR has gained widespread application in urban mapping~\cite{TomoSAR2,SVD,TomoSAR7}, while airborne TomoSAR, constrained by the flight trajectory planning in urban areas, finds its primary utility in natural environments such as forests and glaciers~\cite{TomoSAR6,TomoSAR9,UAVforest,AirbornForest,UAVsnow}.
The advent of unmanned aerial vehicle (UAV), especially small UAV, has markedly accelerated the implementation of airborne TomoSAR in urban scenarios~\cite{DRIVE,ARTINO1,UAV2,UAVzeng,UAVHu}. In contrast to conventional TomoSAR systems employing large platforms like aircraft and satellites, UAV platform offers notable advantages, including cost-effectiveness, swift deployment, and adaptable trajectory planning~\cite{UAV1}.
Characterized by low-altitude, slow-speed flight (e.g., ARTINO system at 200 m altitude and 10-15 m/s speed~\cite{ARTINO1}), signals obtained from small UAV possess limited Doppler bandwidth. Given the inherently narrow swath in UAV-borne SAR imaging, the pursuit of excessively small pulse repetition frequency (PRF) becomes unnecessary. This presents an opportunity for efficient exploitation of PRF redundancy.

In TomoSAR signal processing, SLC images and the reflectivity function along elevation for each azimuth-range pixel form Fourier transform pairs. Therefore, SAR tomography can be regarded a spatial spectrum estimation problem.
The pioneering SAR tomography approach applies the discrete Fourier transform to an interpolated linear array of baselines~\cite{TomoSAR}. Subsequent studies have centered on long-term repeat-pass acquisitions, emphasizing advanced inversion techniques~\cite{TomoSAR1,TSVD,SVD}.
With the development of compressed sensing (CS)~\cite{CSDonoho,CSBaraniuk,CSCandes}, sparse reconstruction techniques have been introduced to TomoSAR signal processing.
These techniques utilize the sparsity in the elevation direction to generate 3-D images with a small number of acquisitions.
Compared to classical multi-baseline InSAR algorithms, CS-based methods demonstrate superior accuracy and enhance the elevation resolution achievable with a given set of baselines, which is particularly crucial for urban areas where layover is prevalent~\cite{review}.
Presently, many mainstream sparse TomoSAR imaging methods~\cite{TomoSAR3,TomoSAR5,l11,l12,l13,l1Han} employ the $\ell _1$ norm regularization model and spatial domain discretization, encountering the grid mismatching problem (i.e. ``off-grid'' effect) and resulting in degraded performance~\cite{off,OGSBIyang}.

To circumvent this issue, atomic norm~\cite{ANori} emerges as a promising mathematical concept, providing theoretical support for sparse reconstruction in continuous domains.
Atomic norm minimization (ANM)~\cite{ANMTang} builds upon this concept to achieve gridless spectral estimation for single measurement vector (SMV) under the conditions of uniform sampling and full noiseless observation.
Expanding the application scope of atomic norm, atomic norm soft thresholding (AST)~\cite{AST} is proposed to solve the denoising tasks for full observations in the presence of independently and identically distributed (i.i.d.) additive white Gaussian noise (AWGN).
Subsequently, the AST technique is further improved, focusing on partial observations~\cite{ASTYang}.
The research of atomic norm continues into the realm of multiple measurement vectors (MMV), including scenarios with partial noiseless observations and full observations in AWGN~\cite{MMVANMULA,MMVNOISELESS,yang2016exact,MMVyang2,MMVSLAULA,MMVyang2019}.
Despite these advances, theoretical analysis of gridless sparse reconstruction from partial noisy observations in MMV case still remains a topic for investigation.

In recent years, atomic norm based techniques, renowned for their excellent performance in spectral estimation and direction of arrival estimation~\cite{FullNoisyChen,MMVFullNoisyWei,recentWu2023}, have found application in the domain of TomoSAR imaging.
The pioneering work~\cite{TomoANM1} suggests utilizing ANM for TomoSAR imaging but lacks a thorough theoretical derivation.
In~\cite{TomoANM2}, a comprehensive theoretical analysis substantiates the feasibility and efficacy of employing atomic norm based techniques in TomoSAR imaging.
Concurrently, the work in~\cite{TomoANM3} extends the application of atomic norm framework to coprime TomoSAR reconstruction.
Moreover, \cite{3D} assumes that adjacent pixels exhibit similar elevation distributions and regard them as i.i.d. samples, which are then constructed into multiple measurement vectors and solved using atomic norm based techniques. This work~\cite{3D} is also limited to the case of uniform baselines, i.e., the previously described full noisy observations.

This paper focuses on UAV-borne array SAR tomography, and proposes a novel gridless 3-D imaging framework with exceptional super-resolution capability and robust noise immunity.
The proposed framework represents the first attempt that allows multiple snapshot data acquisition, or MMV, in TomoSAR imaging.
This innovation is based on the premise that large oversampling rates along azimuth can be readily achieved in UAV flight experiments, which relies on the significantly narrow Doppler bandwidth associated with slow flight speeds (e.g. Doppler bandwidth = 71 Hz and PRF = 1000 Hz in~\cite{MV3DSAR}). Furthermore, the framework accommodates the sparse arrangement of array elements, a common occurrence due to mounting platform constraints.
Central to the proposed framework is the construction of extracted-MMV (abbreviated as EMMV) from oversampled data, followed by the reconstruction of 3-D point clouds through an innovative atomic norm soft thresholding algorithm. 
Our main contributions can be summarize as follows:
\begin{enumerate}[1)]
	\item We have developed a robust gridless 3-D imaging framework for UAV-borne array SAR tomography by constructing EMMV. A comprehensive investigation into estimation accuracy and super-resolution capability has been carried out.
	\item We have proposed a new atomic norm soft thresholding algorithm for partially observed MMV and give the optimal choice for the regularization parameter in the case of i.i.d. AWGN presenting.
	\item To address the challenge of large-scale semi-definite programming (SDP) problems, we present a fast alternative optimization solver utilizing the alternating direction multiplication method (ADMM).
	\item We conduct numerical experiments using both simulated and measured data to validate the efficacy of the proposed EMPAST framework. Notably, we have applied the framework in flight experiments conducted in Tianjin, China, reconstructing high-quality 3-D point clouds of buildings.
\end{enumerate}

This paper is organized as follows. In \secref{sec:Preliminaries}, we present the traditional TomoSAR imaging model, followed by the grid-based and gridless CS methods under this model, respectively. In \secref{sec:Method}, we systematically investigate the proposed EMPAST framework. Experimental results of simulated data are provided in \secref{sec:Simu}. \secref{sec:Real} introduces the flight experiment and presents the 3-D imaging results of measured data. Finally, the conclusions are given in \secref{sec:Con}.

\subsubsection*{Notations}
$\mathbb{R}$ and $\mathbb{C}$ denote the sets of real and complex numbers respectively.
Vectors are denoted by lowercase bold characters (e.g. $\mathbf{v}$) and matrices by uppercase bold characters (e.g. $\mathbf{M})$.
Specifically, $\mathbf{I}$ denotes the identity matrix.
$[\cdot]_{k,l}$ denotes the $(k,l)$-th entry.
The $(\cdot)^*$ denotes the conjugate operator.
$(\cdot)^T$ and $(\cdot)^H$ are the transpose operator and the Hermitian transpose operator respectively. The $\text{tr}(\cdot)$ stands for the trace operator.
$(\cdot)^{-1}$ computes the inverse of square matrix.
$\lVert \cdot\rVert_1$, $\lVert \cdot\rVert_2$ and $\lVert \cdot\rVert_F$ denote the $\ell_1$, $\ell_2$ and Frobenius norms, respectively. Additionally, $|\cdot|$ signifies the absolute value.
The operator $\mathbb{E}[\cdot]$ represents expectation.

\section{Preliminaries}\label{sec:Preliminaries}

\subsection{TomoSAR Imaging}
\begin{figure}[htbp]
	\centering
	\includegraphics[width=0.7\columnwidth]{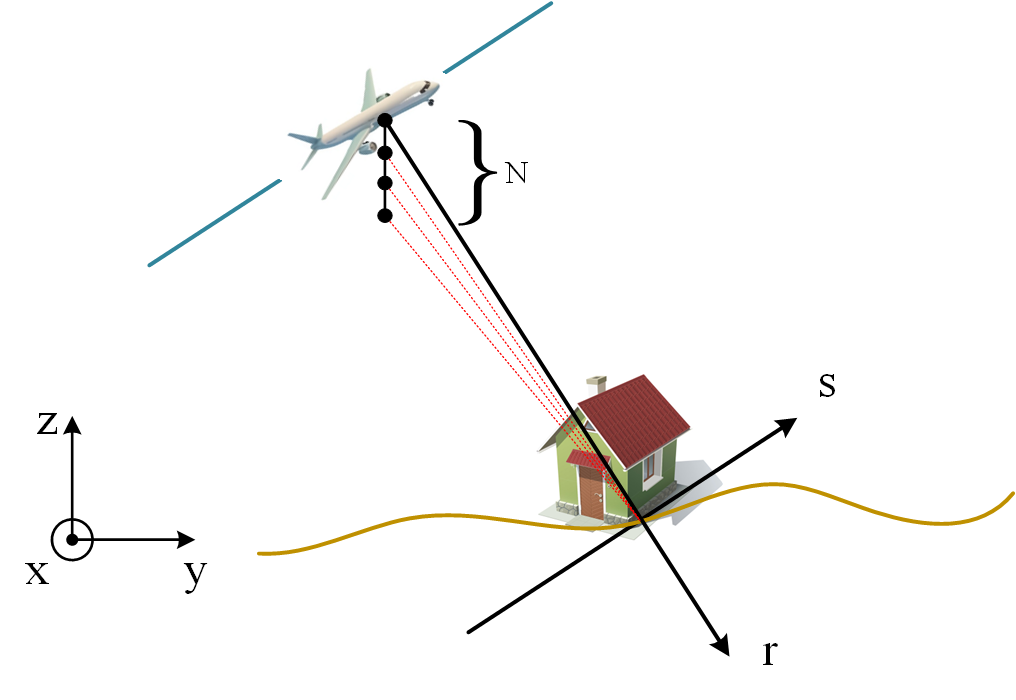}
	\caption{Tomographic SAR imaging model diagram.}
	\label{fig1}
\end{figure}

Traditional SAR imaging systems generate 2-D images in azimuth and range, essentially projecting real 3-D targets onto the azimuth-range plane. When targets exist at varying elevations within the same azimuth-range pixel, their imaging results become superimposed, leading to overlapping. 
A TomoSAR imaging system typically encompasses multiple baselines, each corresponding to either a single-pass acquisition or an element within an antenna array, as illustrated in Fig. \ref{fig1}. 
For each fixed azimuth-range pixel $(x,r)$ within SLC images obtained along the orbit with the $n$-th orthogonal baseline $b_n (n = 1,\cdots,N)$, the noiseless observation $g_n^o$ can be expressed as the integral superposition of the contributions of all scatterers distributed at varying elevations within this pixel:
\begin{equation}
	g_n^o = \int_{-S_1/2}^{+S_1/2} \gamma  (s)\exp \left( { - j2\pi {\zeta _n}s} \right)ds,
\end{equation}
where $S_1$ is the elevation extension of the illuminated scene, $\gamma(s)\in \mathbb{C}$ is the backscattering coefficient as a function of a spatially continuous elevation $s$, the spatial frequency ${\zeta _n} =  - \frac{{2{b_{n}}}}{{\lambda {r}}}$ is proportional to the respective baseline $b_{n}$, and $\lambda$ is the wavelength.
By estimating the spatial spectrum of $\mathbf{g}^o=\begin{bmatrix}
	g_1^o,g_2^o,\cdots,g_N^o\end{bmatrix}^T$, the resolving along elevation can be obtained with the Rayleigh resolution as:
\begin{equation}\label{eq:rhos}
	\rho_s =\frac{\lambda r_0}{2D},
\end{equation}
where $D$ is the elevation aperture and $r_0$ is the reference distance.

Considering that the distribution of targets in the elevation direction usually exhibits sparse properties, $\mathbf{g}^o$ can be regarded as a spectrally sparse signal. In the presence of noise $\mathbf{e}\in \mathbb{C}^{N}$, the sparse SAR tomography model for $K$ targets in the elevation direction can be expressed as:
\begin{equation}\label{sparse}
	\mathbf{g} =\mathbf{g}^o+  \mathbf{e}=\sum_{k=1}^{K}\gamma_k \mathbf{a}(s_k)+  \mathbf{e}=\mathbf{A}\bm{\gamma}+ \mathbf{e},
\end{equation}
where $\gamma_k$ and $s_k$ are the backscattering coefficient and elevation location of the $k$-th target, $\mathbf{a}(s_k)=[\exp \left( { - j2\pi {\zeta_1}s_k} \right), \exp \left( { - j2\pi {\zeta_2}s_k} \right), \cdots, \exp \left( { - j2\pi {\zeta_N}s_k} \right)]^T\in \mathbb{C}^{N}$ is the steering vector of the $k$-th target,  $\mathbf{A}=[\mathbf{a}(s_1),\mathbf{a}(s_2),\cdots,\mathbf{a}(s_K)]\in\mathbb{C}^{N\times K}$ is the steering matrix, and $\bm{\gamma}=[\gamma_1,\gamma_2,\cdots,\gamma_K]^T\in \mathbb{C}^{K}$.

\subsection{Grid-based CS}
The on-grid SAR tomography model discretizes the target along $s\in \left(-\frac{S_1}{2},\frac{S_1}{2}\right)$ and yields the following approximate measurement model:
\begin{equation}\label{grid}
	\mathbf{g} =\mathbf{A}_Q\bar{\bm{\gamma}}+ \mathbf{e},
\end{equation}
where $\mathbf{A}_Q\in\mathbb{C}^{N\times Q}$ is a predefined dictionary whose columns are the steering vectors corresponding to $Q$ predefined elevation grids $\bar{s}_q(q = 1,\cdots,Q)$. A basis set of steering vectors is defined as $\mathcal{A}_Q=\left\{ \boldsymbol{a}( s ) :s\in \left( \bar{s}_1,\bar{s}_2,\cdots ,\bar{s}_Q \right) \right\}$. 
Then the sparsity of target scene can be expressed in terms of $\ell _0$ norm (and relaxed to $\ell _1$ norm) of $\mathbf{g}$ over the basis $\mathcal{A}_Q$ as:
\begin{equation}\label{csl0}
	\lVert \mathbf{g} \rVert _{\mathcal{A}_Q,0}\triangleq\inf\left\{ K:\sum_{k=1}^K{\hat{\gamma}_k\mathbf{a}\left( \hat{s}_k \right) =\mathbf{g}},\mathbf{a}\left( \hat{s}_k \right) \in \mathcal{A}_Q \right\},
\end{equation}
\begin{equation}\label{csl1}
	\lVert \mathbf{g} \rVert _{\mathcal{A}_Q,1}
	\triangleq\inf\left\{ \sum_{k=1}^K|\hat{\gamma}_k|:\sum_{k=1}^K{\hat{\gamma}_k\mathbf{a}\left( \hat{s}_k \right) =\mathbf{g}},\mathbf{a}\left( \hat{s}_k \right) \in \mathcal{A}_Q \right\} .
\end{equation}

In the compressed sensing framework, it is proved that we can recover $\{(\gamma_k, s_k)\}$ via relaxed convex $\ell _1$ norm minimization under some conditions~\cite{rip}. Therefore, the elevation information of targets can be obtained by solving the optimization problem based on $\ell _1$ norm minimization as follows:
\begin{equation}\label{cs}
	\min_{\bar{\bm{\gamma}}} \frac{1}{2}\lVert \mathbf{g}-\mathbf{A}_Q\bar{\bm{\gamma}} \rVert_2^2 +\tau\lVert \mathbf{g} \rVert_{\mathcal{A}_Q,1},
\end{equation}
where $\tau$ is the regularization parameter.
Many well-established grid-based CS (GBCS) algorithms~\cite{BPDN,Rl1,FISTA} can be applied to this inversion problem.
Note that we use $\mathbf{A}_Q$ and $\bar{\bm{\gamma}}$ to emphasize that the models \eqref{sparse} and \eqref{grid} are fundamentally different. 
Since the targets may be located at any elevation position, the columns of $\mathbf{A}$ may not be in $\mathcal{A}_Q$.

\subsection{Gridless CS}\label{subsection:Gridless CS}
GBCS algorithms assume that targets are located on discrete elevation grids, which is inconsistent with the fact that targets are usually off the grids, i.e., $\mathbf{a}\text{(}s_k\text{)}\notin \mathcal{A}_Q$, resulting in the grid mismatch issue. 
The concept of atomic norm~\cite{ANori} addresses this issue by manipulating $Q\rightarrow \infty$, leveraging $\mathcal{A}=\left\{ \mathbf{a}\left( s \right) :\,\,s\in (-\frac{S_1}{2},\frac{S_1}{2}) \right\} $.
Then, the gridless version of \eqref{csl1} can be defined as atomic norm:
\begin{equation}\label{anm}
	\begin{aligned}
		\lVert \mathbf{g} \rVert _{\mathcal{A}} &=\lVert \mathbf{g} \rVert_{\mathcal{A},1}
		\triangleq\inf\left\{ t>0:\mathbf{g}\in t \text{conv}(\mathcal{A})\right\}
		\\&=\inf\left\{ \sum_{k=1}^K{\left| \gamma_k \right|}:\sum_{k=1}^K{\gamma_k\mathbf{a}\left( s_k \right) =\mathbf{g},\mathbf{a}\left( s_k \right) \in \mathcal{A}} \right\},
	\end{aligned}
\end{equation}
where $\text{conv}(\mathcal{A})$ is the convex hull of $\mathcal{A}$.
In the presence of i.i.d. AWGN, $\{(\gamma_k, s_k)\}$ can be estimated by solving following optimization problem, known as atomic norm soft thresholding (AST):
\begin{equation}\label{minanm}
	\min_{\hat{\mathbf{g}}} \frac{1}{2}\lVert \hat{\mathbf{g}}- \mathbf{g}\rVert_2^2+ \tau\lVert \hat{\mathbf{g}} \rVert _{\mathcal{A}},
\end{equation}
which is infinite programming and cannot be easily solved via mature convex optimization methods.
It has been proved that when the atoms are in the form of Vandermonde vectors (e.g. $[\boldsymbol{a}(\omega)]_n=\exp(j\omega n)$), it can be solved via SDP~\cite{AST}.

Considering the case of partial observations, i.e. incomplete data, the atoms are derived from sub-sampled Vandermonde atoms. Here, the set of samples is denoted as $\Omega$, representing an ordered subset within the range $[1, 2, \cdots, N]$, where $M$ signifies the number of samples. Consequently, the set of partially observed atoms is expressed as:
\begin{equation}
	\mathcal{A}_{\Omega} =\left\{ \mathbf{a}_{\Omega}\left( s \right): \mathbf{a}\left( s \right)\in \mathcal{A}\right\},
\end{equation}
where $\mathbf{a}_{\Omega}\left( s \right)$ denotes a subvector of $\mathbf{a}\left( s \right)$ indexed by $\Omega$. Thus, the partially observed atomic norm denoising problem can be formulated as:
\begin{equation}
	\min_{ \hat{\mathbf{g}}_{\Omega}} \quad \frac{1}{2}\lVert \hat{\mathbf{g}}_{\Omega}- \mathbf{g}_{\Omega}\rVert_2^2+ \tau\lVert \hat{\mathbf{g}}_{\Omega} \rVert _{\mathcal{A}_{\Omega}}.
\end{equation}
Similarly, the optimization problem is reformulated as an SDP problem for resolution, accompanied by the establishment of an upper boundary for the regularization parameter $\tau$~\cite{ASTYang}. For conciseness, we shall refer to this approach as partially observed atomic norm soft thresholding (PAST) in the subsequent sections of this paper.

\section{Proposed Framework: EMPAST}\label{sec:Method}
This section introduces the extracted-MMV based partially observed atomic norm soft thresholding (EMPAST), a novel gridless 3-D imaging framework tailored for UAV-borne SAR tomography.
The framework commences by extracting MMV from oversampled data, operating under the fact that achieving large oversampling rates along the azimuth in UAV flight experiments is feasible. Additionally, EMPAST accounts for the sparse arrangement of array elements and introduces an innovative atomic norm soft thresholding algorithm for sparse reconstruction.
\begin{figure*}[bp]
	\centering
	\includegraphics[width=\linewidth]{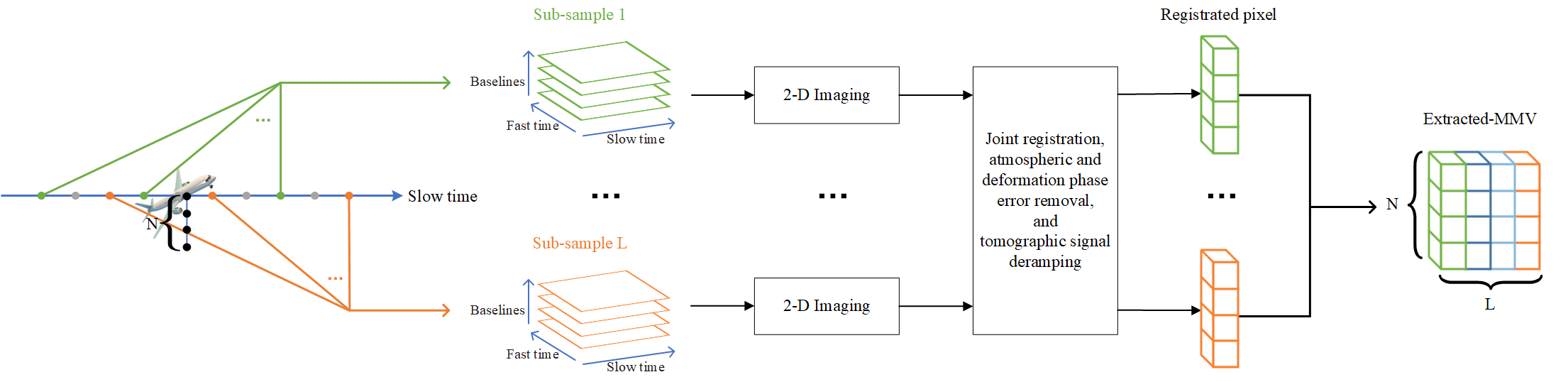}
	\caption{Schematic of EMMV construction.}
	\label{MMVcon}
\end{figure*}
\subsection{EMMV Construction}
In this study, we assume that the array SAR system operates in strip mode, employing a linear frequency modulation signal as the transmitted signal. 
The demodulated SAR baseband signal of the $n$-th baseline is given by:
\begin{equation}
	y_{n}\left(t,\eta\right)=\xi_n\left(t,\eta\right)\otimes h_{imp}\left(t,\eta\right),
\end{equation}
where $t$ and $\eta$ denote the fast time and slow time,
$\xi_n\left(t,\eta\right)$ represents the backscattering coefficients of the illuminated 3-D scene projected onto the azimuth-range plane for the $n$-th baseline, 
$\otimes$ denotes the convolution operator,
and $h_{\text{imp}}\left(t,\eta\right)$ denotes the impulse function of the SAR sensor.

Sampling the signal in both the range direction (fast time) and azimuth direction (slow time) is required to satisfy the Nyquist sampling theorem. Typically, an azimuth oversampling rate between 1.1 and 1.4 is chosen to mitigate azimuth ambiguity. However, due to the slow flight speeds of small UAV platforms, the Doppler bandwidth is often narrow, resulting in significant PRF redundancy. Therefore, we propose constructing multiple measurement data by extracting the slow time.
Assuming the number of EMMV is $L$, the sub-PRF must satisfy:
\begin{equation}
	\text{PRF}_{\text{sub}}=\text{PRF}/L>\Delta f_\text{dop},
\end{equation}
where $\Delta f_\text{dop}$ is the Doppler bandwidth. If the original sampling time related to the original PRF is denoted as $\{\eta_1,\eta_2,\cdots\}$, then in the $l$-th extracted measurement ($ l=1,\cdots L$), the $l$-th sample and every $n$-th sample after the $l$-th sample are retained. Consequently, the sampling time of the $l$-th extracted measurement is $\{\eta_l,\eta_{l+L},\cdots\}$.

Subsequently, classical focusing algorithms can be employed to generate full-resolution SLC images for each baseline per extracted measurement, and all $N\times L$ focused SLC images are then registrated with the reference measurement ($n=1,l=1$), followed by atmospheric and deformation phase error removal and tomographic signal deramping. For each fixed azimuth-range pixel $(x,r)$, the EMMV observation can be succinctly represented as $\mathbf{G}=\begin{bmatrix}
		\mathbf{g}_1,\mathbf{g}_2,\cdots,\mathbf{g}_L\end{bmatrix}\in\mathbb{C}^{N\times L}$. The corresponding sparse SAR tomography model can be expressed as follows:
\begin{equation}\label{MMVsparse}
	\mathbf{G} =\mathbf{G}^o+ \mathbf{E}=\mathbf{A}\bm{\Gamma}+ \mathbf{E},
\end{equation}
where $\mathbf{G}^o$ is the noiseless EMMV observation,
$[\bm{\Gamma}]_{k,l}\in \mathbb{C}$ represents the backscattering coefficient of the $k$-th target at the $l$-th measurement, and $\mathbf{E}\in \mathbb{C}^{N\times L}$ signifies noise. 
The schematic diagram of constructing EMMV is shown in \figref{MMVcon}.

Notably, the MMV construction method proposed in this paper ensures the full resolution of both azimuth and range.

\subsection{Atomic Norm Soft Thresholding for Partially Observed MMV}

\subsubsection{Atomic Norm}
We commence our exploration by introducing an atom for the fully observed MMV model, defined as
	$\mathbf{A}(s,\mathbf{b})=\mathbf{a}\left( s \right) \mathbf{b}^H \in \mathbb{C}^{N\times L}$,
where the complex vector $\mathbf{b}\in \mathbb{C}^L$ satisfies $\lVert \mathbf{b} \rVert_2=1$.
The atom set $\mathcal{A}$ is then defined as
	$\mathcal{A}=\left\{ \mathbf{A}(s,\mathbf{b}):s\in \left(-\frac{S_1}{2},\frac{S_1}{2}\right), \lVert \mathbf{b} \rVert_2=1\right\}$.
The atomic norm based on this atom set is formulated as:
\begin{equation}
	\begin{aligned}
		&\lVert \mathbf{G} \rVert _{\mathcal{A}} =\lVert \mathbf{G} \rVert_{\mathcal{A},1}
		\triangleq\inf\left\{ t>0:\mathbf{G}\in t \text{conv}(\mathcal{A})\right\}\\
		=&\inf\left\{ \sum_{k=1}^K{\left| \gamma_k \right|}:\sum_{k=1}^K{\gamma_k\mathbf{A}(s,\mathbf{b})=\mathbf{G},\mathbf{A}(s,\mathbf{b})\in \mathcal{A}} \right\} \\
		=&\inf_{\mathbf{V}\in\mathbb{C}^{L\times L},\mathbf{u}\in\mathbb{C}^{N}}
		\left\{ 
		\frac{\text{tr}\left(\mathbf{V}\right)}{2}+\frac{\text{tr}\left( \mathcal{T}\left( \mathbf{u} \right) \right)}{2N} :
		\begin{pmatrix}
			\mathcal{T}\left( \mathbf{u} \right) & \mathbf{G} \\
			\mathbf{G}^{H}                 & \mathbf{V}       \\
		\end{pmatrix} \succeq 0
		\right\},
		\end{aligned}
\end{equation}
where $\mathcal{T}( \mathbf{u} )\in\mathbb{C}^{N\times N}$ is a Hermitian Toeplitz matrix uniquely determined by its first column $\mathbf{u}\in\mathbb{C}^{N}$.
The proof of above equivalent SDP characterization can be found in~\cite{yang2016exact}.

In the case of incomplete data $\mathbf{G}_{\Omega} \in\mathbb{C}^{M\times L}$, we consider the atom $ \mathbf{A}_{\Omega}(s,\mathbf{b})=\mathbf{a}_{\Omega}\left( s \right) \mathbf{b}^H \in \mathbb{C}^{M\times L}$ and derive the atom set as
	$\mathcal{A}_{\Omega} =\left\{ \mathbf{A}_{\Omega}(s,\mathbf{b}):\mathbf{A}(s,\mathbf{b})\in \mathcal{A} \right\}$.
This leads to the definition of the atomic norm for the partially observed MMV model:
\begin{equation}
	\begin{aligned}
		&\lVert \mathbf{G}_{\Omega} \rVert_{\mathcal{A}_{\Omega}}=\lVert \mathbf{G}_{\Omega} \rVert_{\mathcal{A}_{\Omega},1}
		\triangleq\inf\left\{ t>0:\mathbf{G}_{\Omega}\in t \text{conv}(\mathcal{A}_{\Omega})\right\}\\
		=& \inf\left\{ \sum_{k=1}^K{\left| \gamma_k \right|}:\sum_{k=1}^K{\gamma_k\mathbf{A}_{\Omega} (s,\mathbf{b})=\mathbf{G}_{\Omega} ,\mathbf{A}_{\Omega} (s,\mathbf{b})\in \mathcal{A}_{\Omega}} \right\}.
	\end{aligned}
\end{equation}
Denoting the complement of $\Omega$ in $\{1, 2, \cdots, N\}$ by $\bar{\Omega}$ yields the following relationship between $\lVert \cdot \rVert_\mathcal{A}$ and $\lVert \cdot \rVert_{\mathcal{A}_{\Omega}}$:
\begin{equation}\label{eq:AandAo}
	\begin{aligned}
		&\lVert \mathbf{G}_{\Omega} \rVert_{\mathcal{A}_{\Omega}}\\
		=& \inf\left\{ \sum_{k=1}^K{\left| \gamma_k \right|}:\sum_{k=1}^K{\gamma_k\mathbf{A}_{\Omega} (s,\mathbf{b})=\mathbf{G}_{\Omega} ,\mathbf{A}_{\Omega} (s,\mathbf{b})\in \mathcal{A}_{\Omega}} \right\}\\
		=&\min_{\mathbf{G}_{\bar{\Omega}}}\inf\left\{ \sum_{k=1}^K{\left| \gamma_k \right|}:\sum_{k=1}^K{\gamma_k\mathbf{A}(s,\mathbf{b})=\mathbf{G},\mathbf{A}(s,\mathbf{b})\in \mathcal{A}} \right\}\\
		=&\min_{\mathbf{G}_{\bar{\Omega}}} \lVert \mathbf{G} \rVert_{\mathcal{A}}.
	\end{aligned}
\end{equation}

\subsubsection{Optimization via SDP}
Supposing the observations contaminated with i.i.d. AWGN, the atomic norm denoising problem for partially observed MMV can be formulated as:
\begin{equation}\label{eq:MPAST1}
	\begin{aligned}
	&\min_{ \hat{\mathbf{G}}_{\Omega}} \quad\frac{1}{2}\lVert \hat{\mathbf{G}}_{\Omega}- \mathbf{G}_{\Omega}\rVert_F^2+ \tau\lVert \hat{\mathbf{G}}_{\Omega} \rVert _{\mathcal{A}_{\Omega}}\\
	=&\min_{ \hat{\mathbf{G}}} \quad\frac{1}{2}\lVert \hat{\mathbf{G}}_{\Omega}- \mathbf{G}_{\Omega}\rVert_F^2+ \tau\lVert \hat{\mathbf{G}} \rVert _{\mathcal{A}},
	\end{aligned}
\end{equation}
where $\hat{\mathbf{G}}$ is the recovered denoised fully observed MMV signal.
Consequently, this optimization problem can be recast as the following SDP:
\begin{equation}\label{mpast}
	\begin{aligned}
		\min_{\hat{\mathbf{G}},\mathbf{V}, \mathbf{u}}\quad& \frac{1}{2}\lVert \hat{\mathbf{G}}_{\Omega}- \mathbf{G}_{\Omega}\rVert_F^2+ \frac{\tau}{2}\left(\text{tr}\left(\mathbf{V}\right)+\frac{1}{N}\text{tr}\left( \mathcal{T}\left( \mathbf{u} \right) \right)\right)\\
		\text{s.t.~~}            \quad
		& \begin{pmatrix}
				\mathcal{T}\left( \mathbf{u} \right) & \hat{\mathbf{G}} \\
				\hat{\mathbf{G}}^{H}                 & \mathbf{V}       \\
			\end{pmatrix} \succeq 0,
	\end{aligned}
\end{equation}
where $\tau$ is a factor related to the noise level.
The elevation locations of targets are embedded within $\mathcal{T}( \mathbf{u} )$ and the estimates can be obtained via Vandermonde decomposition.

\subsubsection{Implementation Using ADMM}
Off-the-shelf solvers, such as SeDuMi~\cite{SeDuMi} and SDPT3~\cite{SDPT3}, are commonly employed for solving SDP problems. However, their efficiency tends to degrade when tackling large-scale problems. To address this challenge, we present a reasonably efficient algorithm based on ADMM~\cite{ADMM}. In the initial step, we introduce auxiliary variable $\mathbf{U}$ and reformulate the SDP problem \eqref{mpast} as follows:
\begin{equation}\label{mpast-admm}
	\begin{aligned}
		\min_{\hat{\mathbf{G}},\mathbf{V}, \mathbf{u},\mathbf{U}\succeq 0}\quad& \frac{1}{2}\lVert \hat{\mathbf{G}}_{\Omega}- \mathbf{G}_{\Omega}\rVert_F^2+ \frac{\tau}{2}\left(\text{tr}\left(\mathbf{V}\right)+\frac{1}{N}\text{tr}\left( \mathcal{T}\left( \mathbf{u} \right) \right)\right)\\
		\text{s.t.~~~}            \quad
		&\mathbf{U}=\left( \begin{matrix}
				\mathcal{T}\left( \mathbf{u} \right) & \hat{\mathbf{G}} \\
				\hat{\mathbf{G}}^{H}                 & \mathbf{V}       \\
			\end{matrix}\right).
	\end{aligned}
\end{equation}

Subsequently, the corresponding augmented Lagrangian function in scaled form is defined as:
\begin{equation}\label{mpast-admm}
	\begin{aligned}
		&\mathcal{L}(\hat{\mathbf{G}},\mathbf{V},\mathbf{u},\mathbf{U},\mathbf{\Lambda})\\
		=&\frac{1}{2}\lVert \hat{\mathbf{G}}_{\Omega}- \mathbf{G}_{\Omega}\rVert_F^2+\frac{\tau}{2}\left(\text{tr}\left(\mathbf{V}\right)+\frac{1}{N}\text{tr}\left( \mathcal{T}\left( \mathbf{u} \right) \right)\right)\\
		&+\frac{\eta}{2}\lVert \mathbf{U}-\begin{pmatrix}
				\mathcal{T}\left( \mathbf{u} \right) & \hat{\mathbf{G}} \\
				\hat{\mathbf{G}}^{H}                 & \mathbf{V}       \\
			\end{pmatrix}+\eta^{-1}\mathbf{\Lambda}\rVert_F^2-\frac{1}{2\eta}\lVert\mathbf{\Lambda}\rVert_F^2+\delta_\text{PSD}(\mathbf{U}),
	\end{aligned}
\end{equation}
where $\mathbf{\Lambda}$ is the Lagrangian multiplier, $\eta>0$ is a penalty parameter, and $\delta_\text{PSD}(\cdot)$ is the indicator function: $\delta_\text{PSD}(\mathbf{U})=0$ if $\mathbf{U}\succeq 0$ or $+\infty$ otherwise. The implementation of ADMM involves the following iterative steps, which lead to convergence towards the optimal solution of \eqref{mpast}.
\begin{enumerate}[Step 1.]
	\item Update $\hat{\mathbf{G}}$, $\mathbf{V}$ and $\mathbf{u}$:
	\begin{align}
&\hat{\mathbf{G}}^{k+1}=(2\eta\mathbf{I}+\mathbf{P}_{\Omega}^H\mathbf{P}_{\Omega})^{-1}(2\eta\mathbf{U}^k_{\mathbf{G}}+2\mathbf{\Lambda}^k_{\mathbf{G}}+\mathbf{P}_{\Omega}^H\mathbf{G}_{\Omega}),\\
&\hat{\mathbf{V}}^{k+1}=\mathbf{U}^k_{\mathbf{V}}+\frac{1}{\eta}\mathbf{\Lambda}^k_{\mathbf{V}}-\frac{\tau}{2\eta}\mathbf{I},\\
&\hat{\mathbf{u}}^{k+1}=\mathcal{T}^{-1}\left( \mathcal{P}_{\mathcal{T}}\left(\mathbf{U}^k_{\mathcal{T}}+\frac{1}{\eta}\mathbf{\Lambda}^k_{\mathcal{T}}-\frac{\tau}{2\eta N}\mathbf{I}\right)\right);
\end{align}

	\item Update $\mathbf{U}$:
	      \begin{equation}
		      \mathbf{U}^{k+1} = \mathcal{P}_{\text{PSD}}\left(\begin{pmatrix}
		      	\mathcal{T}\left( \mathbf{u}^{k+1} \right) & \hat{\mathbf{G}}^{k+1} \\
		      	(\hat{\mathbf{G}}^{k+1})^{H}                 & \mathbf{V}^{k+1}       \\
		      \end{pmatrix}-\frac{1}{\eta}\mathbf{\Lambda}^k\right);
	      \end{equation}
	\item Update $\mathbf{\Lambda}$:
	      \begin{equation}
		      \mathbf{\Lambda}^{k+1} = \mathbf{\Lambda}^k + \eta \left(\mathbf{U}^{k+1} - \left( \begin{matrix} \mathcal{T}\left( \mathbf{u}^{k+1} \right) & \hat{\mathbf{G}}^{k+1} \\ (\hat{\mathbf{G}}^{k+1})^H & \mathbf{V}^{k+1} \end{matrix} \right)\right).
	      \end{equation}
\end{enumerate}

Here, $\mathbf{P}_{\Omega}$ is the sub-sampling matrix with $\Omega$ being the set of indices of observed entries.
The $\mathcal{P}_{\mathcal{T}}(\cdot)$ operator projects onto the nearest Hermitian Toeplitz matrix~\cite{Toeplitz}, i.e., the first column $\mathbf{u}$ of $ \mathcal{P}_{\mathcal{T}}(\mathbf{T})$ satisfying $[\mathbf{u}]_n = \frac{1}{2(N-n+1)}\sum_{j_1-j_2=n-1}\left([\mathbf{T}]_{j_1,j_2}+[\mathbf{T}]^*_{j_2,j_1}\right)$.
The $\mathcal{P}_{\text{PSD}}(\cdot)$ operator projects onto the nearest positive semi-definite (PSD) matrix through eigenvalue decomposition, preserving non-negative eigenvalues and their associated eigenvectors.
For simplicity in formulas, we present matrices $\mathbf{U}^{k}$ and $\mathbf{\Lambda}^{k}$ in chunks as $\begin{pmatrix}\mathbf{U}^k_{\mathcal{T}}&\mathbf{U}^k_{\mathbf{G}}\\(\mathbf{U}^k_{\mathbf{G}})^H&\mathbf{U}^k_{\mathbf{V}}\end{pmatrix}$ and $\begin{pmatrix}\mathbf{\Lambda}^k_{\mathcal{T}}&\mathbf{\Lambda}^k_{\mathbf{G}}\\(\mathbf{\Lambda}^k_{\mathbf{G}})^H&\mathbf{\Lambda}^k_{\mathbf{V}}\end{pmatrix}$ respectively. 

\subsubsection{Choosing the regularization parameter}\label{sec:tau}
The selection of regularization parameter $\tau$ is determined by the characteristics of noise model. Assume that each element in the matrix $\mathbf{E}_{\Omega}$ follows an i.i.d. Gaussian distribution with zero mean and variance $\sigma$.
According to~\cite{AST}, we can choose $\tau\geq \mathbb{E}\left[\lVert \mathbf{E}_{\Omega} \rVert_{\mathcal{A}_{\Omega}}^*\right]$ and then, the expected (per-element) mean squared error of the estimate $\hat{\mathbf{G}}_{\Omega}$ obtained from the solution of the optimization problem \eqref{eq:MPAST1} is bounded as:
\begin{equation}
	\label{mse}
	\frac{1}{ML} \mathbb{E}\left[\lVert \hat{\mathbf{G}}_{\Omega}- \mathbf{G}_{\Omega}^o \rVert_F^2    \right]\leq \frac{\tau}{ML} \lVert \mathbf{G}_{\Omega}^o \rVert_{\mathcal{A}_{\Omega}},
\end{equation}
where $\|\cdot\|_{\mathcal{A}_\Omega}^*$ is the dual norm of $\|\cdot\|_{\mathcal{A}_\Omega}$ and $\mathbf{G}_{\Omega}^o$ is the original noiseless signal.
Consequently, by computing an upper bound on $\mathbb{E}\left[\lVert \mathbf{E}_{\Omega} \rVert_{\mathcal{A}_{\Omega}}^*\right]$, we let
\begin{equation}
	\label{tau}
	\tau=\frac{8\sqrt{\sigma M}}{7-8p^{-1}}\sqrt{2L\log17+\log\left(\pi Np+1\right)+1},
\end{equation}
where $p=4L\log(6L+\log N)$. The detailed derivations are deferred to Appendix.

\begin{figure}[htbp]
	\centering
	\includegraphics[width=\linewidth]{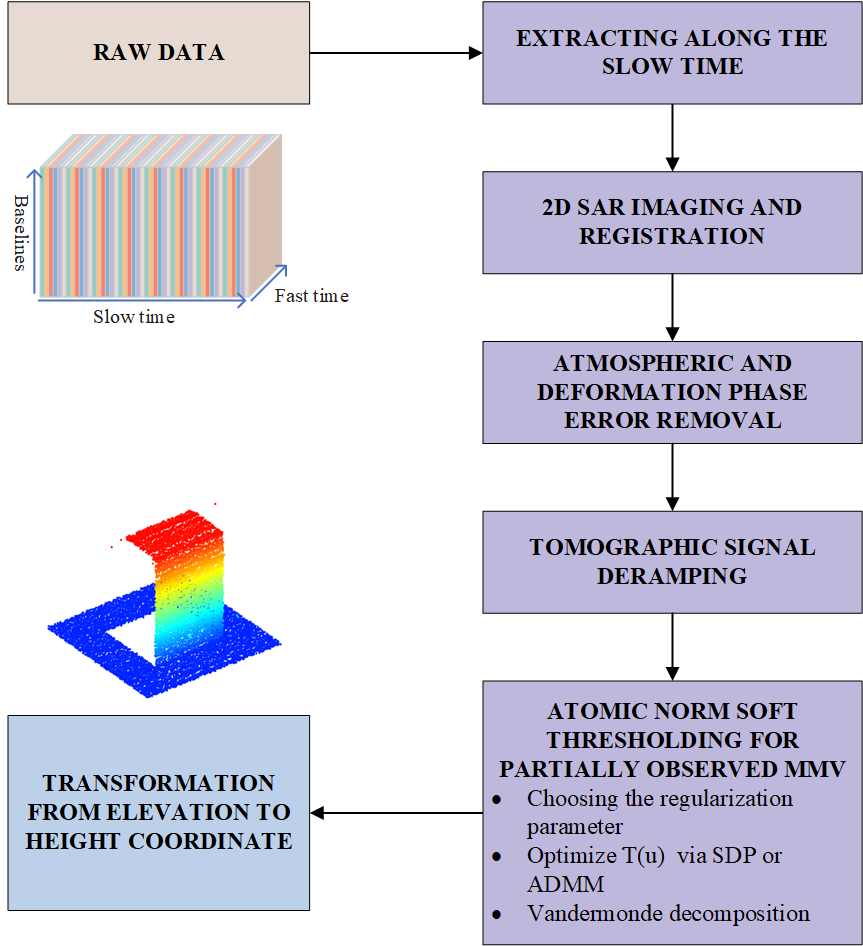}
	\caption{The proposed EMPAST procedure.}
	\label{workflow}
\end{figure}

Through the aforementioned steps, we can determine the elevation positions of targets on each azimuth-range pixel. The complete 3-D point clouds imaging workflow based on our proposed EMPAST framework is illustrated in \figref{workflow}.

\section{Numerical Results With Simulated Data}\label{sec:Simu}

\subsection{Experimental Settings}
To validate the effectiveness of the proposed framework, we use simulated data to compare EMPAST with other state-of-the-art TomoSAR imaging methods, including SVD-Wiener~\cite{SVD}, GBCS~\cite{TomoSAR3}, and PAST~\cite{ASTYang}.
Radar system parameters employed for these experiments are detailed in \tabref{simuPara}.
Typically, the oversampling rate is taken as 1.1 - 1.4 to alleviate azimuth ambiguity, so in our experiments we constrain the number of EMMV $L\leq 8$. Moreover, we set the grid interval to $\rho_s/8$ in GBCS methods. For each Monte Carlo simulation, we construct baselines of partial observations by randomly selecting $M$ array elements from the uniform array.
\begin{table}[htbp]
	\begin{center}
		\caption{Simulated System Parameters.}
		\label{simuPara}
		\begin{tabular}{ l  l  l }
			\toprule
			Name                                & Symbol                & Value    \\
			\midrule
			Carrier frequency                   & $f_c$                 & 15.2 GHz \\
			Pulse repetition frequency          & $\text{PRF}$          & 1.0 kHz  \\
			Doppler bandwidth                   & $\Delta f_\text{dop}$ & 100.0 Hz \\
			Reference distance                  & $r_0$                 & 500.0 m  \\
			Maximal elevation aperture          & $D$                   & 1.1 m    \\
			Number of uniform array elements    & $N$                   & 12       \\
			Number of array elements            & $M$          & 8        \\
			Rayleigh resolution along elevation & $\rho_s$              & 4.5 m    \\
			\bottomrule
		\end{tabular}
	\end{center}
\end{table}

To address the innovative points of the proposed framework, we quantitatively analyze the reconstruction performance from four aspects: accuracy and robustness, super-resolution capability, 3-D point clouds reconstruction performance and efficiency.

\subsection{Accuracy and Robustness}
\begin{figure}[htbp]
	\centering
	\subfloat[]{\label{simuOnetimeon}\includegraphics[width=\linewidth]{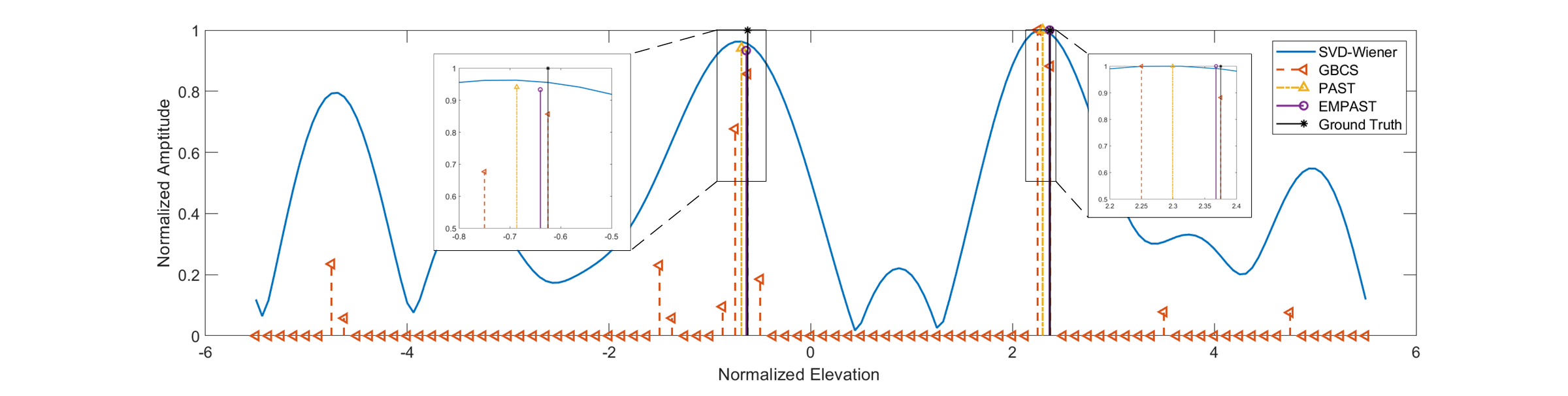}}
	\vspace {0em} 
	\subfloat[]{\label{simuOnetimeoff}\includegraphics[width=\linewidth]{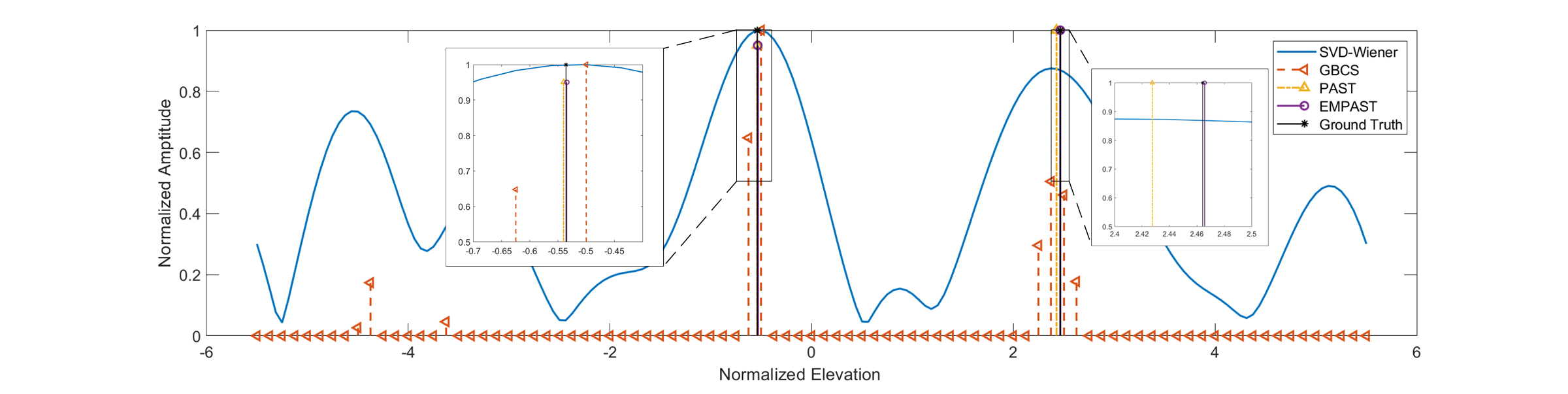}}
	\caption{Reconstruction results for double well-separated scatterers with $\Delta s = 3\rho_s$ using SVD-Wiener, GBCS, PAST and EMPAST ($L = 8$) under SNR = 10 dB. (a) On-grid scatterers. (b) Off-grid scatterers. The regions surrounding the ground truth are zoomed in for better illustration.}
	\label{simuOnetime}
\end{figure}

We initiate our experimental exploration with scenarios featuring double well-separated scatterers, aiming to provide a visual representation of the elevation reconstruction results produced by distinct methods. Specifically, we simulate a scenario with double scatterers exhibiting equal amplitudes and randomized phases of backscattering coefficients, the elevation distance set to $\Delta s = 3\rho_s$, and SNR set at 10 dB. 
Comparisons of elevation reconstruction results of different TomoSAR imaging methods are shown in \figref{simuOnetime} with two scatterers set to be on-grid and off-grid, respectively.
In this figure, the black solid lines denote the simulated elevations, while the blue solid line corresponds to SVD-Wiener, the red dashed lines to GBCS, the yellow dashed lines to PAST, and the purple solid lines represent our proposed EMPAST. For better illustration, the regions surrounding the ground truth are magnified within the delineated black boxes.

In \figref{simuOnetimeon}, the experimental results reveal that for on-grid targets, all four methods successfully reconstruct them in close proximity to simulated locations, demonstrating accurate estimates of backscattering coefficients. Notably, for off-grid targets in \figref{simuOnetimeoff}, the reconstruction performance of SVD-Wiener, PAST, and EMPAST remains consistently robust, with EMPAST providing the closest estimates to the ground truth, followed by PAST and then SVD-Wiener. Conversely, GBCS exhibits susceptibility to the ``off-grid'' effect, manifesting as the energy of a single scatterer spreading across multiple grids in the vicinity. 
This results in a notable disparity between the estimated and simulated backscattering coefficients, increasing the risk of false detection and leaking detection. The inversion results highlight the efficacy of atomic norm based methods in enhancing sparsity and ensuring the accurate localization of scatterers, with the proposed EMPAST standing out particularly.

To further quantitatively assess the accuracy and robustness of the proposed framework, Monte Carlo simulations are performed.
For evaluating accuracy of elevation reconstruction, we utilize the normalized root mean squared error of elevation estimation, denoted as $\sigma_s$, which gauges the disparity between the estimated elevation and the ground truth.
Additionally, we gauge robustness of elevation reconstruction using the probability of detection, denoted as $P_D$. The computation of these two evaluation metrics is outlined below:
\begin{equation}
	\sigma_s = \frac{1}{\rho_s}\sqrt{\frac{1}{N_{\text{mc}}}\sum_{n_{\text{mc}}=1}^{N_{\text{mc}}}\frac{1}{K}\sum_{k=1}^{K}(\hat{s}_{n_{\text{mc}},k}-s_{n_{\text{mc}},k})^2},
\end{equation}
\begin{equation}\label{sr}
	P_D= P(|\hat{s}_{n_{\text{mc}},k}-s_{n_{\text{mc}},k}|\leq T\text{ for all } k),
\end{equation}
where $N_{\text{mc}}$ denotes the number of Monte Carlo simulations and $K$ denotes the number of targets. 
$s_{n_{\text{mc}},k}$ is the ground-truth elevation of the $k$-th target during the $n_{\text{mc}}$-th simulation, and $\hat{s}_{n_{\text{mc}},k}$ is the corresponding estimated elevation.
The notation $P(\cdot)$ denotes the probability of satisfying the condition in parentheses. Threshold $T$ is a small constant and is set to $T=\rho_s/8$ in the following experiments.

\begin{figure}[htbp]
	\centering
	\subfloat[]{\includegraphics[width=0.7\linewidth]{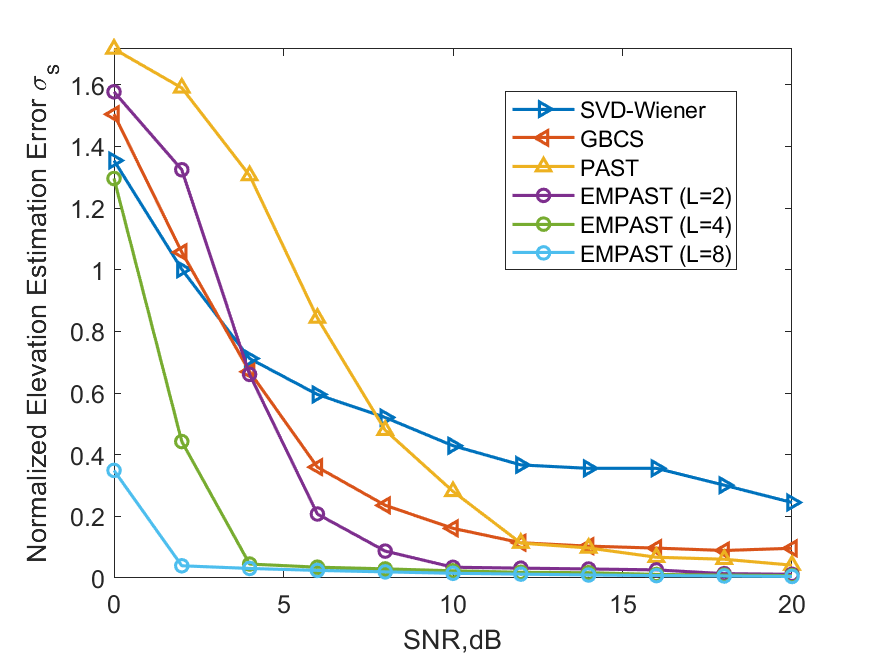}}
	\vspace{0em}
	\subfloat[]{\includegraphics[width=0.7\linewidth]{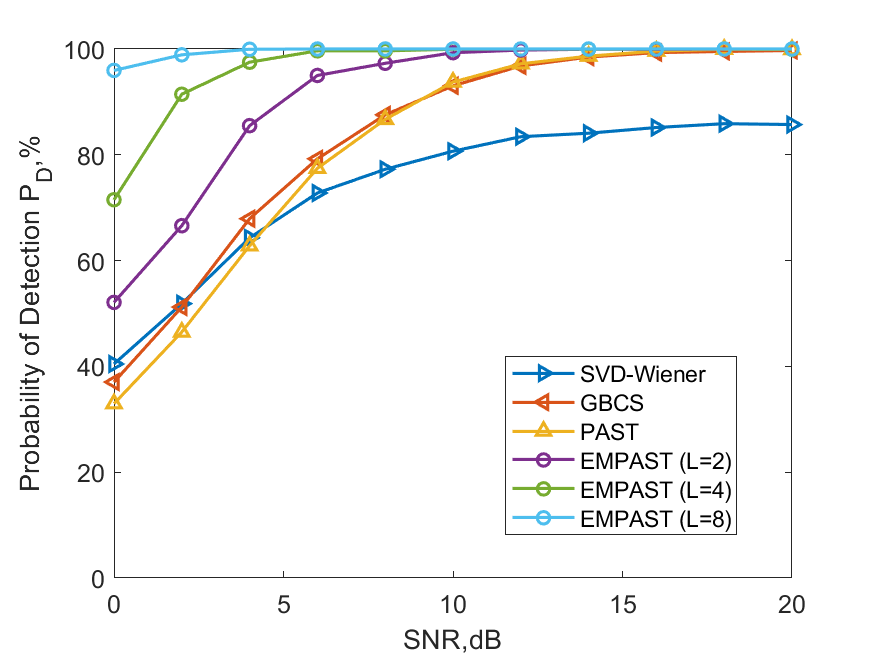}}
	\caption{Reconstruction performance comparison for a single scatterer or double well-seperated scatterers using SVD-Wiener, GBCS, PAST and EMPAST as a function of SNR. (a) Normalized elevation estimation error $\sigma_s$. (b) Probability of detection $P_D$.}
	\label{simuSingle}
\end{figure}

In this simulation, we compare the reconstruction performance of our proposed EMPAST with the SVD-Wiener, GBCS, and PAST methods through 2000 Monte Carlo simulations with the number of scatterers $K$ randomly set to either 1 or 2 at each SNR $\in [0,20]$ dB. When $K = 1$, it is assumed that the elevation location and backscattering coefficient of the scatterer both are randomized. 
When $K = 2$, assume that elevation locations of double targets are randomly distributed between $-\frac{S_1}{2}$ and $\frac{S_1}{2}$ with elevation distance $\Delta s> \frac{4}{N-1}S_1$. Scatterers are with equal amplitudes for backscattering coefficients, and random phases ranging from $0$ to $2\pi$. The normalized elevation estimation error $\sigma_s$ and probability of detection $P_D$ versus SNR are shown in \figref{simuSingle}.

From the experimental results, it can be observed that with an increase in SNR, the normalized elevation estimation error $\sigma_s$ of all methods undergoes a rapid descent, followed by stabilization. Concurrently, the probability of detection $P_D$ of all methods also exhibits an increasing trend. 
Across diverse SNR, EMPAST ($L\geq 4$) consistently outperforms other methods. 
Notably, compared to the gridless method PAST, EMPAST demonstrates robust noise resilience. Moreover, the reconstruction performance of EMPAST demonstrates improvement with an augmented number of EMMV $L$.
In contrast to GBCS, EMPAST is immune to the ``off-grid" effect. When SNR exceeds 5 dB, $\sigma_s$ of EMPAST consistently remains markedly smaller than those of PAST, GBCS, and SVD-Wiener. 
Specifically, for $L=8$, EMPAST maintains $\sigma_s$ consistently below 0.4, with $P_D$ exceeding 90\%. These experimental findings validate the efficacy of the proposed EMPAST framework in terms of accuracy and robustness.

\subsection{Super-resolution Capability}
In this section, we perform a series of experiments to evaluate the performance of the proposed EMPAST framework for reconstructing closely-spaced double scatterers, and compare it with other state-of-the-art algorithms such as SVD-Wiener, GBCS and PAST.

\subsubsection{Spacing Sensitivity}
\begin{figure*}[htbp]
	\centering
	\subfloat[]{\includegraphics[width=0.25\linewidth]{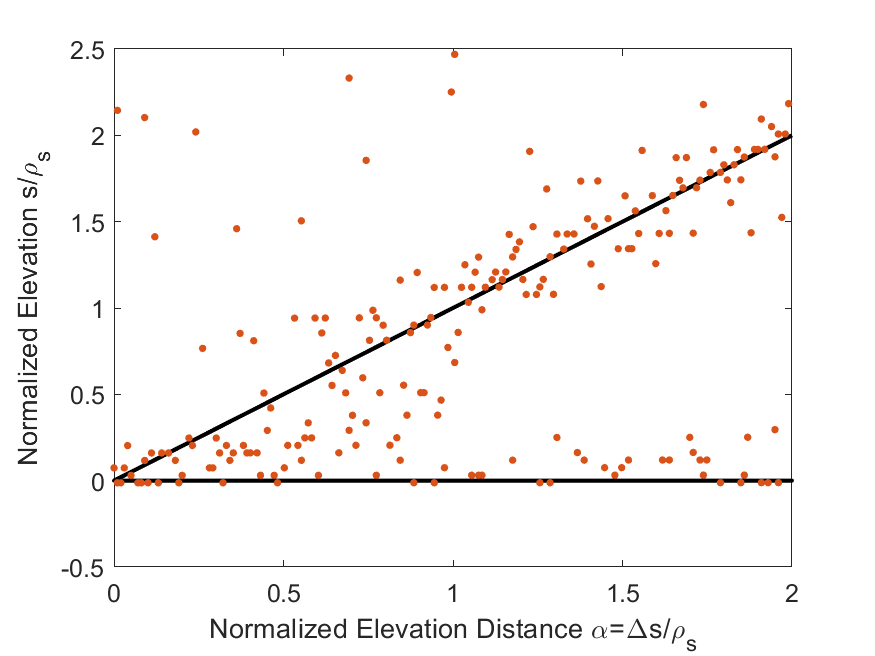}}
	\subfloat[]{\includegraphics[width=0.25\linewidth]{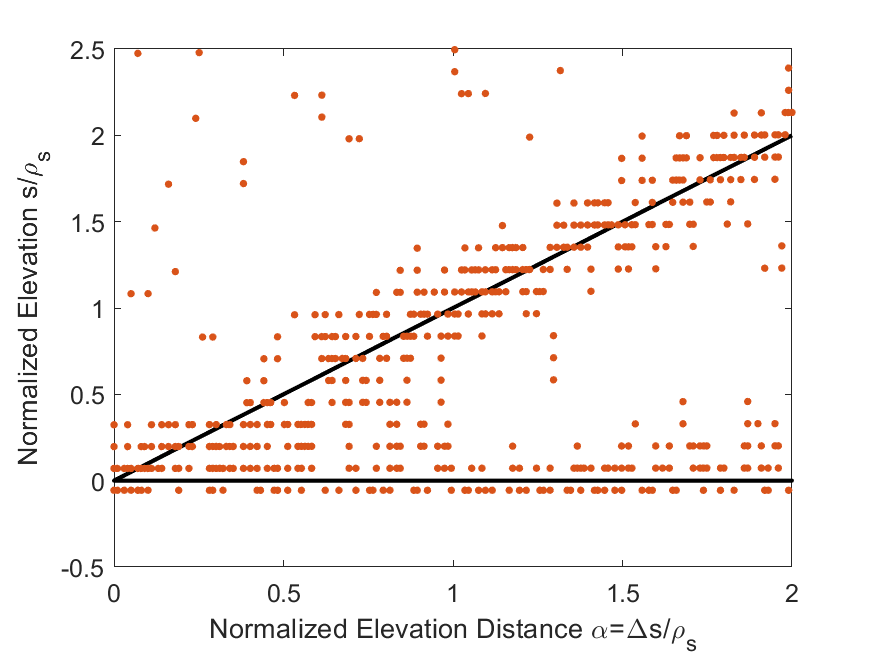}}
	\subfloat[]{\includegraphics[width=0.25\linewidth]{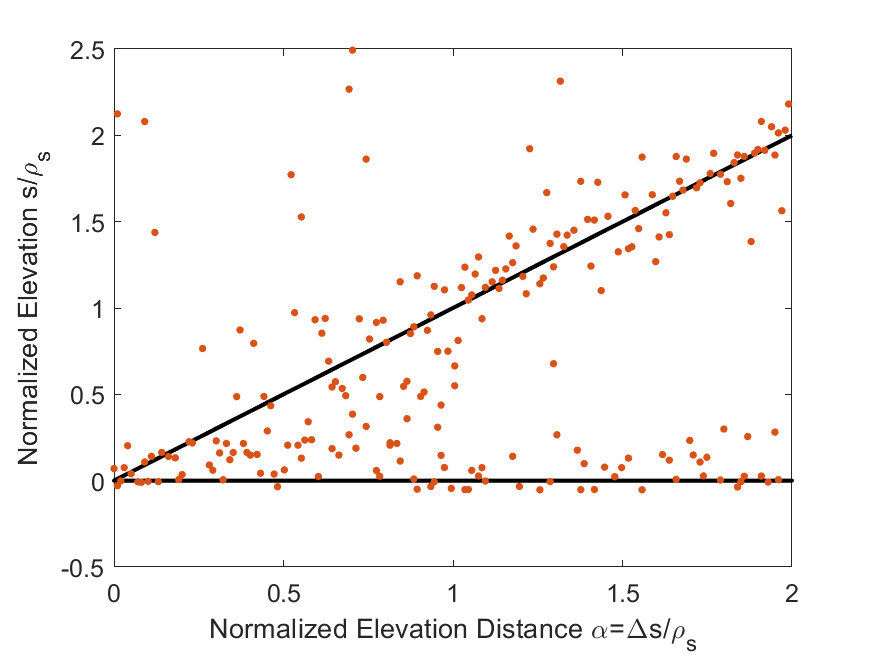}}
	\subfloat[]{\includegraphics[width=0.25\linewidth]{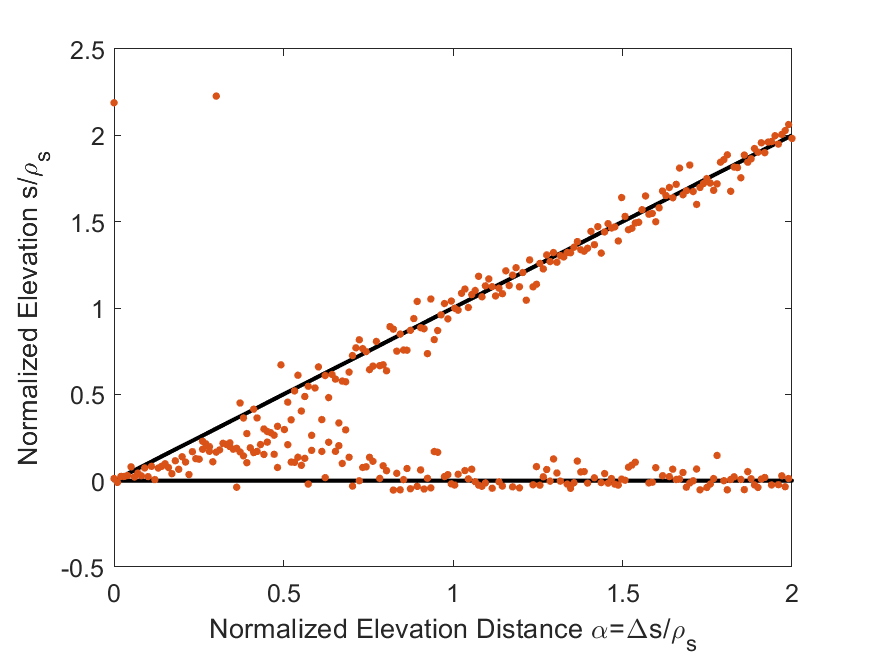}}
	\vspace{-1em}
	\subfloat[]{\includegraphics[width=0.25\linewidth]{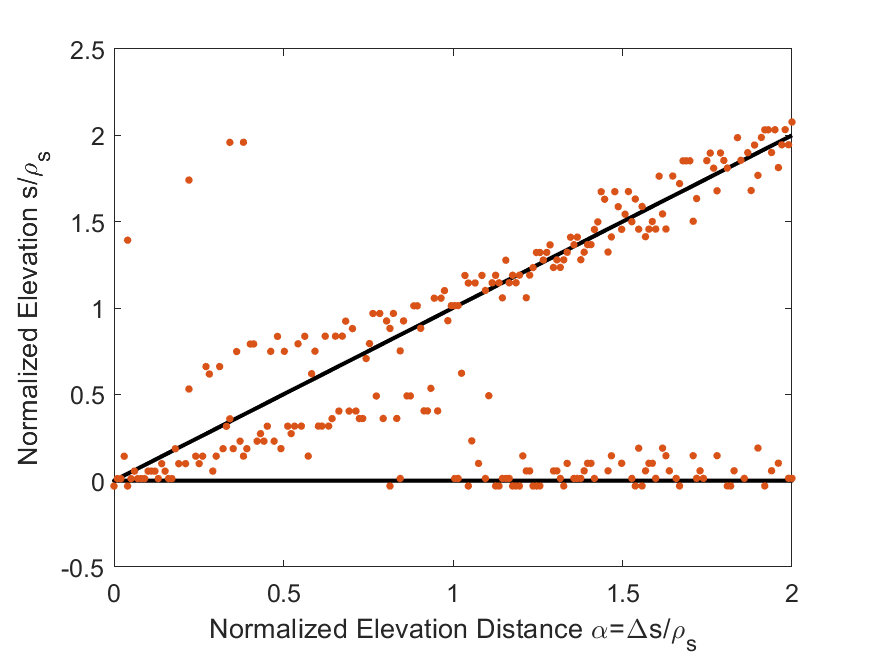}}
	\subfloat[]{\includegraphics[width=0.25\linewidth]{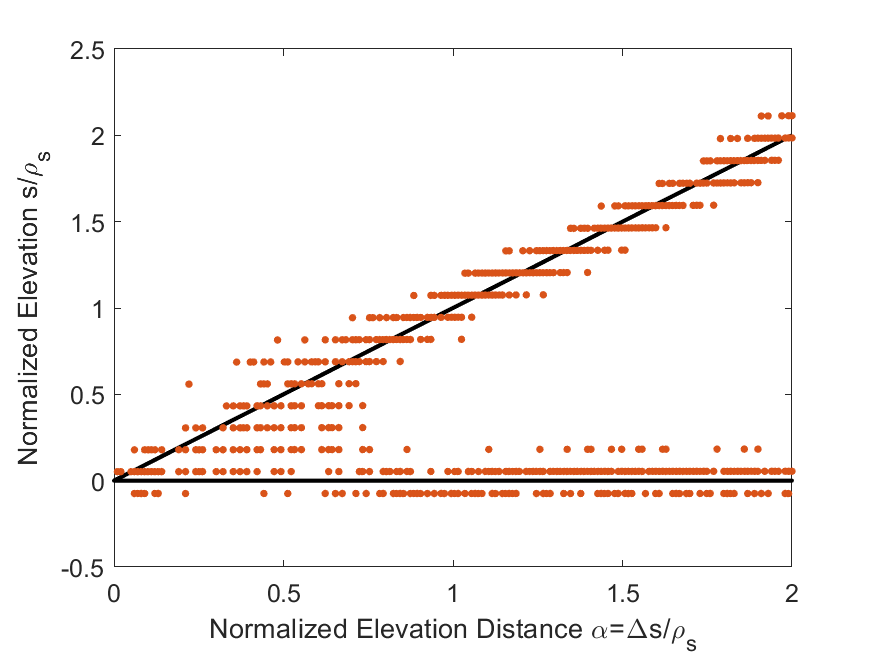}}
	\subfloat[]{\includegraphics[width=0.25\linewidth]{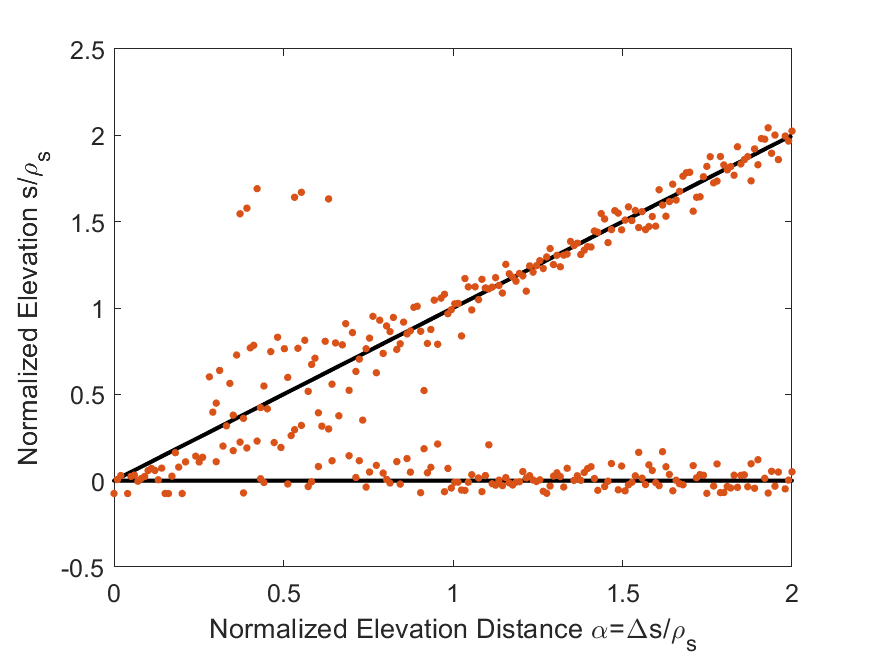}}
	\subfloat[]{\includegraphics[width=0.25\linewidth]{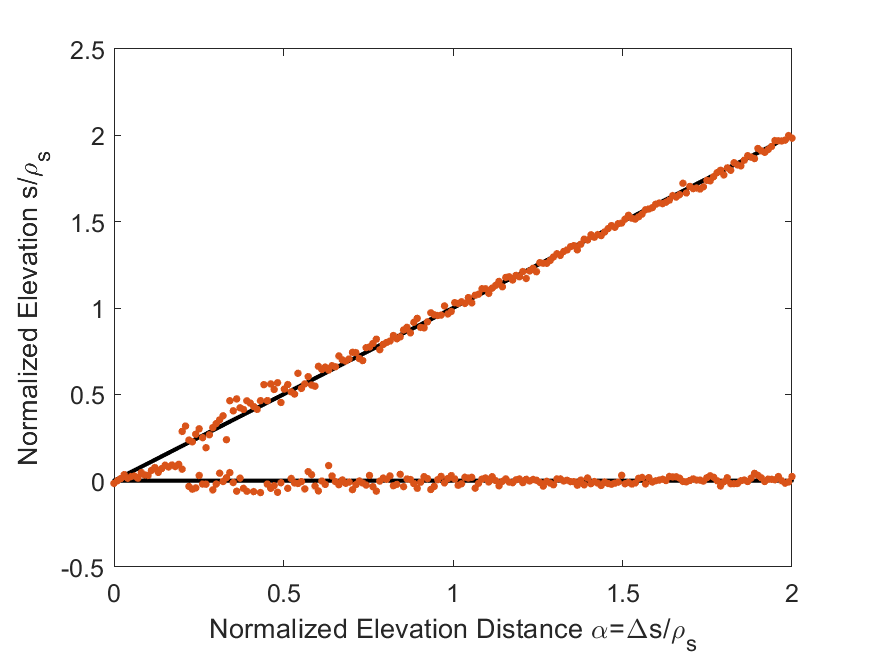}}
	\vspace{-1em}
	\subfloat[]{\includegraphics[width=0.25\linewidth]{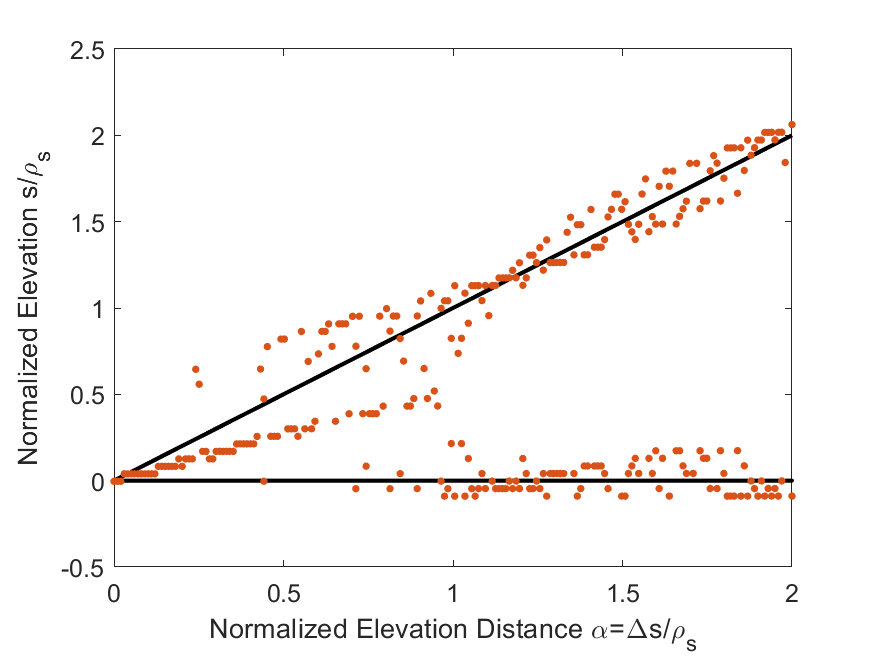}}
	\subfloat[]{\includegraphics[width=0.25\linewidth]{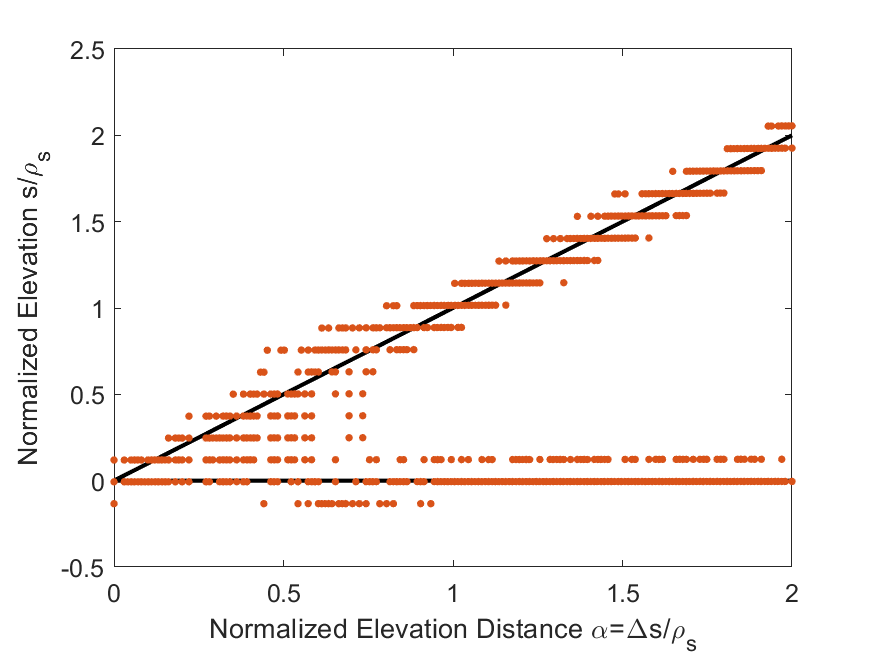}}
	\subfloat[]{\includegraphics[width=0.25\linewidth]{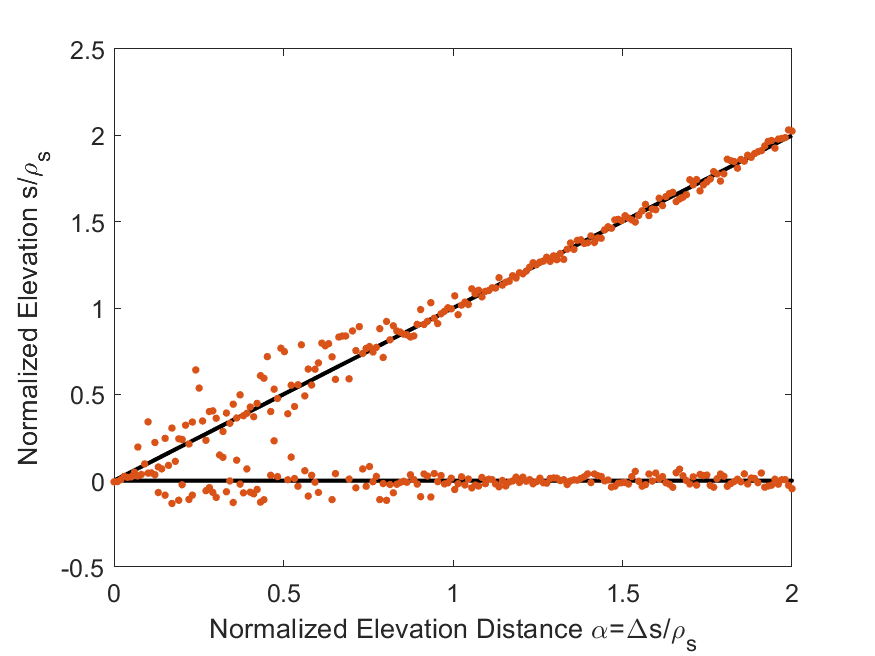}}
	\subfloat[]{\includegraphics[width=0.25\linewidth]{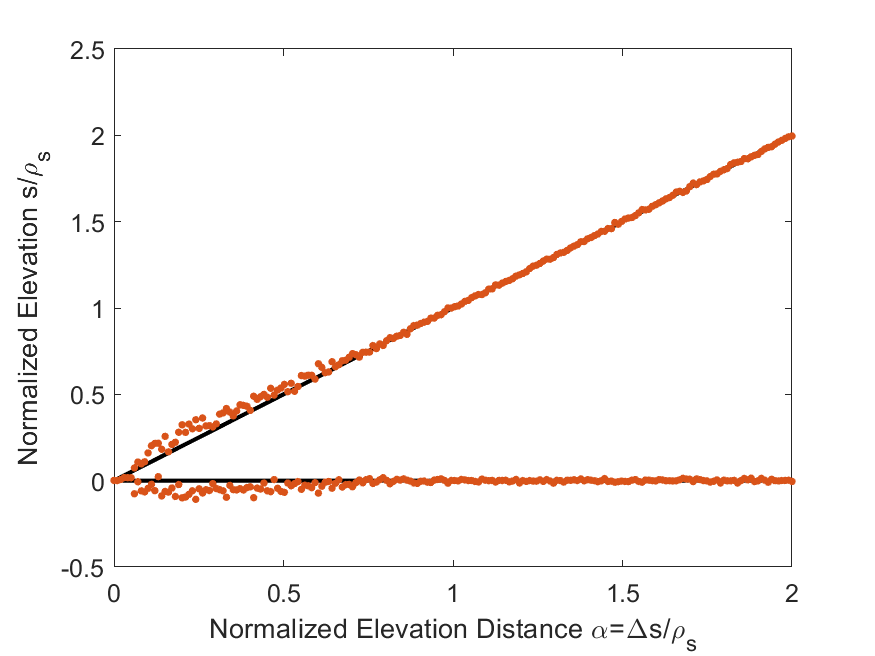}}
	\caption{Reconstruction results for double scatterers with (Top row) SNR = 0 dB, (Middle row) SNR = 10 dB, (Bottom row) SNR = 20 dB.
		(a), (e) and (i) SVD-Wiener. (b), (f) and (j) GBCS. (c), (g) and (k) PAST. (d), (h) and (l) EMPAST ($L = 8$).  Black: simulated normalized elevation. Red: estimated normalized elevation.}
	\label{simu1}
\end{figure*}

In this simulation, we compare the reconstruction results of our proposed EMPAST with the SVD-Wiener, GBCS, and PAST methods at a fixed SNR for double scatterers with different elevation spacings, aiming to gain a preliminary understanding of the sensitivity to scatterers spacing.
We consider scenarios of $K = 2$ scatterers with varying elevation distance $\Delta s \in (0,2\rho_s]$ at SNR = $\{0, 10, 20\}$ dB. Similar to the previous simulation, targets remain at the equal amplitudes and random phases for backscattering coefficients.
The simulation results are presented in \figref{simu1}. The black solid line represents the true location of targets, while the red solid dots depict the reconstructed normalized elevation. Each column corresponds to a specific method, and the rows present results for varying SNR: the top row represents SNR = 0 dB, the middle row 10 dB, and the bottom row 20 dB.

Comparing the results in the first row of \figref{simu1}, we can clearly observe that the proposed EMPAST reconstructs the highest point density and is closest to the ground truth. As SNR increases, the reconstructed points from all four methods converge towards the ground truth. When SNR reaches 20 dB, due to algorithm limitations, SVD-Wiener and GBCS struggle to stably separate double scatterers closer than $\rho_s$, the former due to the lack of super-resolution capability and the latter affected by the ``off-grid'' effect. In contrast, both PAST and EMPAST exhibit excellent super-resolution capability, and at the same noise level, EMPAST achieves a finer minimum resolvable separation than PAST and is closer to the ground truth.

\subsubsection{Monte Carlo Simulations}
\begin{figure}[htbp]
	\centering
	\subfloat[]{\includegraphics[width=0.7\linewidth]{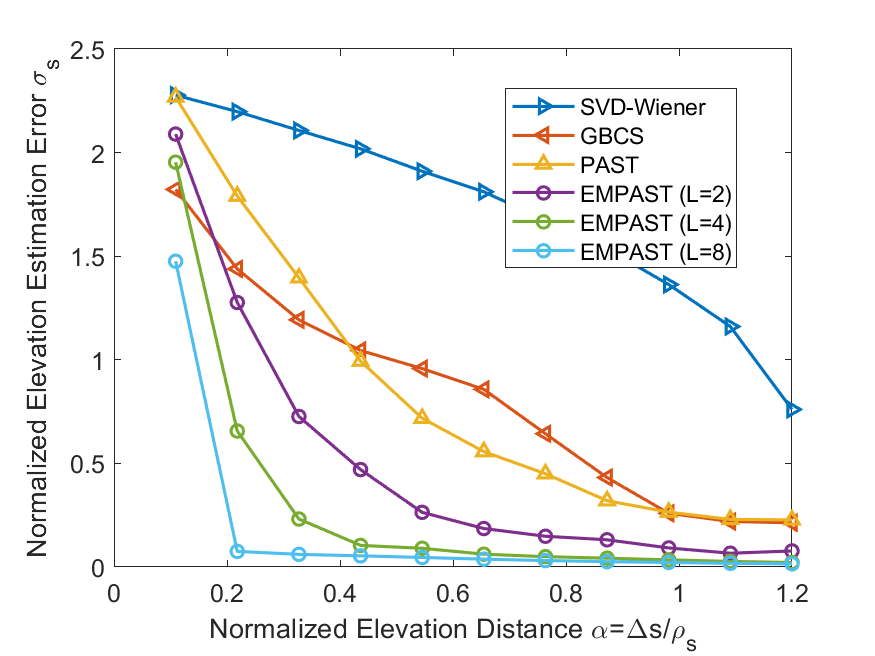}}
	\vspace{0em}
	\subfloat[]{\includegraphics[width=0.7\linewidth]{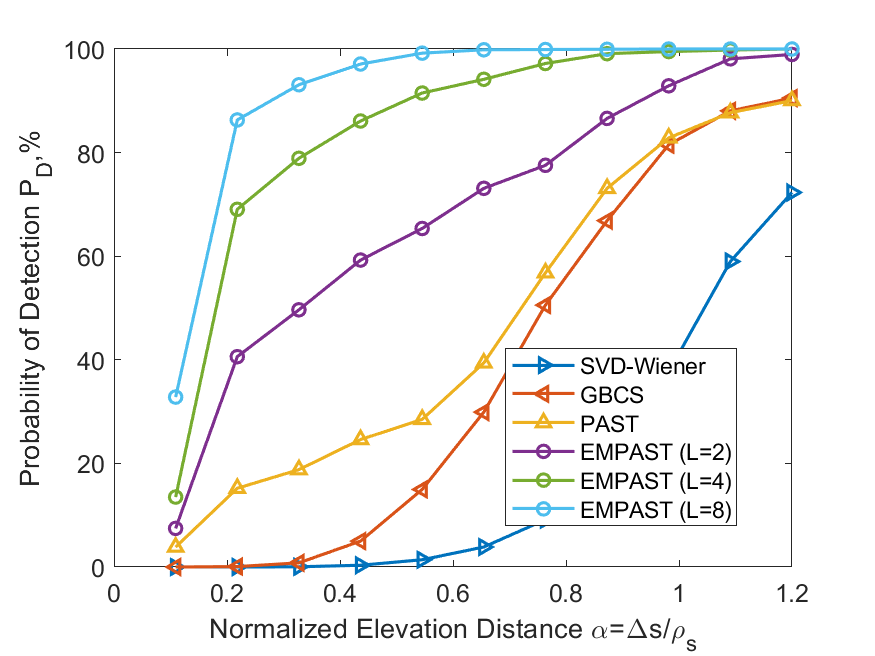}}
	\caption{Reconstruction performance comparison for double scatterers using SVD-Wiener, GBCS, PAST and EMPAST as a function of normalized elevation distance $\alpha=\Delta s/\rho_s$ under SNR = 10 dB. (a) Normalized elevation estimation error $\sigma_s$. (b) Probability of detection $P_D$.}
	\label{simuLoc}
\end{figure}

In this simulation, 2000 Monte Carlo experiments are conducted at a fixed SNR to gain insight into the reliability and stability of EMPAST. Statistical analyses of its reconstruction performance for double scatterers over multiple experiments are performed, allowing for a comparative evaluation with other algorithms.
We vary the elevation distance $\Delta s \in (0,1.2\rho_s]$ , with SNR = 10 dB and the backscattering coefficients of scatterers equal in amplitude and random in phase. The results of reconstruction performance comparison of SVD-Wiener, GBCS, PAST and EMPAST are shown in \figref{simuLoc}.

From the experimental results, it can be seen that, both from the perspective of $\sigma_s$ and $P_D$, the reconstruction performance of SVD-Wiener for closely-spaced scatterers is the worst among four methods, with $\sigma_s>0.7$ and $P_D<80\%$.
In terms of estimation error, GBCS outperforms EMPAST ($L = 8$) under all tested $\alpha$, surpasses PAST when $\alpha\leq 0.4$, and falls behind EMPAST and PAST when $\alpha > 0.4$. $\sigma_s$ of GBCS stabilizes at 0.2 when $\alpha$ reaches 1.2. Across all tested $\alpha$, $P_D$ of the GBCS algorithm consistently lags behind both EMPAST and PAST.
As $\alpha$ increased, $\sigma_s$ of PAST and EMPAST decreased rapidly and $P_D$ increased rapidly, and then both stabilized.
Moreover, PAST consistently performs worse than EMPAST across all tested $\alpha$. At $\alpha=1.2$, PAST maintains a stable $P_D$ of $90\%$, whereas EMPAST ($L = 8$) achieves $P_D=90\%$ at $\alpha=0.3$, and EMPAST ($L = 2$) achieves $P_D=90\%$ at $\alpha=0.9$.
This statistical experiment verifies the super-resolution capability of the proposed EMPAST framework.

\subsubsection{Super-resolution Factor}
\begin{figure}[htbp]
	\centering
	\includegraphics[width=0.7\linewidth]{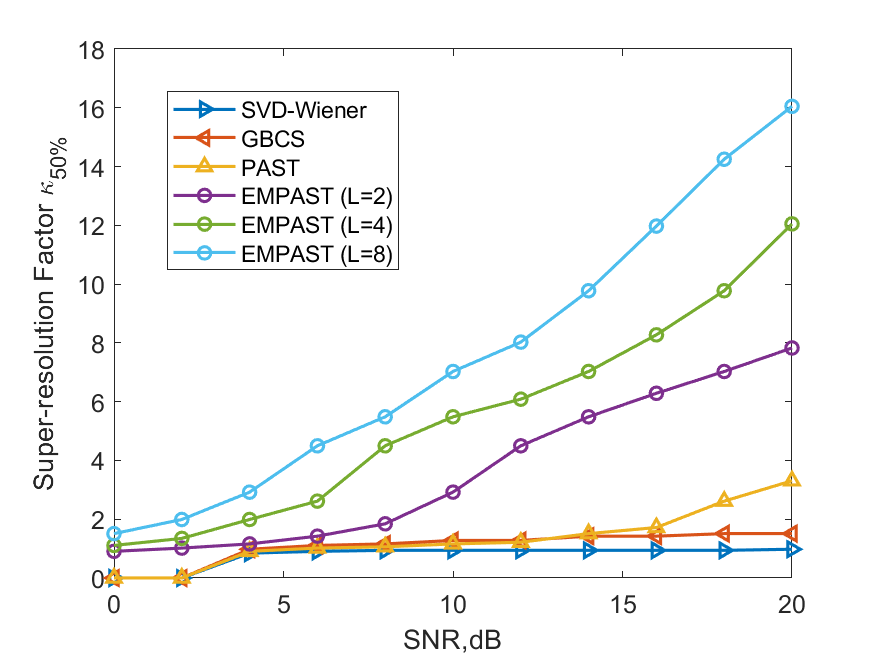}
	\caption{Resolution comparison in terms of super-resolution factor $\kappa$ of SVD-Wiener, GBCS, PAST and EMPAST as a function of SNR with $P_D=50\%$.}
	\label{simuSR}
\end{figure}

In this simulation, a super-resolution factor $\kappa$ is defined as \eqref{eq:SR}~\cite{SRfactor} to evaluate the super-resolution capability at different SNR of different TomoSAR imaging methods.
\begin{equation}
	\kappa_{P_D} = \frac{\rho_s}{\rho_{P_D}},
	\label{eq:SR}
\end{equation}
where $\rho_{P_D}$ is the minimum distance between double scatterers that are separable at a prespecified $P_D$.
We vary the elevation distance $\Delta s \in (0,2\rho_s]$, with the backscattering coefficients of scatterers kept equal in amplitude and random in phase. For each SNR, 10000 Monte Carlo experiments are performed to obtain $\rho_{P_D}$, which further yields the variation of $\kappa$ versus SNR. The experiment result with $P_D=50\%$ is illustrated in \figref{simuSR}.

The experimental result indicates that the super-resolution factor $\kappa_{50\%}$ of SVD-Wiener reaches 0.9 at SNR = 6 dB and maintains stability.
For the GBCS algorithm, $\kappa_{50\%}$ reaches 1.0 at SNR = 4 dB and gradually increases to 1.5 with the rise of SNR. PAST achieves $\kappa_{50\%}=1.0$ at SNR = 6 dB and continues to increase with rising SNR, reaching $\kappa_{50\%}=3.3$ at SNR = 20 dB. In contrast, EMPAST ($L=8$) already achieves $\kappa_{50\%}=1.5$ at SNR = 0 dB and continues to rise with increasing SNR. As SNR increases to 20 dB, EMPAST ($L=8$) achieves $\kappa_{50\%}=16.0$, while EMPAST ($L=2$) achieves $\kappa_{50\%}=7.8$.
This experiment, by quantifying the super-resolution capability, demonstrates that EMPAST has superior resolution in elevation reconstruction compared to other state-of-the-art algorithms, particularly in challenging low SNR scenarios.

\subsubsection{Parameter Sensitivity}
\begin{figure}[htbp]
	\centering
	\subfloat[]{\includegraphics[width=0.5\linewidth]{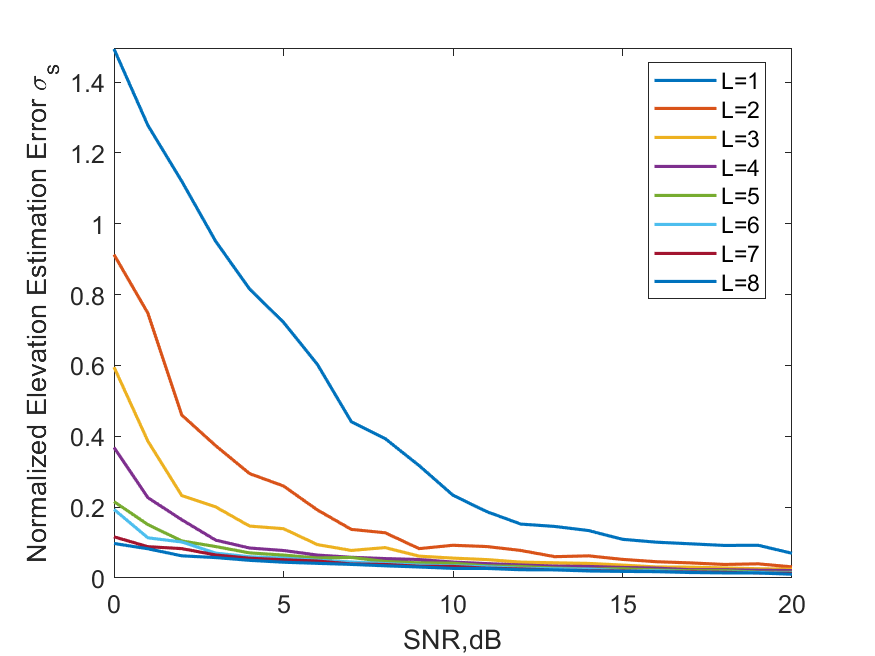}}
	\subfloat[]{\includegraphics[width=0.5\linewidth]{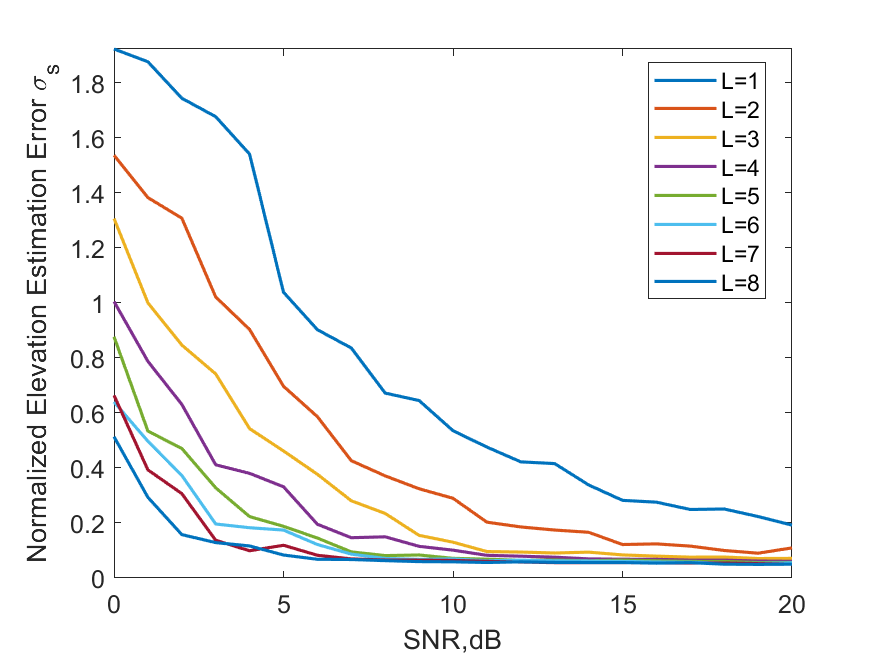}}
	\vspace{-1em}
	\subfloat[]{\includegraphics[width=0.5\linewidth]{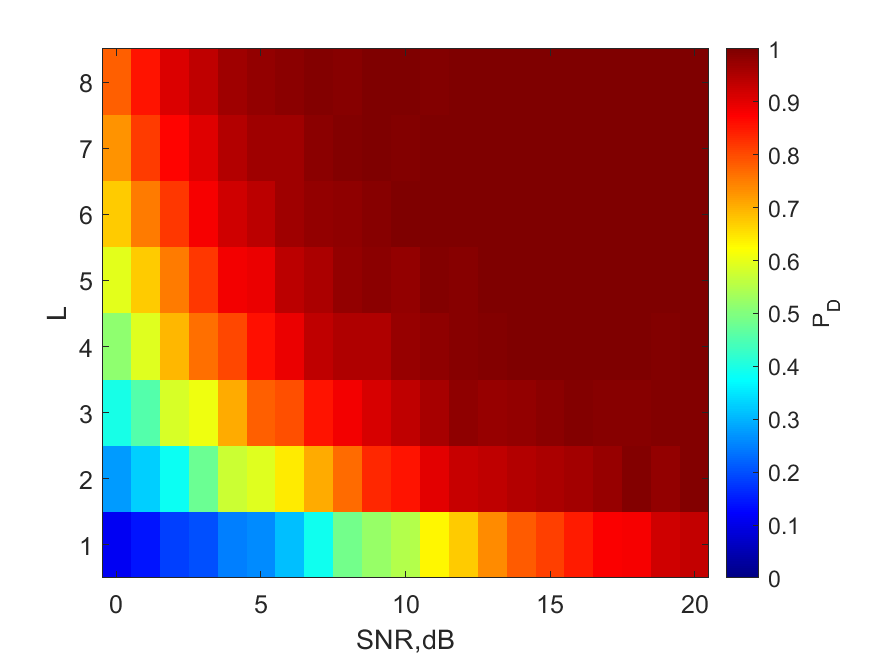}}
	\subfloat[]{\includegraphics[width=0.5\linewidth]{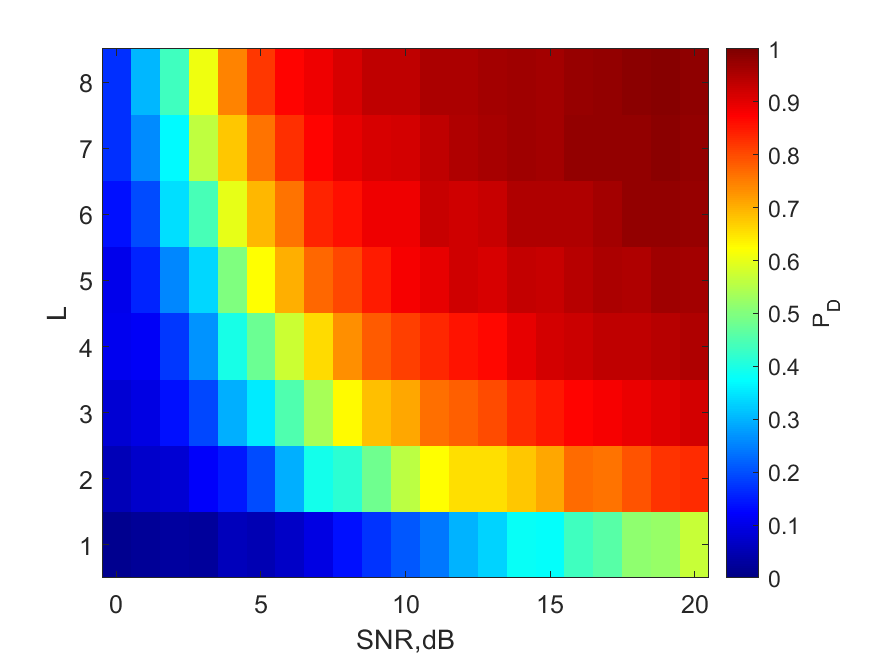}}
	\caption{Reconstruction performance of the proposed EMPAST (with different $L$) as a function of SNR for double scatterers with different
		normalized elevation distance $\alpha$. (a) and (c) $\alpha=1$. (b) and (d) $\alpha=1/2$. The top row depicts the normalized elevation estimation error $\sigma_s$, while the bottom row illustrates the probability of detection $P_D$.}
	\label{simuL}
\end{figure}

Focusing specifically on the proposed EMPAST framework, the effect of $L$ on its ability to reconstruct double scatterers is investigated through Monte Carlo experiments. The double scatterers exhibit same amplitudes and random phases of backscattering coefficients.
We set the normalized elevation distance $\alpha$ to $1$ and $1/2$ respectively, and set SNR varied from 0 dB to 20 dB. Performing 2000 Monte Carlo experiments in each case, the reconstruction performances are shown in \figref{simuL}.

To visually present experimental results, we utilize phase transition diagram to showcase the probability of detection. The horizontal axis corresponds to SNR, the vertical axis represents $L$, and the color at each point indicates the value of $P_D$. This graphical representation provides an intuitive way to observe how $P_D$ varies with noise level and EMMV number, aiding in a comprehensive understanding of reconstruction performance across diverse scenarios.
Analysis of the experimental outcomes reveals that, when $\alpha=1$, the influence of $L$ on reconstruction performance is most pronounced at SNR below 5 dB. Larger $L$ result in superior reconstruction performance. Nevertheless, as SNR surpasses 10 dB, the results for $L\leq 2$ exhibit similarity and tend to stabilize.
Similarly, when $\alpha=1/2$, the observed trend echoes that of $\alpha=1$. However, stable reconstruction is unattainable with all tested $L$ at SNR below 3 dB. As SNR exceeds 3 dB, the impact of $L$ on reconstruction performance becomes discernible. Notably, at SNR above 15 dB, the results for $L\leq 2$ become comparable.
Experimental results demonstrate that the reconstruction performance is significantly influenced by $L$, with varying impacts observed across different SNR. 
The judicious selection of parameter $L$ serves as a pivotal guide in shaping the design considerations for PRF of radar systems, and contributes to the efficient utilization of resources. 

\subsection{3-D Point Clouds Simulation}
\begin{figure}[htbp]\vspace{-1em}
	\centering
	\includegraphics[width=0.6\linewidth]{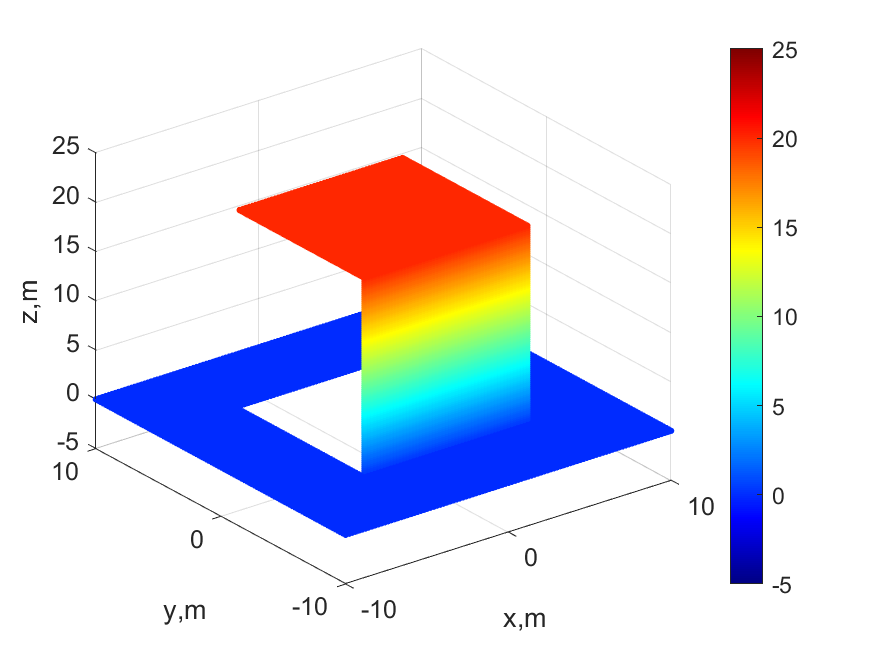}
	\caption{3-D point clouds simulation scenario.}
	\label{simuModel3d}
\end{figure}

In this section,  we simulate a straightforward building scenario comprising planes representing the ground, wall, and roof, as illustrated in \figref{simuModel3d}. The building has a height of 20 m, with the ground situated at an altitude of 0 m. Leveraging the simulated scenario, we generate three-dimensional raw data and introduced varying levels of noise. Subsequently, we employ four TomoSAR imaging methods for 3-D point clouds reconstruction.

\begin{figure*}[htbp]
	\centering
	\subfloat[]{\includegraphics[width=0.33\linewidth]{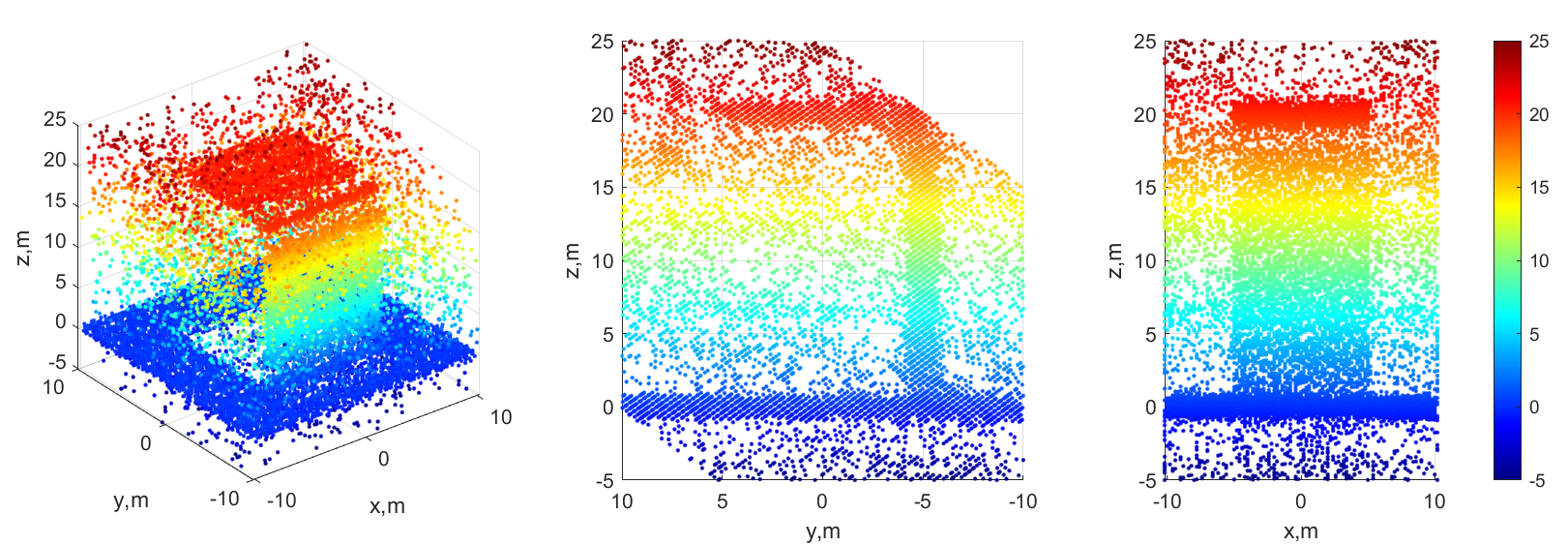}}
	\subfloat[]{\includegraphics[width=0.33\linewidth]{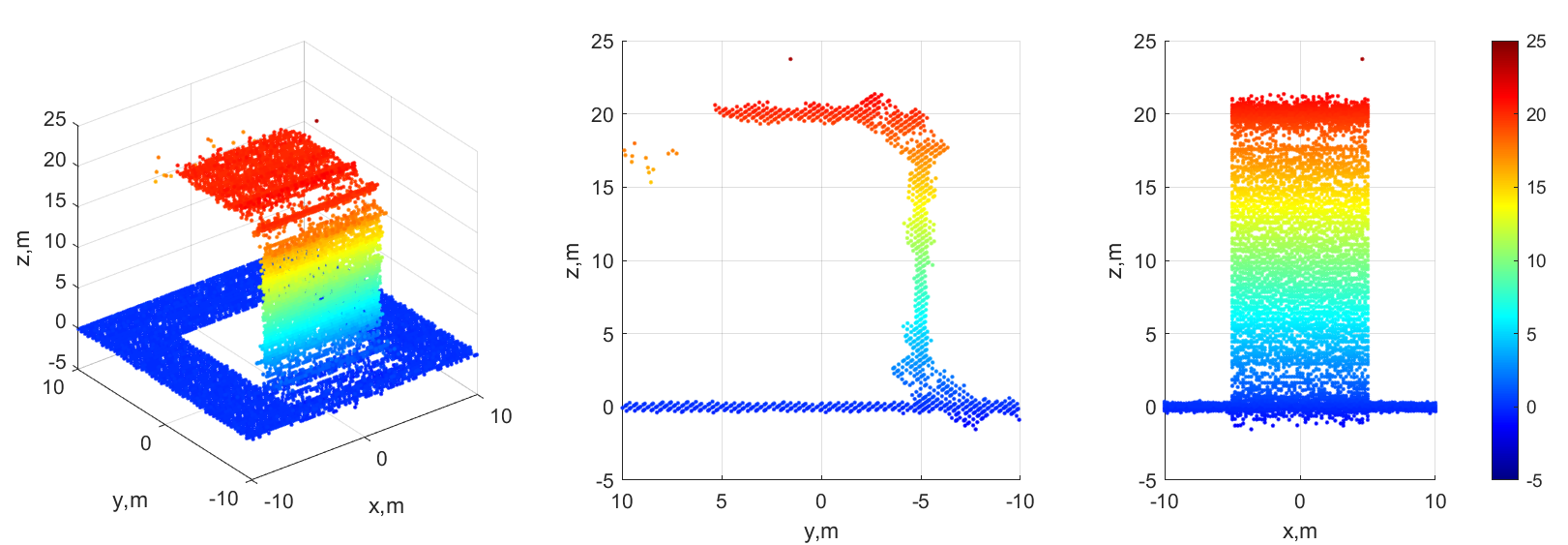}}
	\subfloat[]{\includegraphics[width=0.33\linewidth]{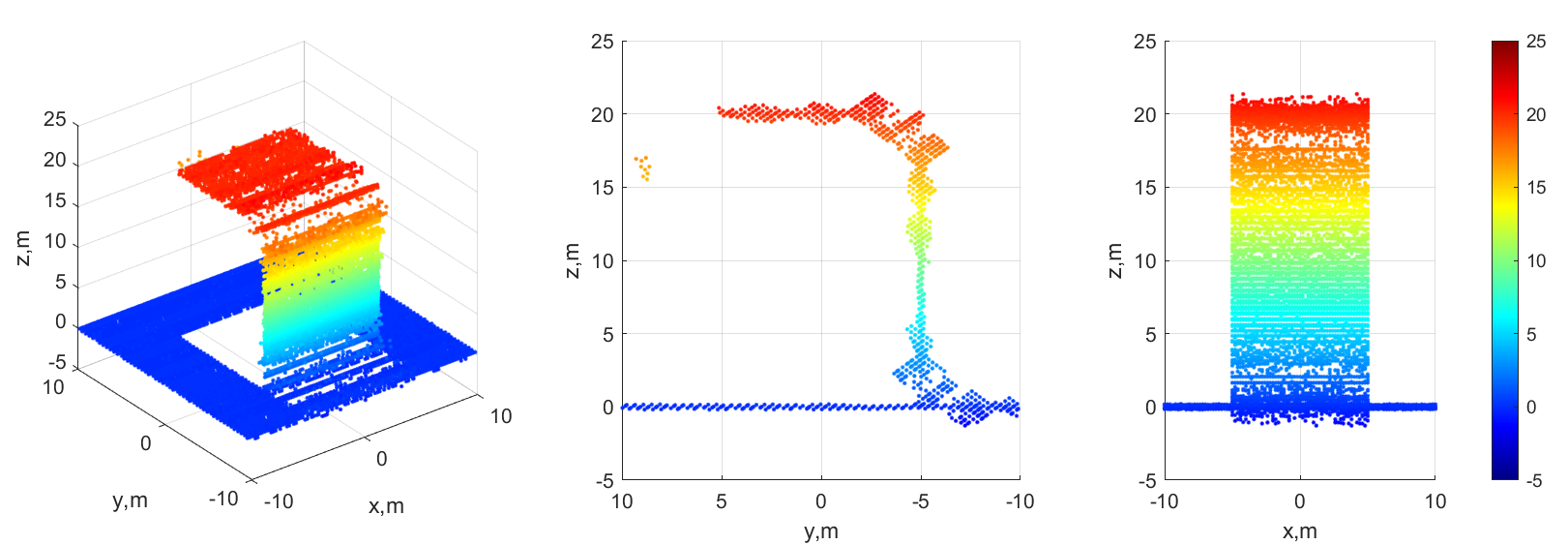}}
	\vspace{-1em}
	\subfloat[]{\includegraphics[width=0.33\linewidth]{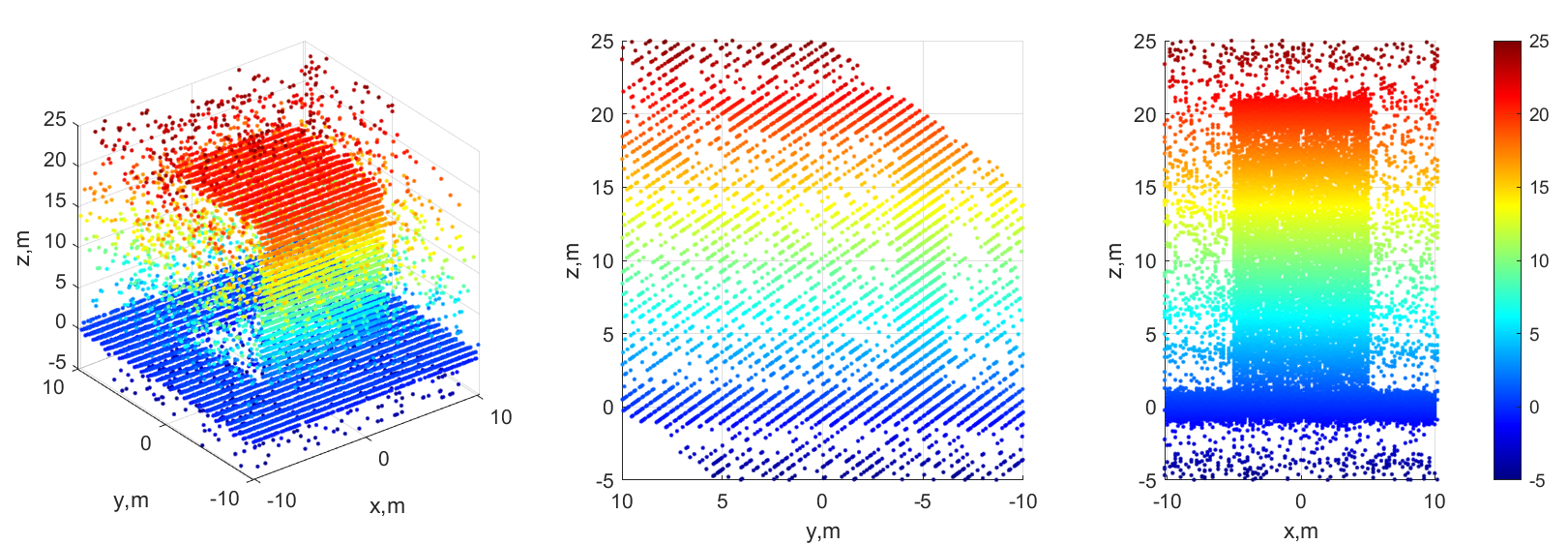}}
	\subfloat[]{\includegraphics[width=0.33\linewidth]{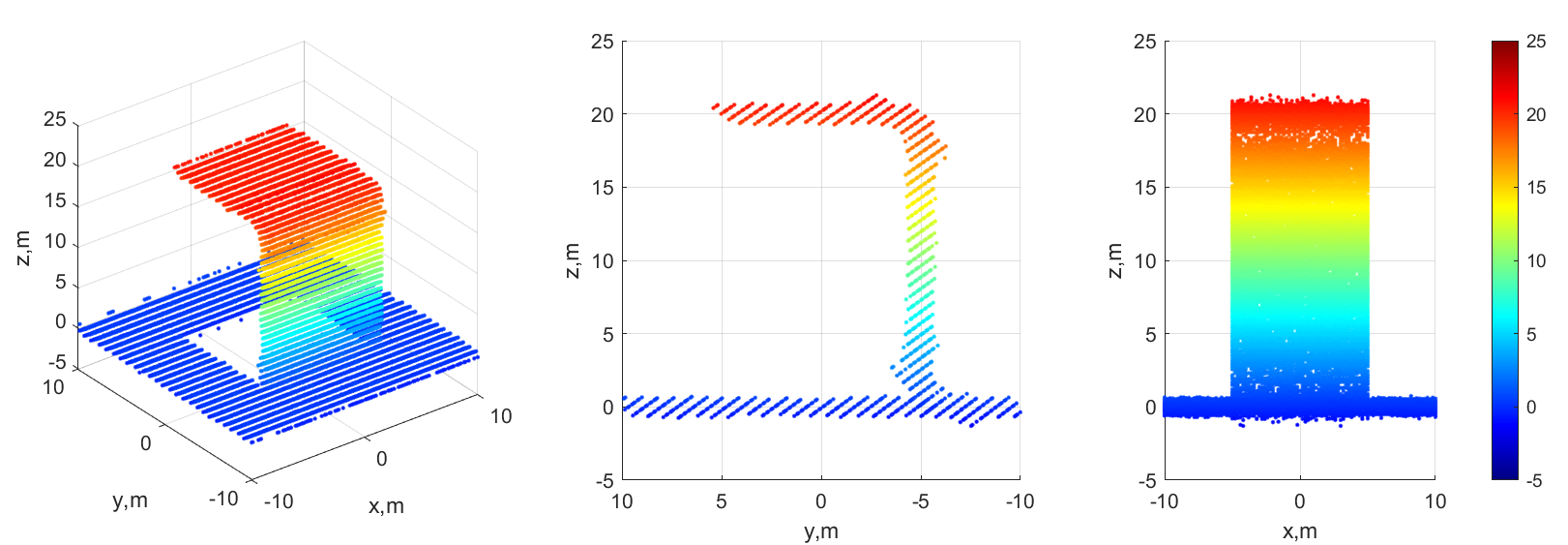}}
	\subfloat[]{\includegraphics[width=0.33\linewidth]{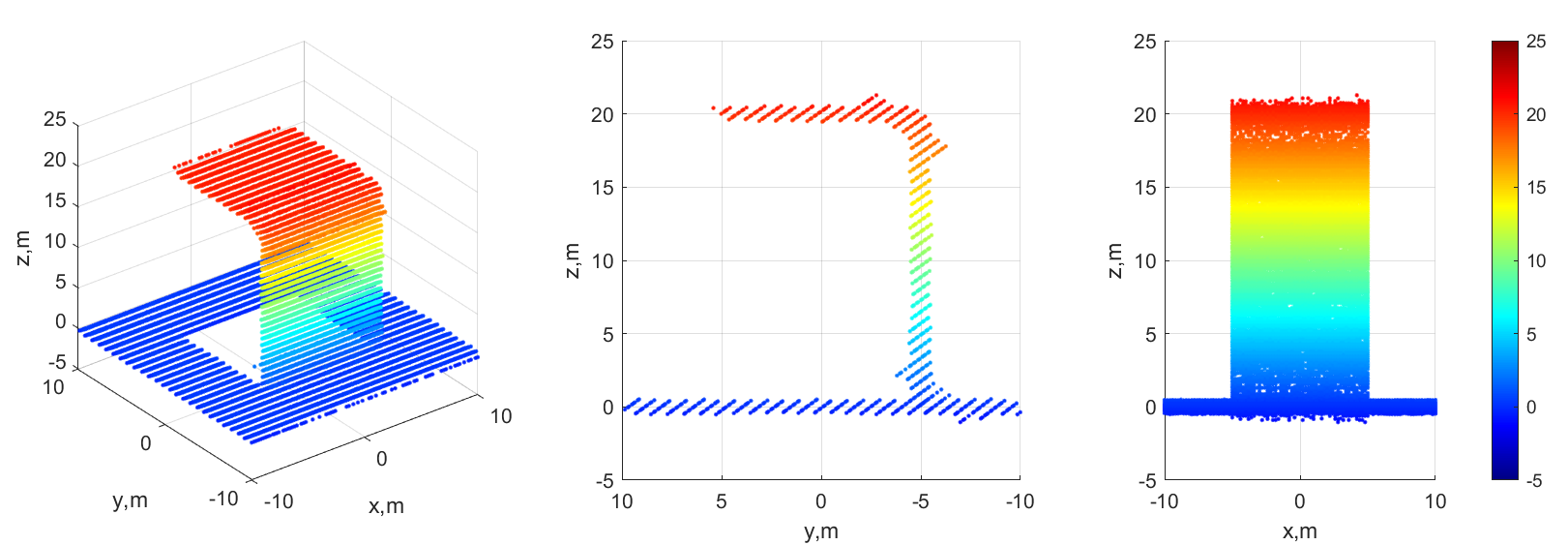}}
	\vspace{-1em}
	\subfloat[]{\includegraphics[width=0.33\linewidth]{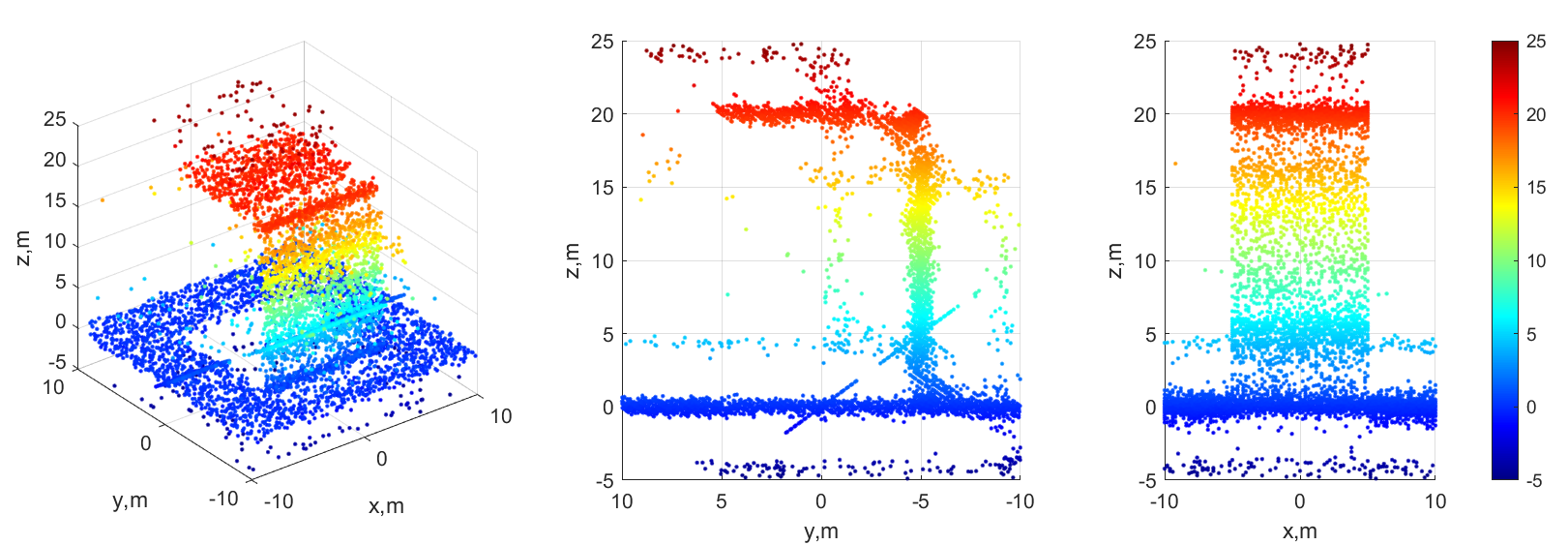}}
	\subfloat[]{\includegraphics[width=0.33\linewidth]{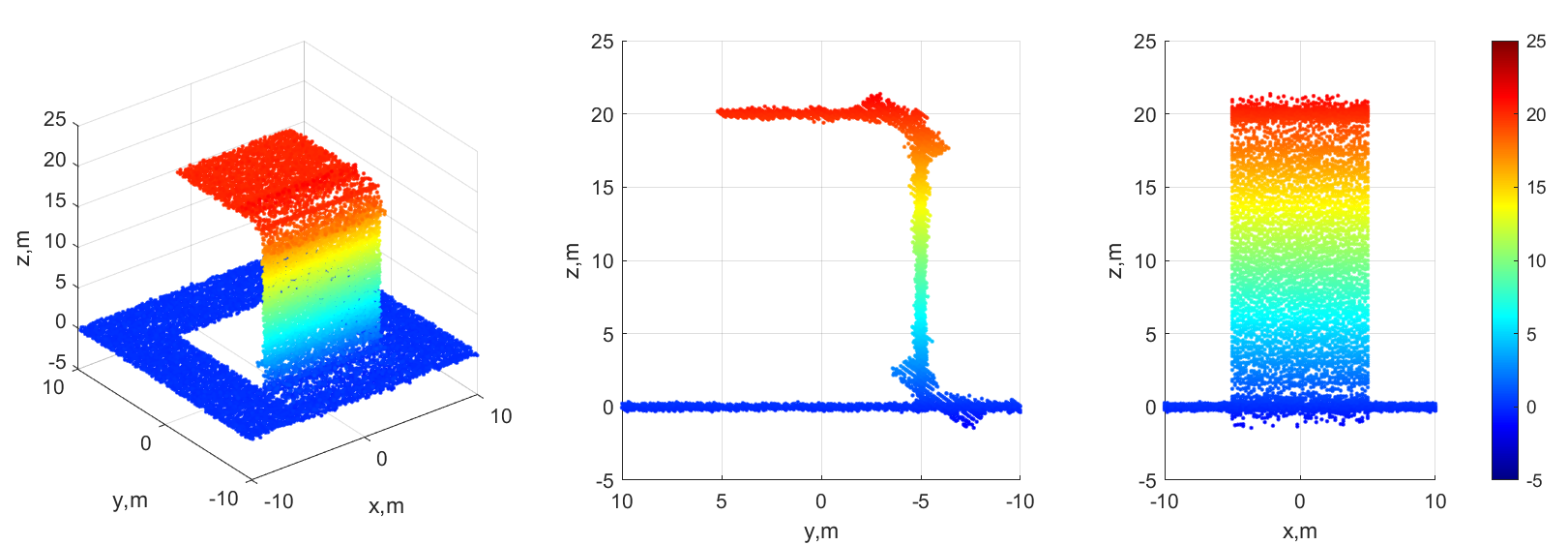}}
	\subfloat[]{\includegraphics[width=0.33\linewidth]{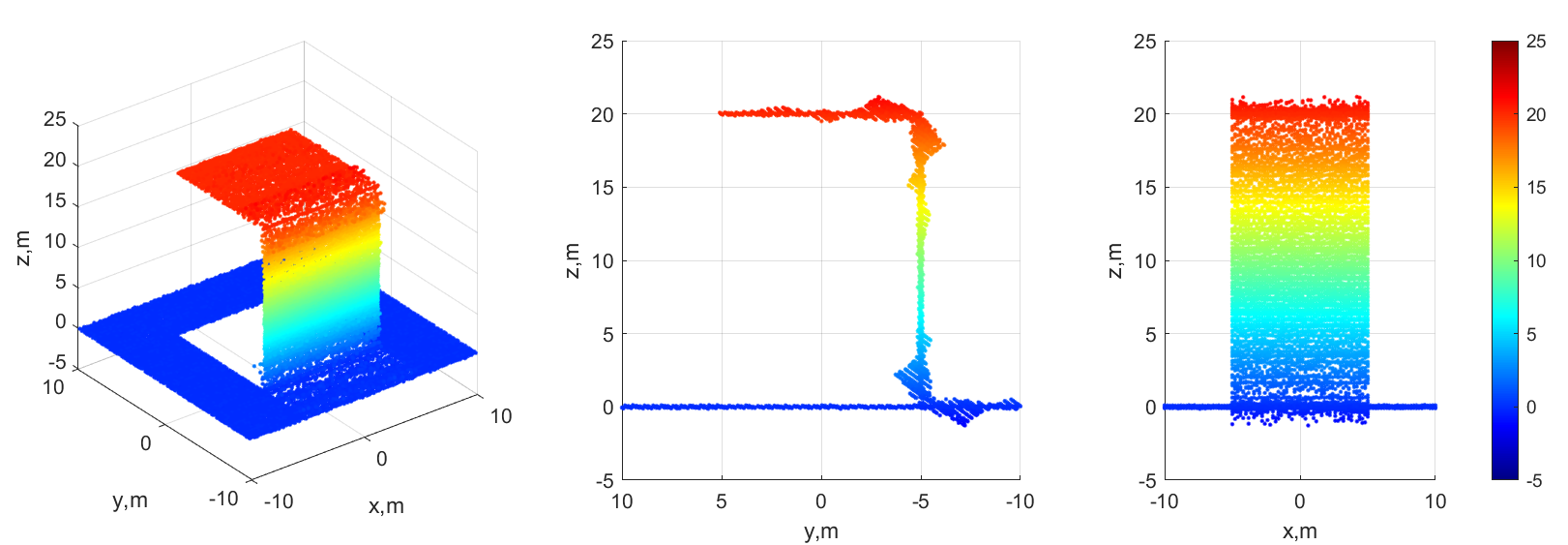}}
	\vspace{-1em}
	\subfloat[]{\includegraphics[width=0.33\linewidth]{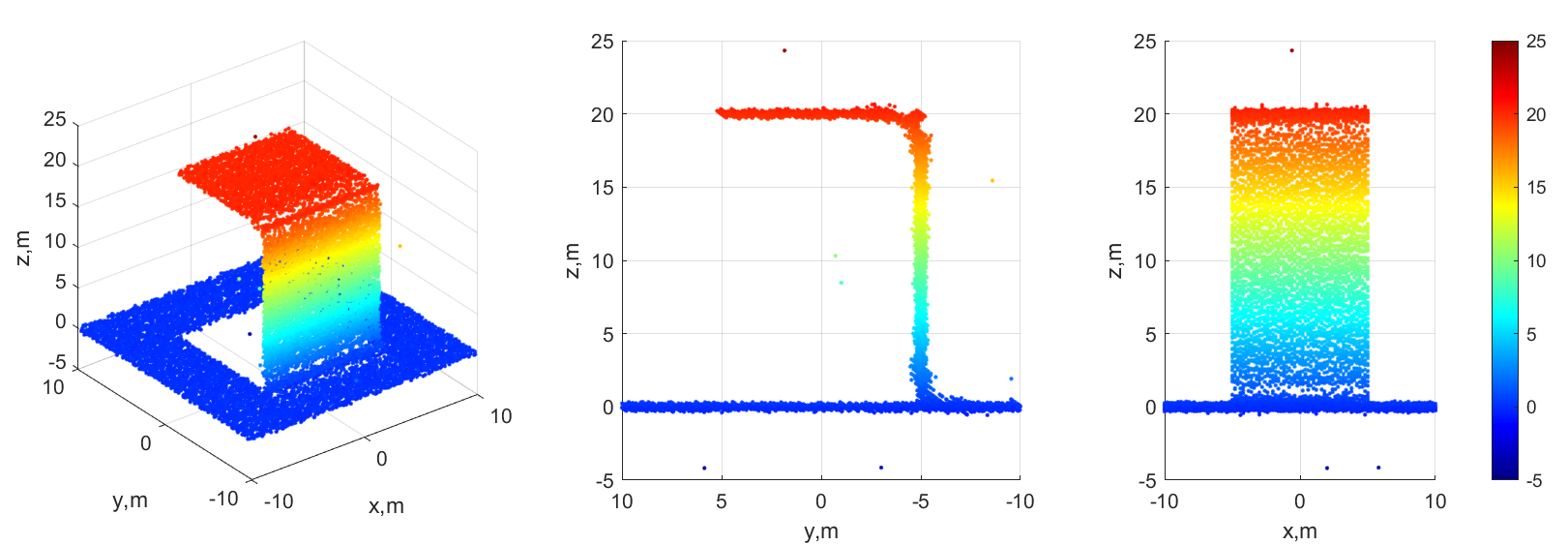}}
	\subfloat[]{\includegraphics[width=0.33\linewidth]{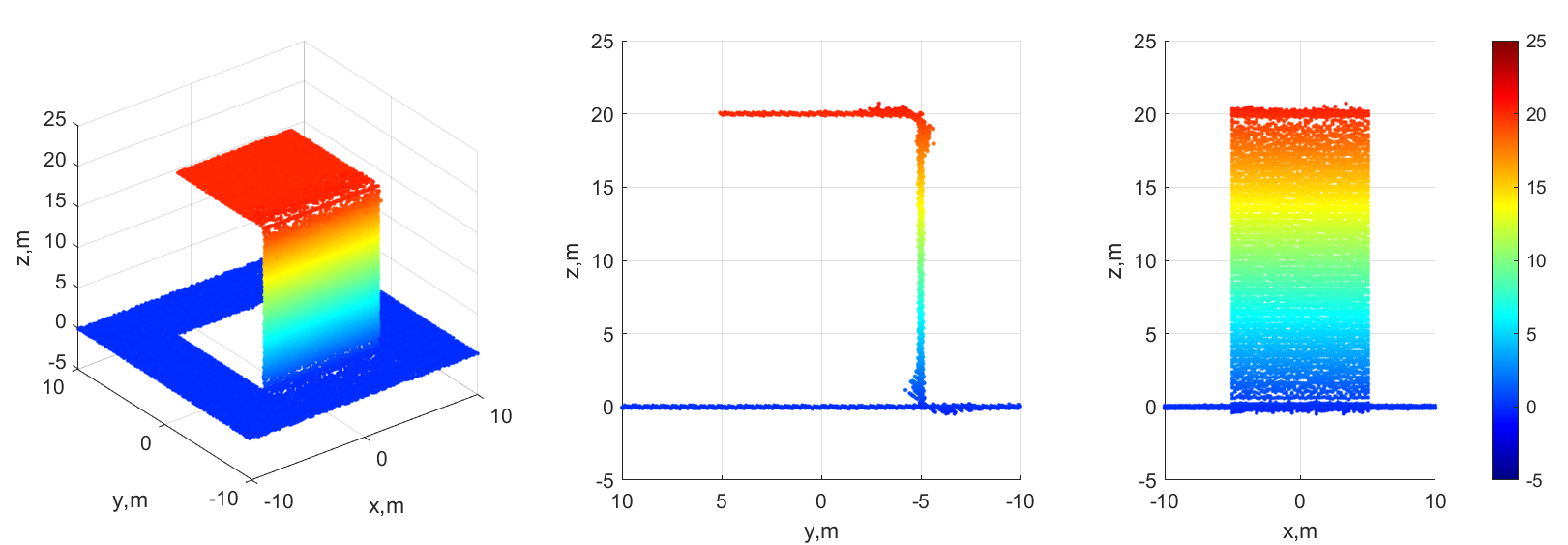}}
	\subfloat[]{\includegraphics[width=0.33\linewidth]{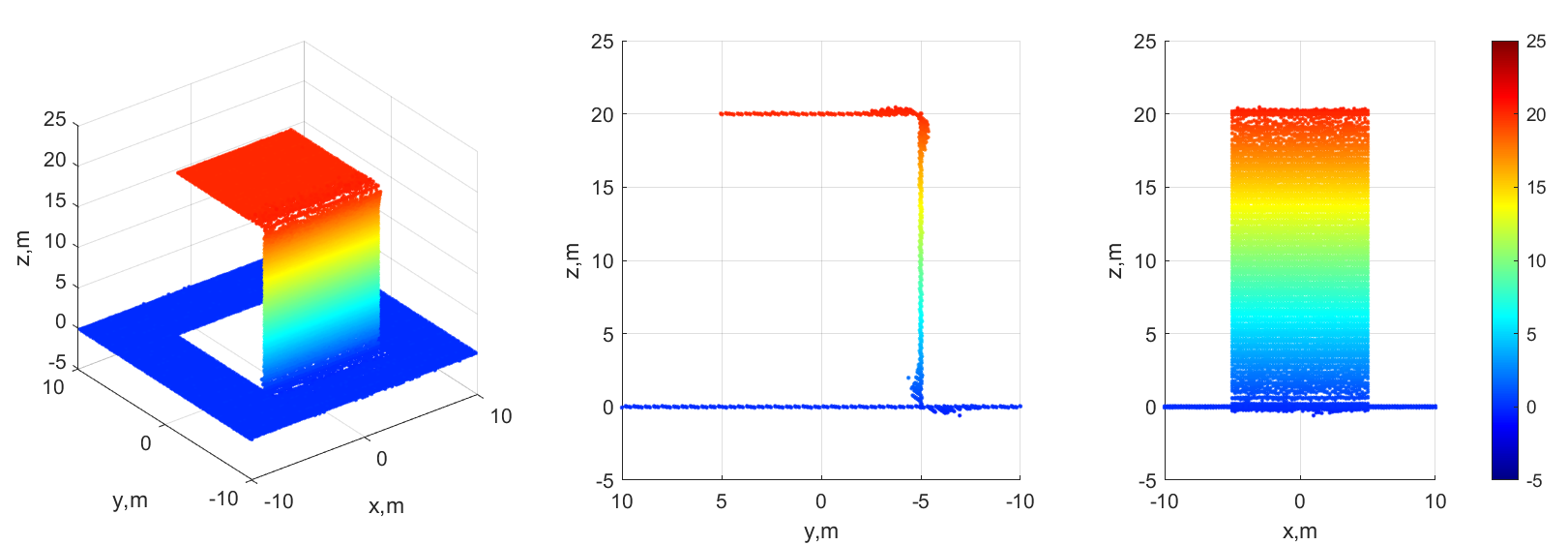}}
	\caption{Three views of 3-D point clouds reconstruction results of (Top row) SVD-Wiener, (Second row) GBCS, (Third row) PAST and (Bottom row) EMPAST ($L=8$) with different SNR. (a), (d), (g) and (j) SNR = 0 dB. (b), (e), (h) and (k) SNR = 10 dB. (c), (f), (i) and (l) SNR = 20 dB.}
	\label{simuResult3d}
\end{figure*}

The reconstructed results are presented from three different views in \figref{simuResult3d}. In this $4\times 3$ figure, each row corresponds to a distinct tomographic imaging method, and each column delineates different SNR levels.
At the lowest SNR = 0 dB, the SVD-Wiener, GBCS, and PAST algorithms show notable discrepancies from the simulated scene, manifesting an exceptionally high occurrence of false targets. This is particularly evident in SVD-Wiener and PAST, struggling to delineate the building contours effectively. In stark contrast, our proposed framework offers a distinct advantage, delivering clear and precise reconstructions while demonstrating robust noise immunity.
With increasing SNR, all four methods exhibit gradual enhancements. However, our EMPAST framework consistently maintains a competitive edge, ensuring clear and accurate reconstructions across varying noise levels.
Noteworthy is the improved performance of the proposed EMPAST as SNR increases, allowing for the reconstruction of thinner planes, bringing it more similar to the simulated point clouds.

For a rigorous quantitative assessment of the reconstruction performance, we fit the ground and the roof of the building by random sample consensus (RANSAC)~\cite{RANSAC}, and the mean absolute error $\mu$ and the standard deviation $\sigma$ of the estimated heights with respect to the simulated heights are calculated and given in \tabref{simuResultplanetable}. From the quantitative comparisons, EMPAST demonstrates superior performance compared to the other three methods, exhibiting significantly lower mean errors in ground and roof height estimates, particularly under low SNR conditions. Visually, a smaller standard deviation $\sigma$ corresponds to a thinner plane, suggesting the stronger robustness of our proposed EMPAST in accurate height estimation.
\begin{table}[htbp]
	\begin{center}
		\caption{Quantitative comparison of height estimation error (in meters)}
		\label{simuResultplanetable}
		\begin{tabular}{cccccc}
			\toprule
			\multirow{2}{*}{SNR}  & \multirow{2}{*}{Method} & \multicolumn{2}{c}{Ground} & \multicolumn{2}{c}{Roof}                        \\
			\cmidrule(lr){3-4}\cmidrule(lr){5-6}
			                      &                         & $\mu$                      & $\sigma$                 & $\mu$    & $\sigma$  \\
			\midrule
			\multirow{4}{*}{0 dB}  & SVD-Wiener              & 0.13                       & 0.61                     & 0.06     & 0.91      \\
			                      & GBCS                    & 0.11                       & 0.60                     & 0.10     & 0.73      \\
			                      & PAST                    & 0.09                       & 0.57                     & 0.13     & 0.62      \\
			                      & EMPAST                  & \bf   0.02                 & \bf   0.16               & \bf 0.03 & \bf  0.16 \\
			\midrule
			\multirow{4}{*}{10 dB} & SVD-Wiener              & 0.14                       & 0.55                     & 0.08     & 0.46      \\
			                      & GBCS                    & 0.13                       & 0.58                     & 0.10     & 0.47      \\
			                      & PAST                    & 0.17                       & 0.43                     & 0.03     & 0.24      \\
			                      & EMPAST                  & \bf   0.01                 & \bf    0.10              & \bf 0.01 & \bf 0.09  \\
			\midrule
			\multirow{4}{*}{20 dB} & SVD-Wiener              & 0.11                       & 0.45                     & 0.05     & 0.40      \\
			                      & GBCS                    & 0.05                       & 0.36                     & 0.06     & 0.35      \\
			                      & PAST                    & 0.05             & 0.29                     & 0.03 & 0.18      \\
			                      & EMPAST                  & \bf   0.01                 & \bf  0.06                & \bf 0.01 & \bf 0.07  \\
			\bottomrule
		\end{tabular}
	\end{center}
\end{table}

Furthermore, in the context of 3-D point clouds evaluation with available ground truth, two key metrics, accuracy and completeness, collectively characterize the quality of reconstructed point clouds~\cite{Nonlocal}. Accuracy measures the spatial fidelity of the reconstruction by assessing how well each reconstructed point aligns with its corresponding point in the ground truth. Completeness, on the other hand, evaluates how well the reconstructed point clouds capture the entirety of the ground truth.
Suppose $P_i$ and $G_j$ represent the $i$-th point in reconstructed point clouds and the $j$-th point in the ground truth point clouds, and the term $\text{dist}(\cdot,\cdot)$ denotes the Euclidean distance between them.
Then the accuracy evaluation metric can be defined as the mean absolute error $\mu$ and standard deviation $\sigma$ of $\min_j \text{dist}\left(P_i,G_j\right)$, and completeness can be computed from $\min_i \text{dist}\left(P_i,G_j\right)$.
\tabref{simuResult3dtable} shows the quantitative evaluations of reconstructed 3-D point clouds. Our proposed framework demonstrates superior performance in both accuracy and completeness compared to other algorithms. Specifically, at 0 dB, the good completeness of GBCS can be attributed to the prevalence of false targets, resulting in densely packed false targets throughout the space and causing the existence of $P_i$ located extremely close to each $G_j$.
\begin{table}[htbp]
	\begin{center}
		\caption{Quantitative comparison of 3-D reconstruction (in meters)}
		\label{simuResult3dtable}
		\begin{tabular}{cccccc}
			\toprule
			\multirow{2}{*}{SNR}  & \multirow{2}{*}{Method} & \multicolumn{2}{c}{Accuracy} & \multicolumn{2}{c}{Completeness}                       \\
			\cmidrule(lr){3-4}\cmidrule(lr){5-6}
			                      &                         & $\mu$                        & $\sigma$                         & $\mu$    & $\sigma$ \\
			\midrule
			\multirow{4}{*}{0 dB}  & SVD-Wiener              & 4.40                        & 5.26                            & 0.11     & 0.11     \\
			                      & GBCS                    & 2.33                        & 4.15                            & \bf 0.07 & 0.10     \\
			                      & PAST                    & 0.76                        & 1.56                            & 0.26     & 0.11 \\
			                      & EMPAST                  & \bf 0.13                     & \bf 0.27                         & 0.12     & \bf 0.06     \\
			\midrule
			\multirow{4}{*}{10 dB} & SVD-Wiener              & 0.28                         & 0.99                             & 0.15     & 0.10     \\
			                      & GBCS                    & 0.25                         & 0.24                             & 0.08     & 0.09     \\
			                      & PAST                    & 0.16                         & 0.43                            & 0.12     & 0.07     \\
			                      & EMPAST                  & \bf  0.06                    & \bf 0.05                         & \bf 0.06 & \bf 0.04 \\
			\midrule
			\multirow{4}{*}{20 dB} & SVD-Wiener              & 0.22                         & 0.76                             & 0.13     & 0.11     \\
			                      & GBCS                    & 0.23                         & 0.16                             & 0.08     & 0.09     \\
			                      & PAST                    & 0.09                         & 0.28                             & 0.07     & 0.06     \\
			                      & EMPAST                  & \bf 0.04                     & \bf 0.04                         & \bf 0.04 & \bf 0.03 \\
			\bottomrule
		\end{tabular}
	\end{center}
\end{table}

\subsection{Efficiency}
In this section, we conduct experiments to assess the efficiency of the proposed EMPAST framework employing both SDP solver (SDPT3 is used in this experiments) and ADMM.

\begin{figure}[htbp]
	\centering
	\subfloat[]{\includegraphics[width=0.7\linewidth]{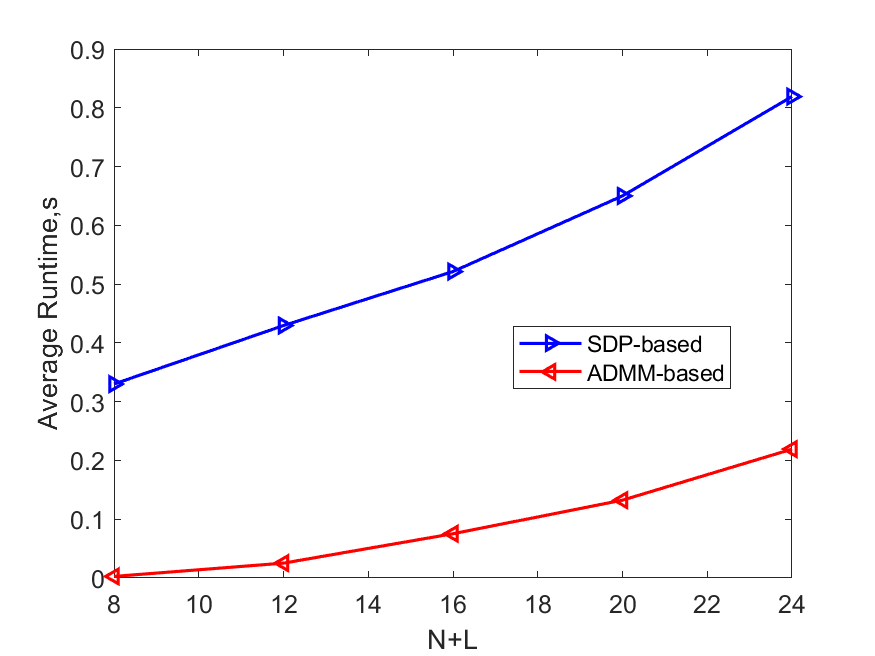}}
	\caption{Average runtime comparison for EMPAST based on SDP and ADMM as a function of the PSD matrix size $N+L$. }
	\label{simuTime}
\end{figure}

\begin{figure}[htbp]
	\centering
	\subfloat[]{\includegraphics[width=0.7\linewidth]{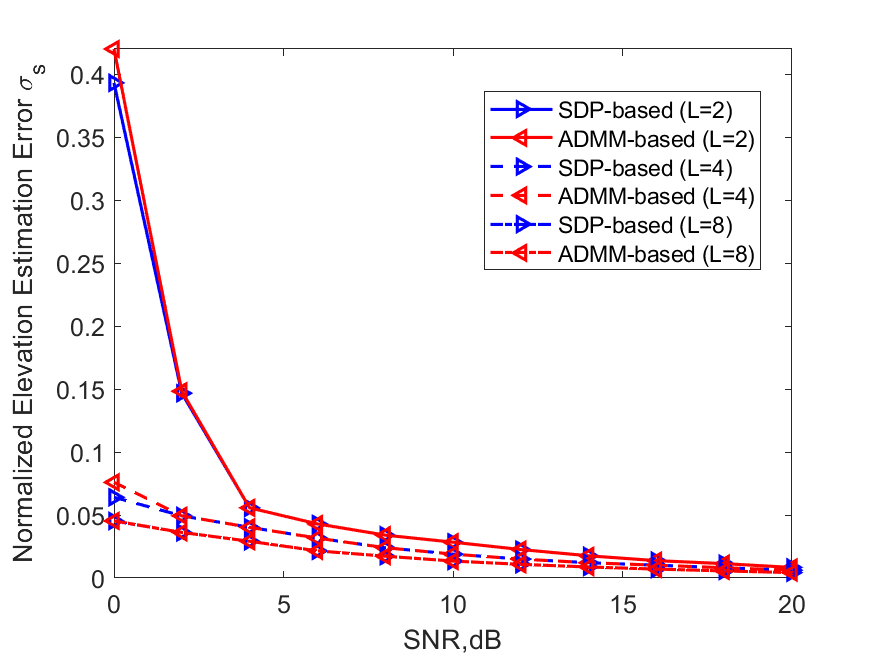}}
	\vspace{0em}
	\subfloat[]{\includegraphics[width=0.7\linewidth]{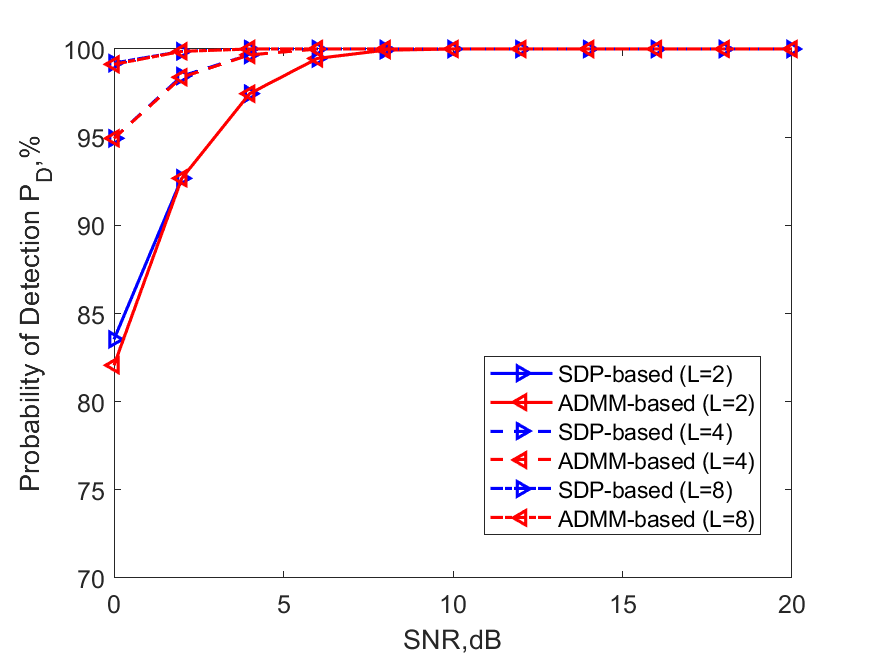}}
	\caption{Reconstruction performance comparison of EMPAST based on SDP and ADMM as a function of SNR for a single scatterer. (a) Normalized elevation estimation error $\sigma_s$. (b) Probability of detection $P_D$.}
	\label{simuADMM}
\end{figure}

The computational efficiency of EMPAST is critically influenced by the scale of the optimization PSD matrix. To comprehensively analyze its performance, we utilize Monte Carlo simulations to collect statistics on the average runtime required for successful reconstruction under various $N+L$. The experimental scenarios encompass both single scatterer and two well-separated scatterers instances.
\figref{simuTime} illustrates the average runtime comparison between EMPAST based on SDP and ADMM. The results clearly indicate that the runtime of ADMM solver is significantly smaller than SDP solver. Furthermore, as $N+L$ increases, the discernible difference in runtime between the two methodologies becomes more pronounced.
In addition to efficiency, a comparative analysis of reconstruction performance between the two solvers is essential. We set SNR from 0 dB to 20 dB, with radar parameters detailed in \tabref{simuPara}. The accuracy and robustness of both solvers are illustrated in \figref{simuADMM}. The experimental outcomes indicate that the SDP and ADMM solvers exhibit comparable performance, with SDP outperforming ADMM under extremely low SNR. 
Theoretically, both solvers converge to a consistent optimal solution.
The observed discrepancy in this experiment arises from the convergence characteristics of ADMM, where it converges rapidly to an acceptable solution. Nevertheless, in the pursuit of very high accuracy at low SNR, the iteration count increases significantly, leading to reduced efficiency. Consequently, the selection of ADMM parameters is crucial for the trade-off between efficiency and accuracy, necessitating a problem-specific determination based on the practical considerations.

In summary, our proposed framework consistently exhibits superior performance across varying SNR, particularly excelling under high noise level. The framework demonstrates strong noise resilience and super-resolution capability, effectively addressing challenges posed by complex real-world scenarios. Moreover, it showcases superior computational efficiency.
\section{Practical Demonstration with MV3DSAR Data}\label{sec:Real}
\subsection{Dataset}
\begin{figure}[htbp]
	\centering
	\subfloat[]{\includegraphics[height=0.3\linewidth]{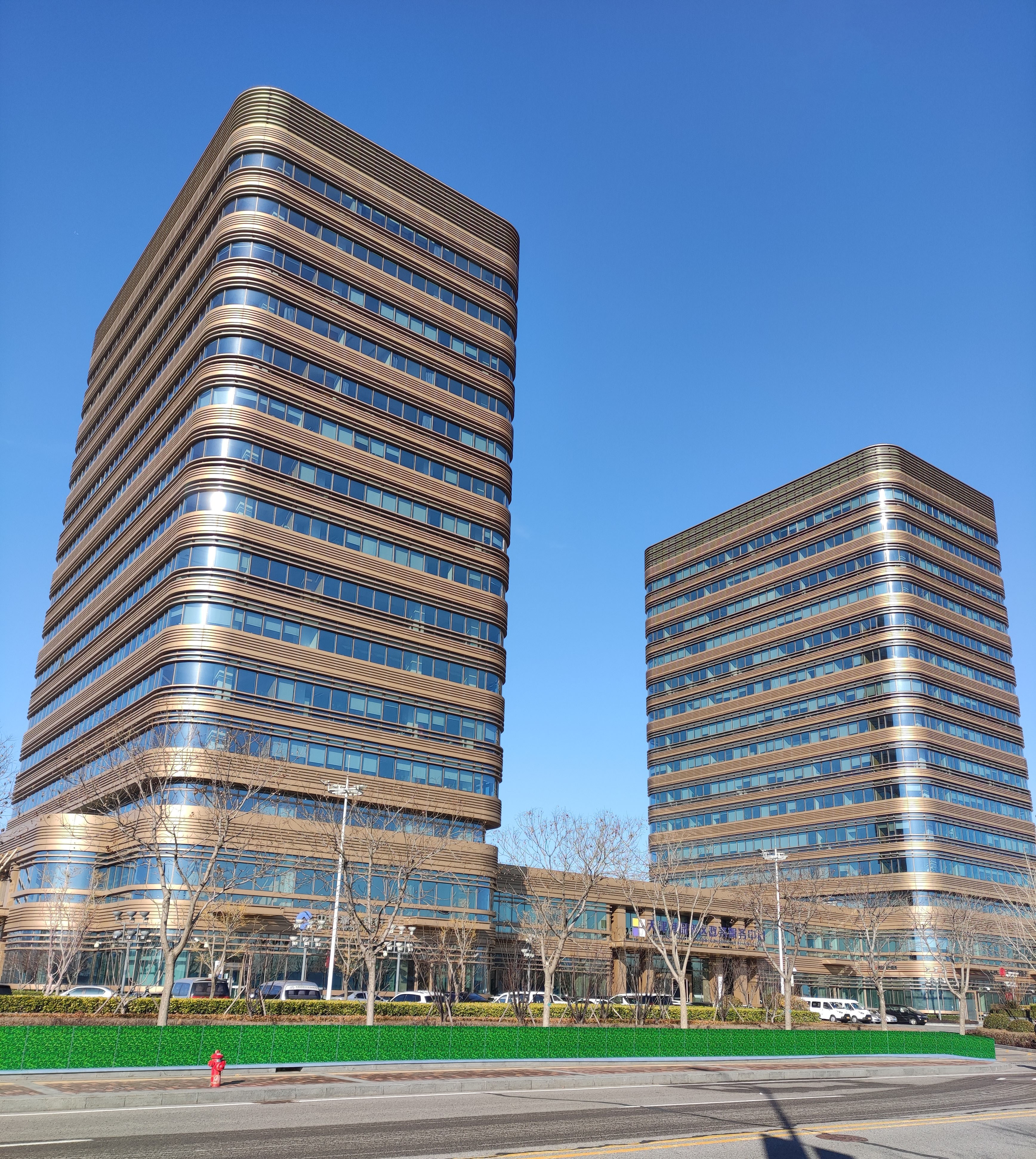}}
	\hfill
	\subfloat[]{\includegraphics[height=0.3\linewidth]{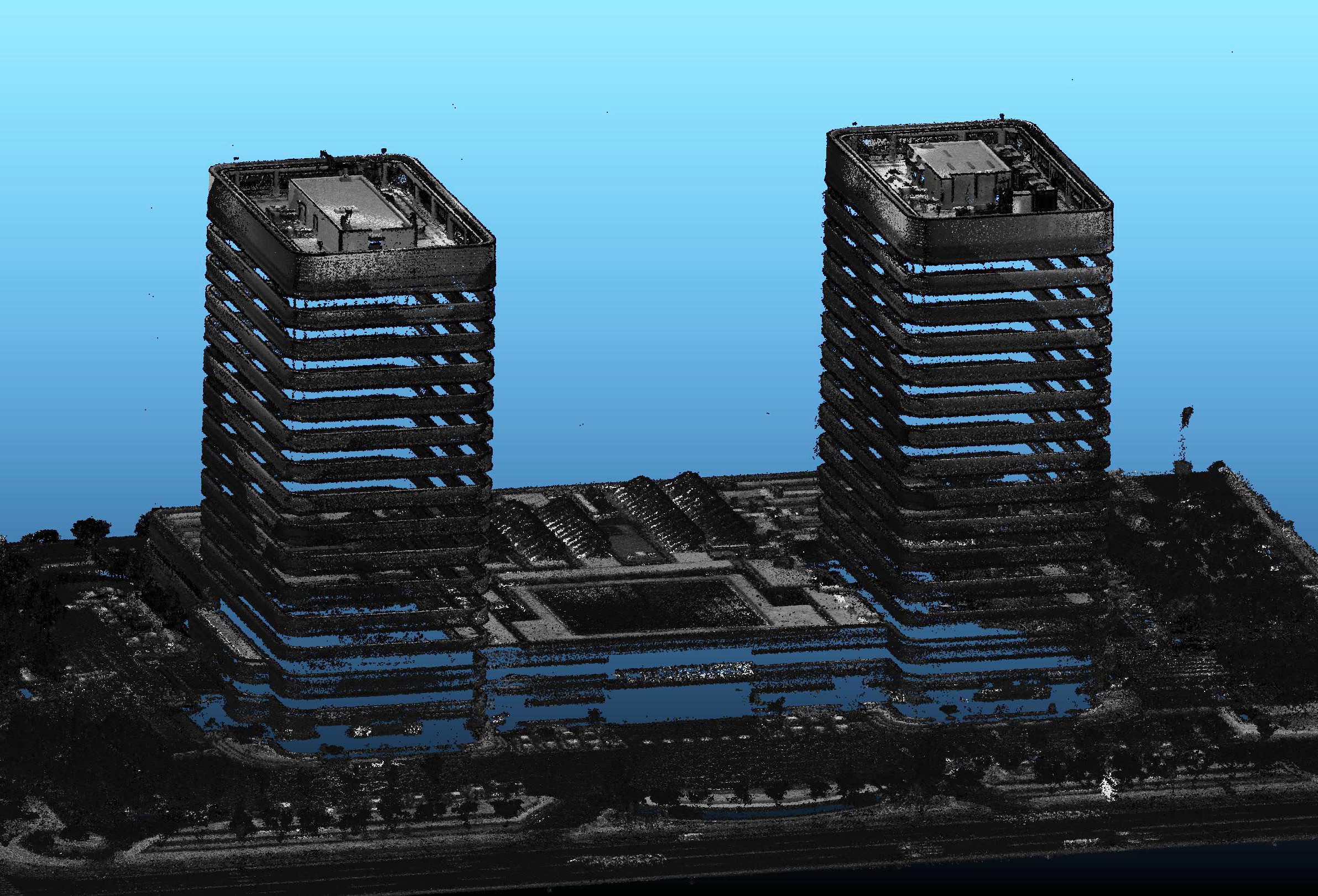}}
	\hfill
	\subfloat[]{\includegraphics[height=0.3\linewidth]{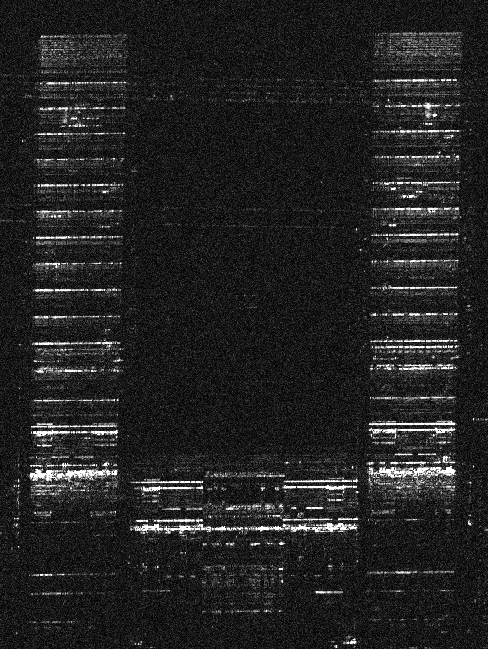}}
	\caption{The illuminated scenario of MV3DSAR flight experiment. (a) Optical image. (b) LiDAR point clouds. (c) SAR image. }
	\label{realSAR}
\end{figure}

The dataset used in our experiments is acquired using the Microwave-Vision 3-D SAR (MV3DSAR) experimental system, a small UAV-borne array InSAR system developed by the Aerospace Information Research Institute, Chinese Academy of Sciences\cite{MV3DSAR}.
We design flight experiments and collect measured data using this platform to validate effectiveness of the proposed EMPAST framework.
The illuminated scenario, shown in \figref{realSAR}, comprises two 15-story high buildings with height of 69.2 m, length of 44.5 m, and width of 27.5 m, separated by a distance of 57.6 m.

\begin{figure}[htbp]
	\centering
	\includegraphics[width=0.8\linewidth]{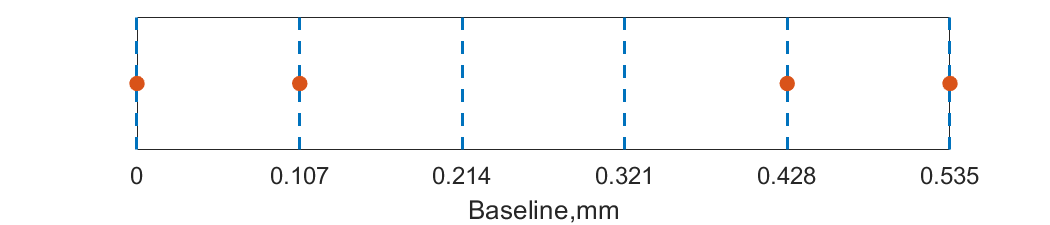}
	\caption{Baseline configuration of MV3DSAR flight experiment. Red dots: antenna phase centers.}
	\label{realbaseline}
\end{figure}

This UAV operates at a flight altitude of 400 m with a squint angle of 45°. The array utilizes 2 transmit and 2 receive antennas, forming 4 equivalent antenna phase centers (APC). The baseline configuration is illustrated in \figref{realbaseline}, and MV3DSAR system parameters employed for this flight experiment are detailed in \tabref{parareal}.
\begin{table}[htbp]
	\begin{center}
		\caption{MV3DSAR System Parameters.}
		\label{parareal}
		\begin{tabular}{ l  l  l }
			\toprule
			Name                                & Symbol                & Value     \\
			\midrule
			Carrier frequency                   & $f_c$                 & 15.20 GHz \\
			Pulse repetition frequency          & $\text{PRF}$          & 1.00 kHz  \\
			Transmitting (chirp) bandwidth      & $B_r$                 & 1.20 GHz  \\
			Platform velocity                   & $v$                   & 8.05 m/s  \\
			Platform altitude                   & $h$                   & 403.18 m  \\
			Reference distance                  & $r_0$                 & 538.40 m  \\
			Maximal elevation aperture          & $D$                   & 0.54 m    \\
			Number of APCs                      & $M$          & 4         \\
			Doppler bandwidth                   & $\Delta f_\text{dop}$ & 71.11 Hz  \\
			Rayleigh resolution along elevation & $\rho_s$              & 9.92 m    \\
			\bottomrule
		\end{tabular}
	\end{center}
\end{table}

\subsection{Experimental Results}
In EMPAST framework, given that PRF significantly exceeds Doppler bandwidth $\Delta f_\text{dop}$, we extract raw data along the slow time direction with $L=8$. Following a sequence of processing steps outlined in \secref{sec:Method}, we ultimately obtain 32 frames ($M=4, L=8$) of registrated SLC images. EMPAST performs tomography imaging from the extracted 8 measurement vectors data, while the other three methods perform tomography imaging directly from the 4 registrated SLC images obtained from raw data. For grid-based methods, we set the grid interval to $\rho_s/8$ as in the simulation experiments.

\begin{figure*}[htbp]
	\centering
	\subfloat[]{\includegraphics[width=0.49\linewidth]{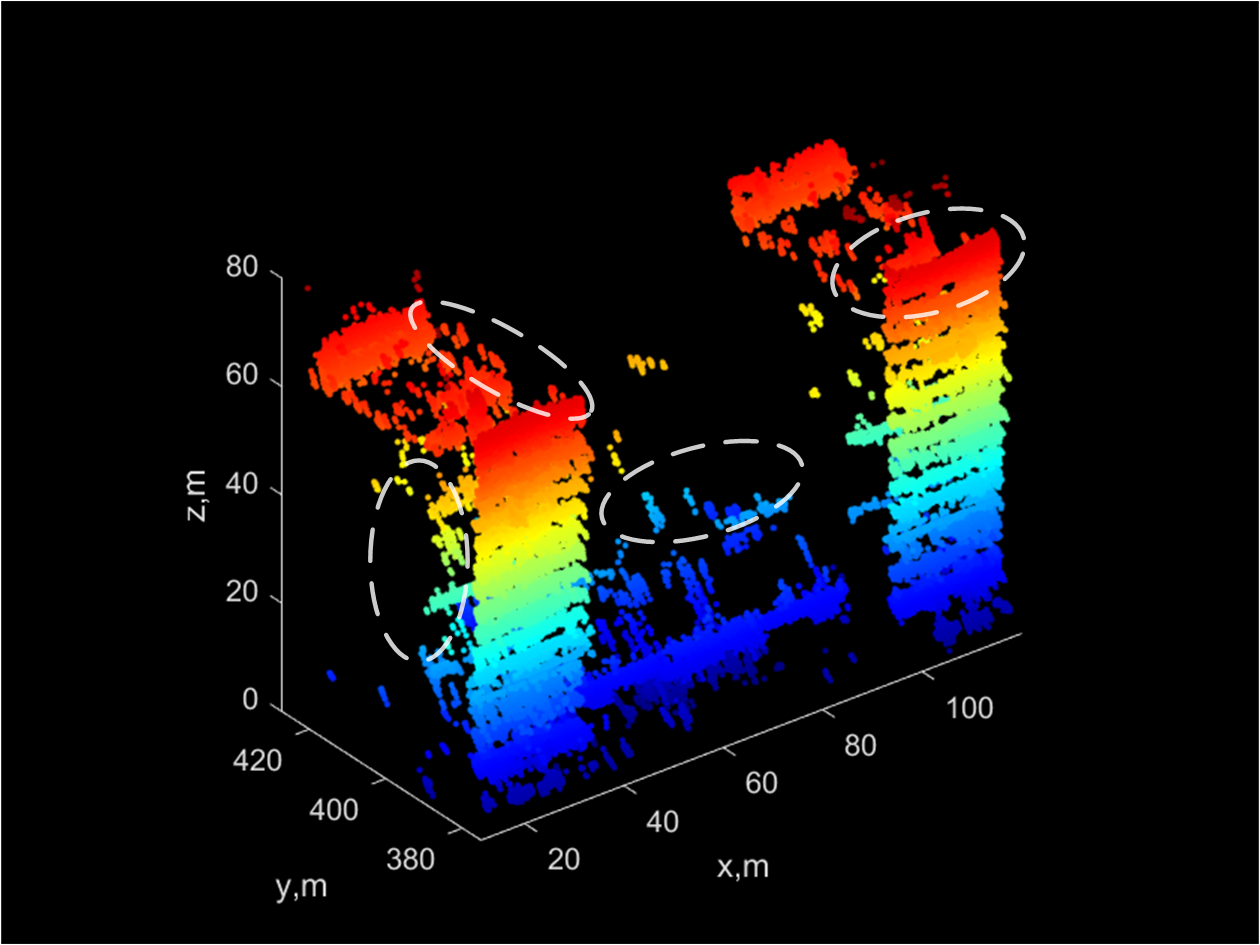}}\hfill
	\subfloat[]{\includegraphics[width=0.49\linewidth]{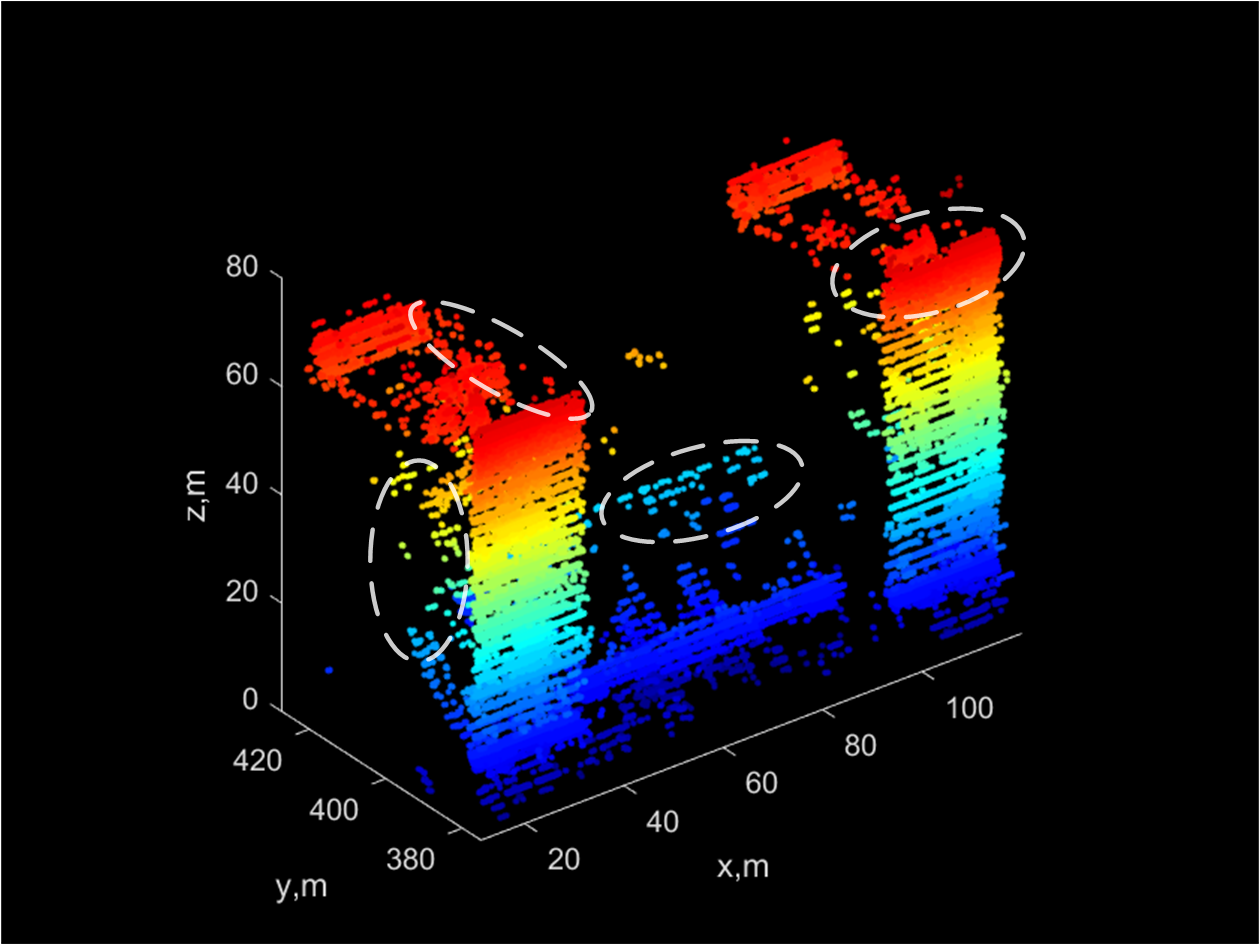}}\vspace{-1em}
	\subfloat[]{\includegraphics[width=0.49\linewidth]{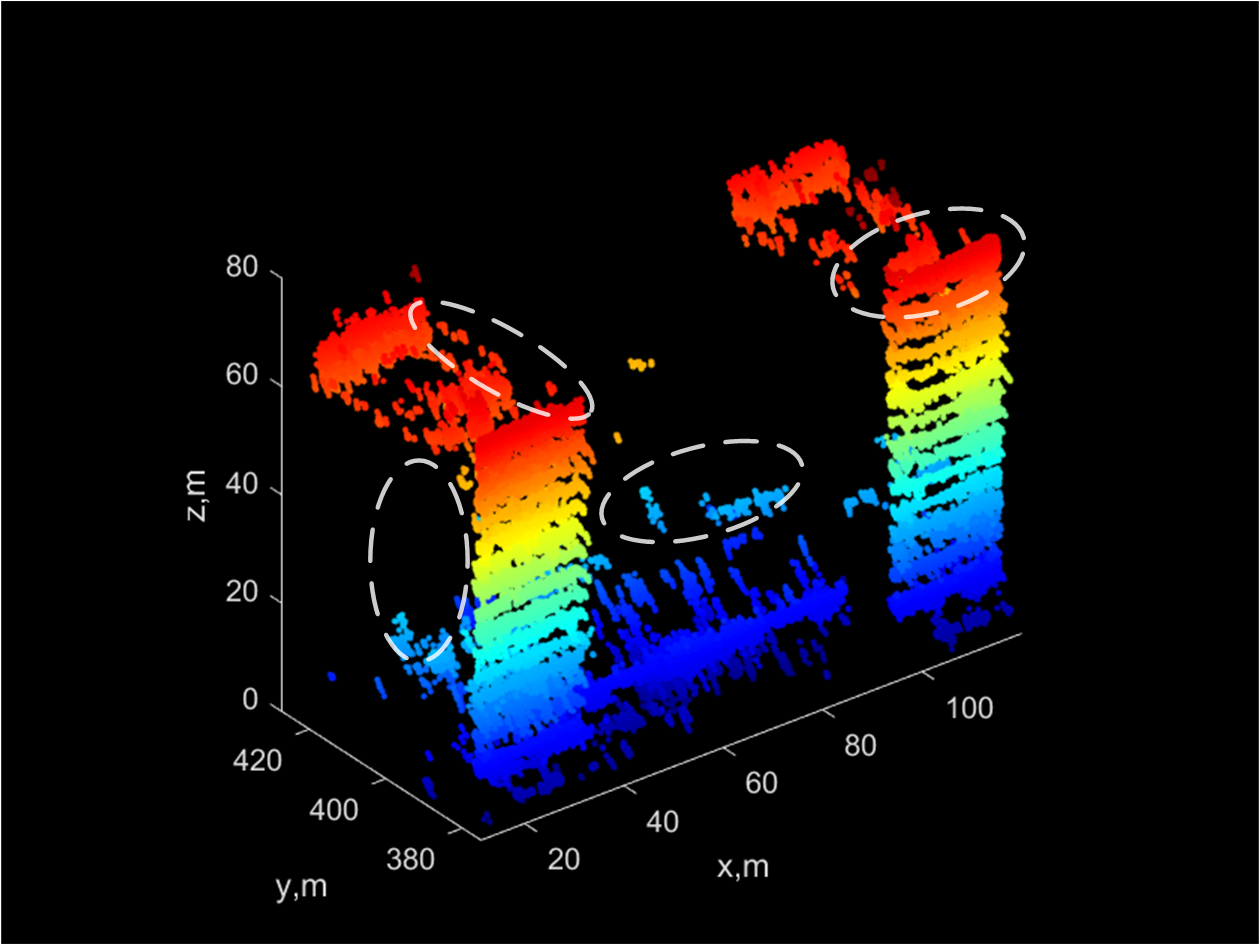}}\hfill
	\subfloat[]{\includegraphics[width=0.49\linewidth]{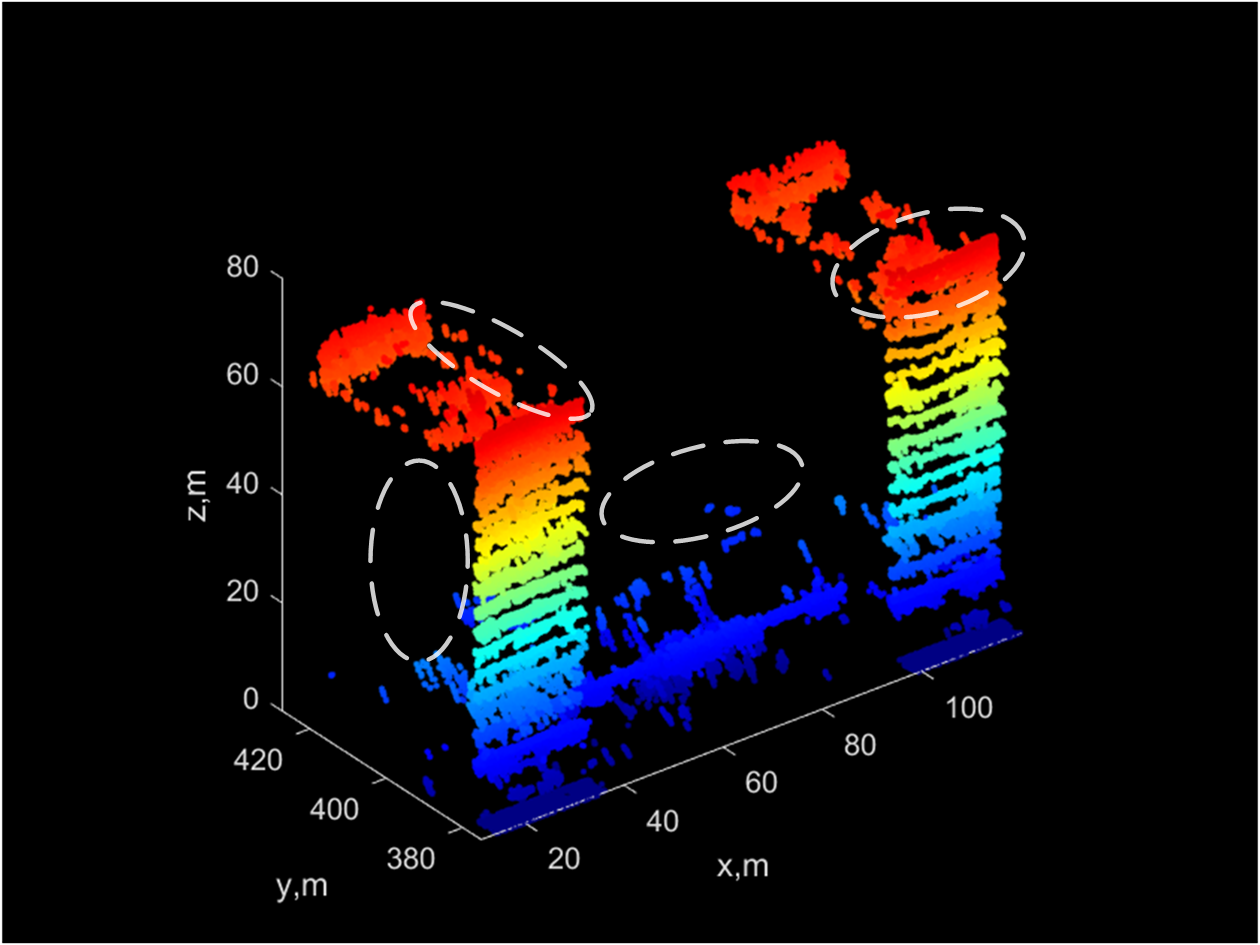}}
	\caption{3-D point clouds reconstruction results of MV3DSAR flight experiment. (a) SVD-Wiener. (b) GBCS. (c) PAST. (d) EMPAST.}
	\label{realTotal}
\end{figure*}

The final 3-D point clouds reconstruction results are shown in \figref{realTotal}, with point colors rendered by height. In terms of the overall structure of the buildings, the SVD-Wiener method exhibits the highest number of false targets reconstructed outside the building contours, particularly behind the walls and above the roofs. GBCS improves the reconstruction results above the roofs but still retains numerous false targets on the backside of walls. PAST further reduces false targets on the backside of the walls. In contrast, the 3-D imaging result corresponding to EMPAST is more organized and concise, with almost no outliers outside the building contours, resulting in a clearer depiction of the building outlines.

\begin{figure*}[htbp]
	\centering
	\subfloat[]{\includegraphics[width=0.24\linewidth]{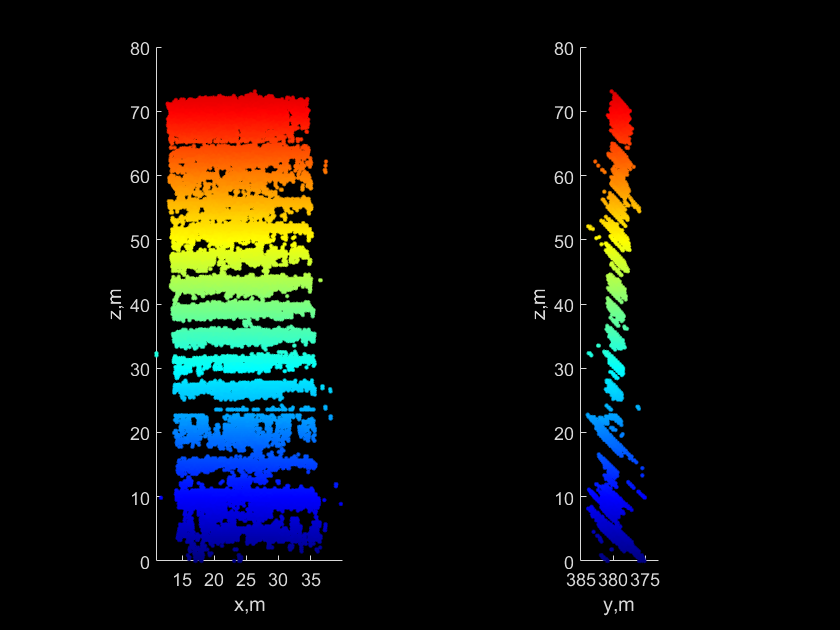}}\hfill
	\subfloat[]{\includegraphics[width=0.24\linewidth]{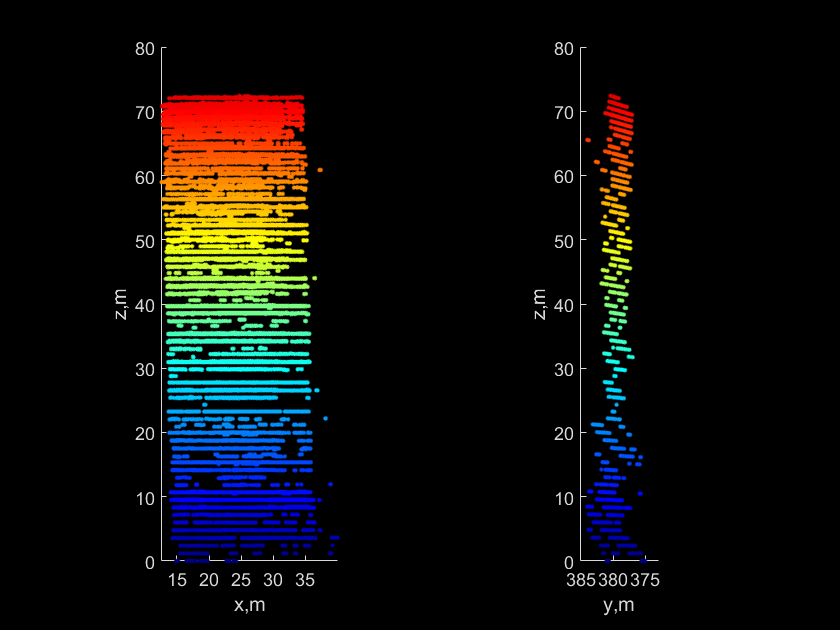}}\hfill
	\subfloat[]{\includegraphics[width=0.24\linewidth]{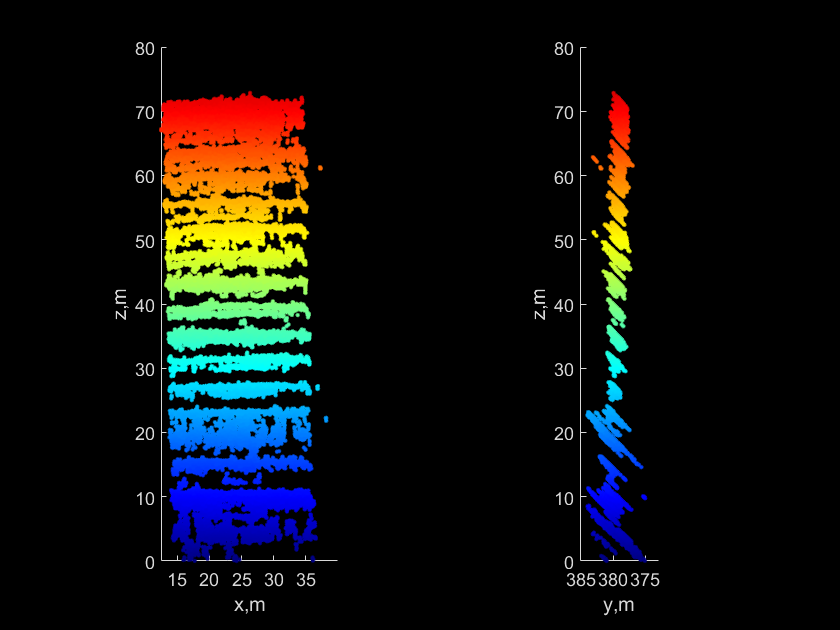}}\hfill
	\subfloat[]{\includegraphics[width=0.24\linewidth]{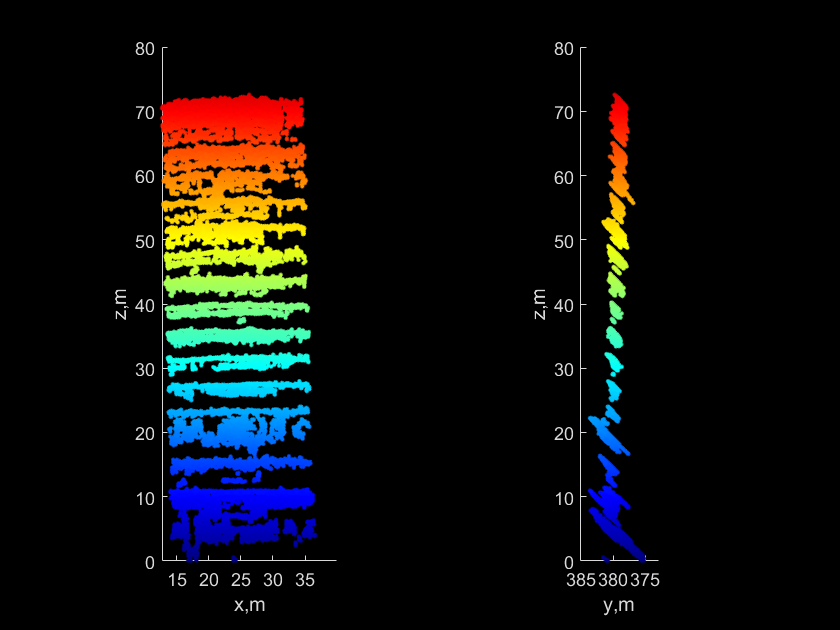}}
	\vspace{-1em}
	\subfloat[]{\includegraphics[width=0.24\linewidth]{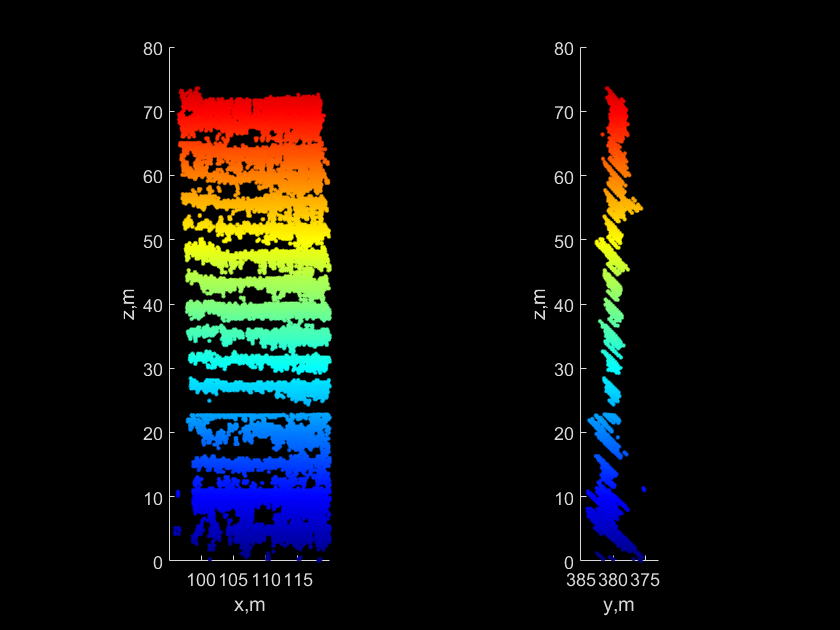}}\hfill
	\subfloat[]{\includegraphics[width=0.24\linewidth]{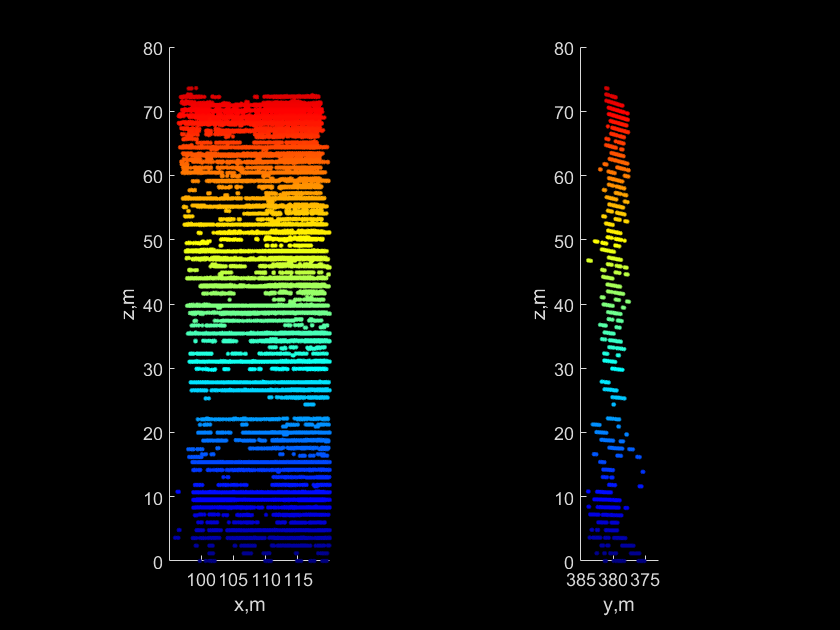}}\hfill
	\subfloat[]{\includegraphics[width=0.24\linewidth]{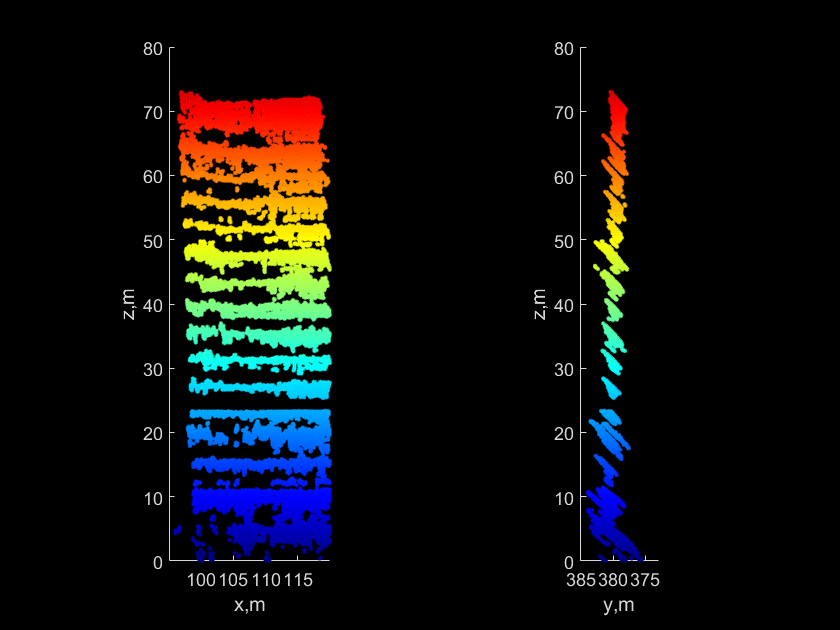}}\hfill
	\subfloat[]{\includegraphics[width=0.24\linewidth]{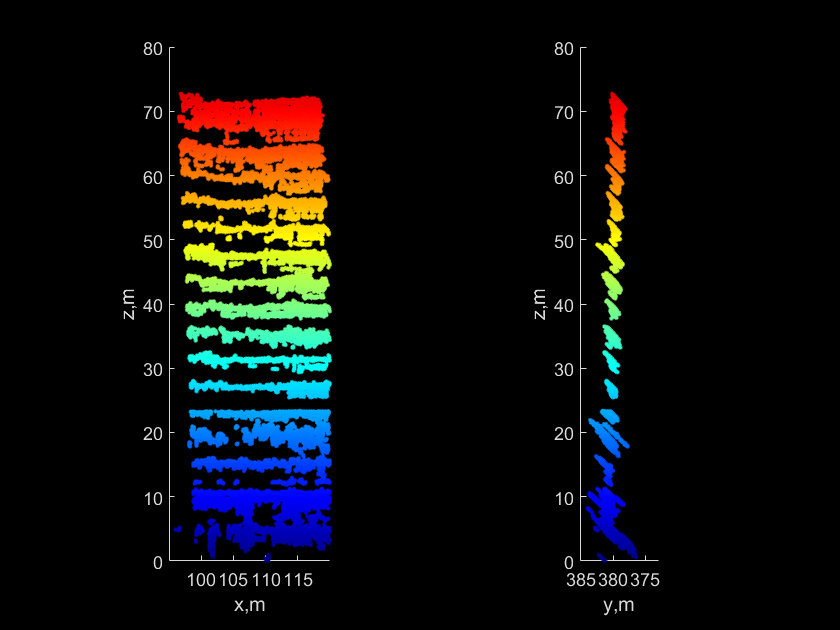}}
	\caption{Two views of reconstruction results for (Top row) the wall of left building, (Bottom row) the wall of right building. (a) and (e) SVD-Wiener. (b) and (f) GBCS. (c) and (g) PAST. (d) and (h) EMPAST.}
	\label{realwall}
\end{figure*}

The walls of the buildings are extracted from the reconstructed 3-D point clouds by RANSAC and projected to the x-z plane and y-z plane, respectively, as shown in \figref{realwall}. 
The top row corresponds to the left building and the bottom row corresponds to the right building.
Besides, we calculate the mean $\mu$ and standard deviation $\sigma$ of the reconstructed walls along the y-axis in \tabref{tabwall}.
\begin{table}[htbp]
	\begin{center}
		\caption{Quantitative comparison of wall estimates along the y-axis (in meters)}
		\label{tabwall}
		\begin{tabular}{ ccccc }
			\toprule
			\multirow{2}{*}{Method} & \multicolumn{2}{c}{Left wall} & \multicolumn{2}{c}{Right wall}                                              \\
			\cmidrule(lr){2-3}\cmidrule(lr){4-5}
			& $\mu$                      & $\sigma$                 & $\mu$    & $\sigma$\\
			\midrule
			SVD-Wiener              & 379.73    & 1.09 &380.11& 1.12   \\
			GBCS                    & 380.04     & 1.21&379.63&1.30      \\
			PAST                     & 380.07     & 0.81 &379.68&  0.93 \\
			EMPAST                  & 380.08 & \bf  0.69&379.66&\bf 0.72 \\
			\bottomrule
		\end{tabular}
	\end{center}
\end{table}
Comparing the results of walls projection onto the x-z plane, it is evident that all four methods reconstruct the entire wall but the GBCS method lacks the floor structures. In contrast, SVD-Wiener, PAST and EMPAST provide distinct reconstructions for each floor, with EMPAST exhibiting the highest clarity, followed by PAST, and then SVD-Wiener. When analyzing the results projected onto the y-z plane, the GBCS reconstruction exhibits a noticeable ``off-grid" effect, while the SVD-Wiener reconstruction demonstrates poor noise robustness, leading to thicker walls. In contrast, the walls reconstructed by PAST and EMPAST are thinner with a higher point density, with EMPAST performing slightly better. The numerical results are consistent with the analysis above, with EMPAST having the smallest standard deviation $\sigma$.

\begin{figure*}[htbp]
	\centering
	\subfloat[]{\includegraphics[width=0.24\linewidth]{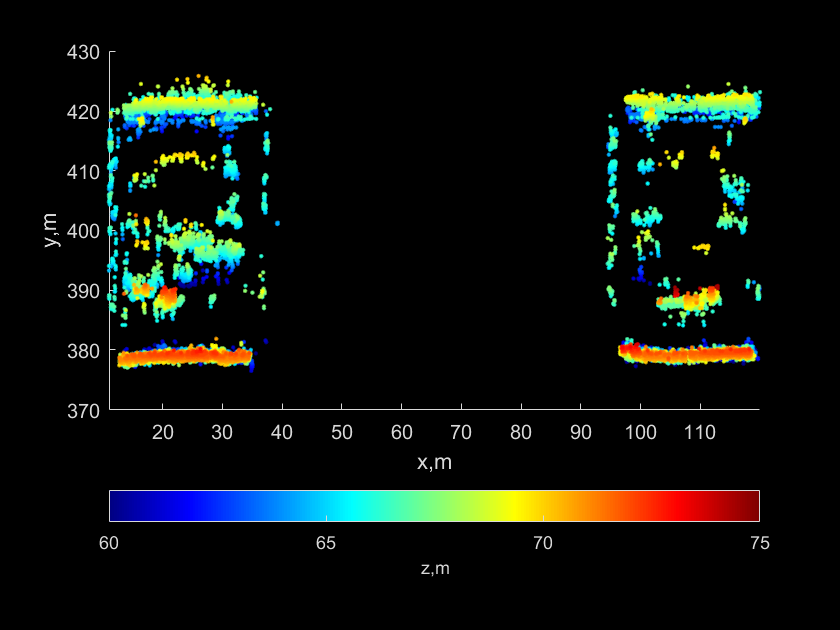}}
	\hfill
	\subfloat[]{\includegraphics[width=0.24\linewidth]{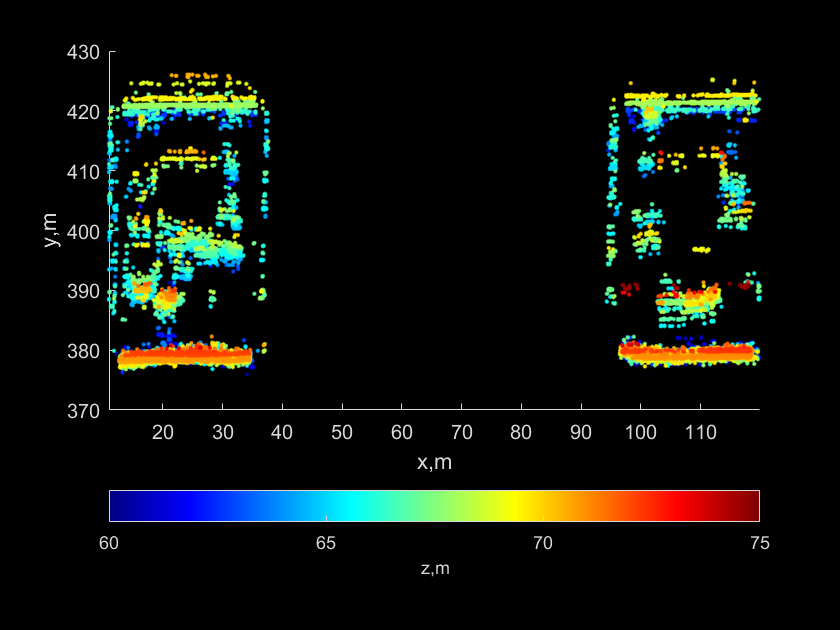}}
	\hfill
	\subfloat[]{\includegraphics[width=0.24\linewidth]{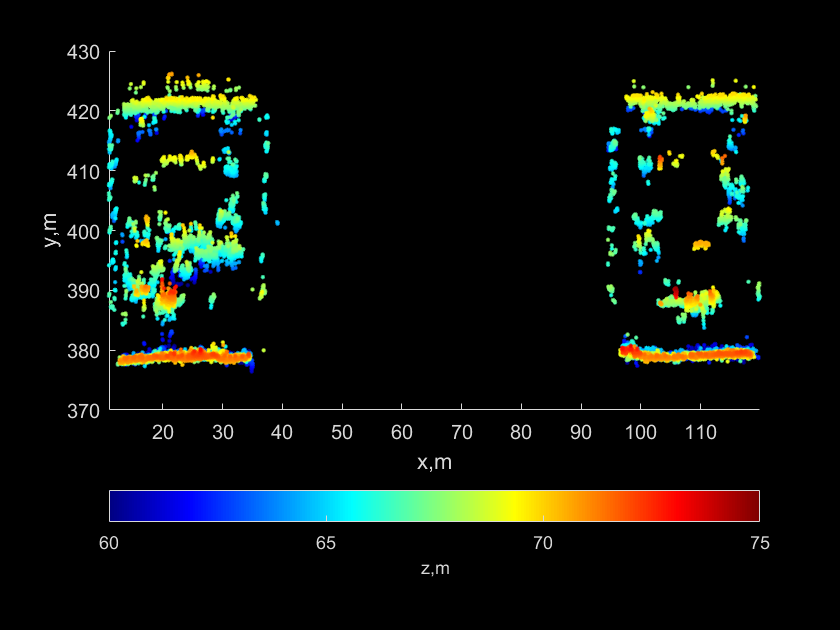}}
	\hfill
	\subfloat[]{\includegraphics[width=0.24\linewidth]{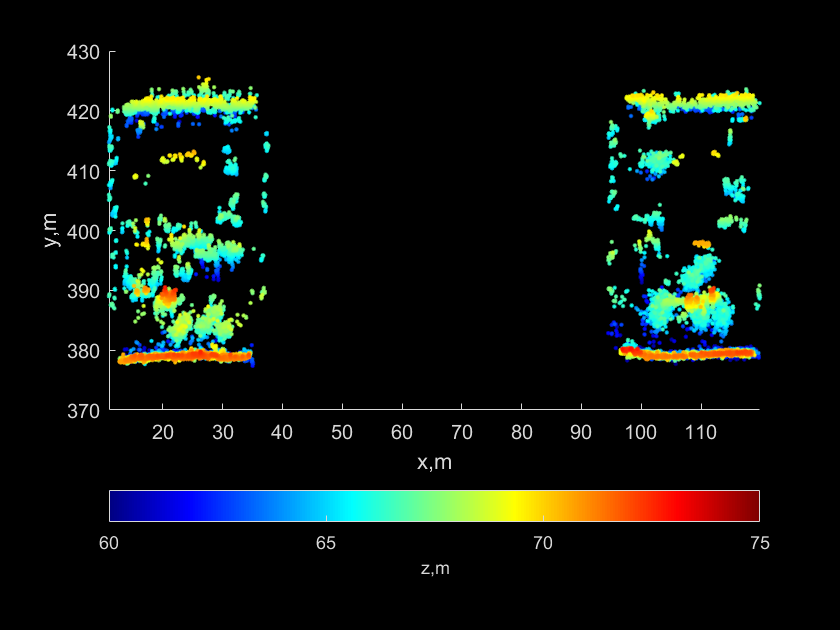}}
	\hfill
	\caption{Reconstruction results for roofs. (a) SVD-Wiener. (b) GBCS. (c) PAST. (d) EMPAST.}
	\label{realroof}
\end{figure*}

As before, we extract the roofs by RANSAC and show the results of projecting them to the x-y plane in \figref{realroof}, with points rendered according to the colorbar below. Besides, we calculate the average height $\mu$ and the standard deviation $\sigma$ of the reconstructed roofs as in \tabref{tabroof}.
\begin{table}[htbp]
	\begin{center}
		\caption{Quantitative comparison of roof height estimates (in meters)}
		\label{tabroof}
		\begin{tabular}{ ccc }
			\toprule
			Method & $\mu$                      & $\sigma$   \\
			\midrule
			SVD-Wiener              & 67.10                       & 2.87    \\
			GBCS                    & 67.31                      &2.59       \\
			PAST                    & 67.14                     & 2.35      \\
			EMPAST                  & 67.32                 & \bf  1.67    \\
			\bottomrule
		\end{tabular}
	\end{center}
\end{table}
From the experimental results, it is evident that all four methods reconstruct the outline of roofs. However, SVD-Wiener exhibits numerous gaps in the reconstruction of roofs within 10 m from walls, due to its poor resolution. Both GBCS and PAST show improvements at the junction of roofs and walls. Notably, EMPAST reconstructs more roof structures, achieving the highest point density at the junction of roofs and walls among the four methods. The standard deviation $\sigma$ of the reconstructed roof heights of EMPAST is also the smallest in terms of numerical results.

\begin{figure*}[htbp]
	\centering
	\subfloat[]{\includegraphics[width=0.24\linewidth]{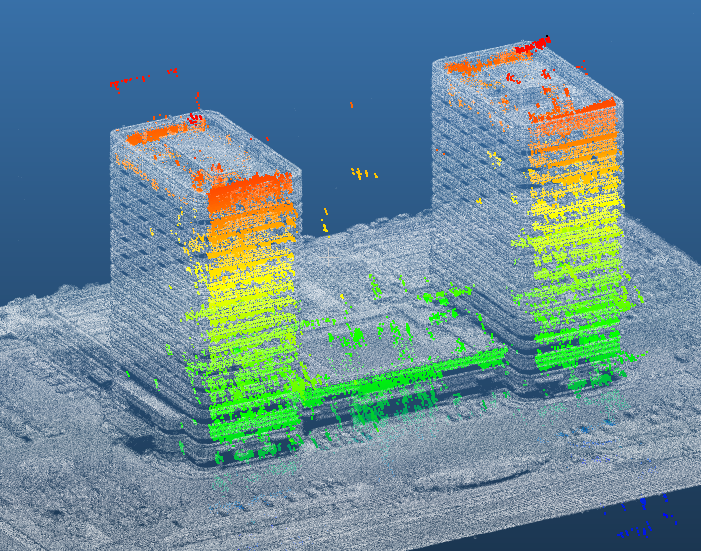}}
	\hfill
	\subfloat[]{\includegraphics[width=0.24\linewidth]{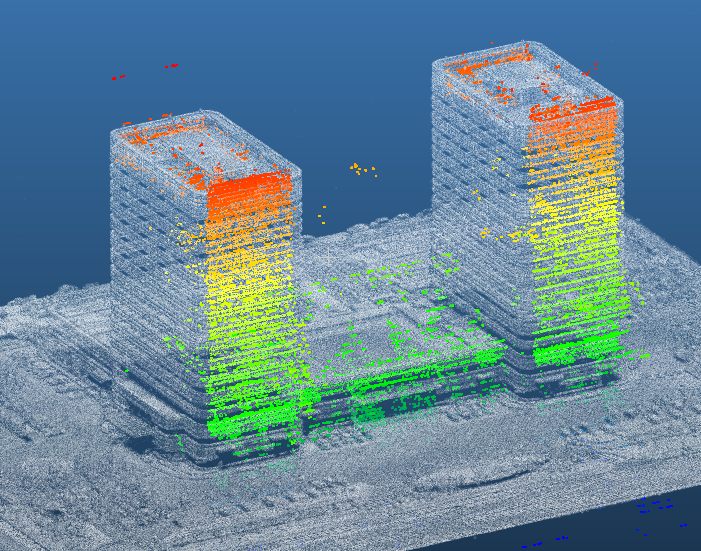}}
	\hfill
	\subfloat[]{\includegraphics[width=0.24\linewidth]{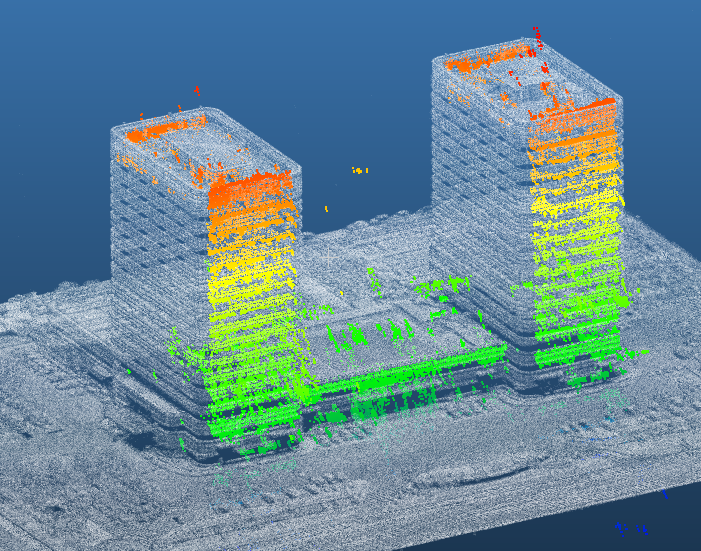}}
	\hfill
	\subfloat[]{\includegraphics[width=0.24\linewidth]{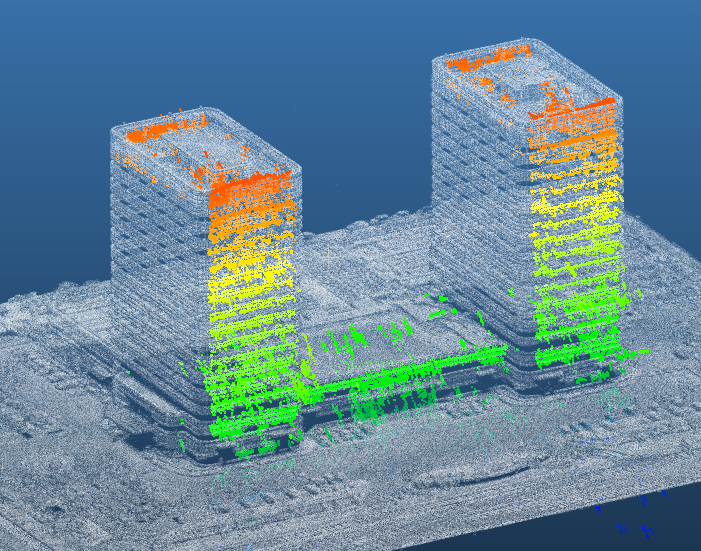}}
	\hfill
	\caption{Fusion illustration of LiDAR point clouds and TomoSAR point clouds. (a) SVD-Wiener. (b) GBCS. (c) PAST. (d) EMPAST.}
	\label{realLAS}
\end{figure*}

Finally, to facilitate a more intuitive evaluation of the 3-D reconstruction results, we registrate and fuse the TomoSAR point clouds with LiDAR point clouds, as illustrated in \figref{realLAS}. 
From the fusion results, we can see that EMPAST reconstructs the most accurate floor structure and has the clearest results for reconstruction of the connecting corridor between two buildings. In conclusion, this experiment demonstrate the performance of the proposed EMPAST framework through comparisons with state-of-the-art methods, using UAV-borne tomographic SAR imaging data.

\section{Conclusion}\label{sec:Con}
This paper proposes a gridless 3-D imaging framework with superior super-resolution capability and robust noise immunity for UAV-borne SAR tomography. The improved performance is achieved via the following contributions.
First, we propose to construct EMMV from oversampled data by utilizing the unique characteristics of UAV, which improves the noise immunity of elevation reconstruction without loss of azimuth and range resolution.
In addition, considering the sparse arrangement of array elements under platform mounting constraints, we propose a new atomic norm soft thresholding algorithm for partially observed MMV. Applying this algorithm to SAR tomography effectively resolves the ``off-grid'' effect common in grid-based CS algorithms and improves the accuracy and resolution along elevation.
Finally, an alternative optimization solver is introduced to improve efficiency.
The proposed framework has been validated via simulation and MV3DSAR flight experiments. The results show that our proposed framework outperforms the existing methods in several aspects, including the reconstruction accuracy, noise immunity, and super-resolution capability.

{\appendix[Derivations of \eqref{tau}]
Inspired by~\cite{AST,wu2024multichannel}, we derive the choice of $\tau$ in \eqref{tau} by providing an upper bound on the mean of the dual atomic norm of the noise $\mathbf{E}_\Omega$. 
For the generality of derivation, we use $\omega\in\mathbb{T}=[0,1)$ to represent the normalized spatial frequency instead of real elevation $s$ by $\omega=\frac{2\Delta b}{\lambda r}s\mod1$, where $\mod$denotes the modulo operation and $\Delta b$ is the antenna spacing of the fully sampled uniform array.

The dual atomic norm is given by
\begin{equation}
	\label{norm-dual}
	\begin{aligned}
		\|\mathbf{E}_\Omega\|_{\mathcal{A}_\Omega}^*&=\sup_{\|\mathbf{V}\|_{\mathcal{A}_\Omega}\leq1}\langle\mathbf{E}_\Omega,\mathbf{V}\rangle\\
		&=\sup_{\omega\in\mathbb{T},\|\mathbf{b}\|_2=1}\langle\mathbf{E}_\Omega,\mathbf{a}_\Omega(\omega)\mathbf{b}^H\rangle\\
		&=\sup_{\omega\in\mathbb{T},\|\mathbf{b}\|_2=1}|\mathbf{b}^H\mathbf{E}_\Omega^H\mathbf{a}_\Omega(\omega)|.
	\end{aligned}
\end{equation}

It is seen in \cite[Corollary 4.2.13]{vershynin2018high} that for any $\varepsilon\in(0,1)$, there exists $\mathcal{N}_1\subset\{\mathbf{b}:\|\mathbf{b}\|_2\leq1,\mathbf{b}\in\mathbb{C}^L\}$ with cardinality $|\mathcal{N}_1|\leq\left(\frac{2}{\varepsilon}+1\right)^{2L}$ such that each $\mathbf{b}$ satisfying $\|\mathbf{b}\|_2=1$ belongs to a Euclidean ball centered at some point $\mathbf{b}_0\in\mathcal{N}_1$ and with radius of $\varepsilon$, i.e., $\|\mathbf{b}-\mathbf{b}_0\|_2\leq\varepsilon$.

Let $\beta$ be a positive number that is great enough and specified later and $\mathcal{N}_2=\{0,\frac{1}{\beta},\frac{2}{\beta},\dots,\frac{\beta-1}{\beta},1\}\subset\mathbb{T}$. Then for any $\omega\in[0,1)$, there exists $\omega_0\in\mathcal{N}_2$ such that $|\omega-\omega_0|\leq\frac{1}{2\beta}$.

Define
\begin{equation}
	W(e^{i2\pi\omega})=\mathbf{b}^H\mathbf{E}_\Omega^H\mathbf{a}_\Omega(\omega).
\end{equation}
Note that $W(e^{i2\pi\omega})$ is a polynomial with respect to $e^{i2\pi\omega}$, and hence,  according to \cite[Theorem 3]{AST}, we obtain that
\begin{equation}
	\label{poly}
	\sup_{\omega\in\mathbb{T}}\left|W'(e^{i2\pi\omega})\right|\leq N\sup_{\omega\in\mathbb{T}}\left|W(e^{i2\pi\omega})\right|.
\end{equation}
It follows immediately that
\begin{equation}
	\begin{aligned}
		&|\mathbf{b}^H\mathbf{E}_\Omega^H\mathbf{a}_\Omega(\omega)|-|\mathbf{b}^H\mathbf{E}_\Omega^H\mathbf{a}_\Omega(\omega_0)|\\
		\leq&|e^{i2\pi\omega}-e^{i2\pi\omega_0}|\sup_{\omega\in\mathbb{T}}\left|W'(e^{i2\pi\omega})\right|\\
		=&\left|e^{i\pi(\omega+\omega_0)}\left(e^{i\pi(\omega-\omega_0)}-e^{i\pi(-\omega+\omega_0)}\right)\right|\sup_{\omega\in\mathbb{T}}\left|W'(e^{i2\pi\omega})\right|\\
		\leq&2\pi|\omega-\omega_0|N\sup_{\omega\in\mathbb{T}}\left|W(e^{i2\pi\omega})\right|\\
		\leq&\frac{\pi N}{\beta}\sup_{\omega\in\mathbb{T}}\left|W(e^{i2\pi\omega})\right|.
	\end{aligned}
\end{equation}
Further, we have
\begin{equation}
	\label{prop}
	\begin{aligned}
		&|\mathbf{b}^H\mathbf{E}_\Omega^H\mathbf{a}_\Omega(\omega)|-|\mathbf{b}_0^H\mathbf{E}_\Omega^H\mathbf{a}_\Omega(\omega_0)|\\
		\leq&|\mathbf{b}^H\mathbf{E}_\Omega^H\mathbf{a}_\Omega(\omega)|-|\mathbf{b}^H\mathbf{E}_\Omega^H\mathbf{a}_\Omega(\omega_0)|+|(\mathbf{b}-\mathbf{b}_0)^H\mathbf{E}_\Omega^H\mathbf{a}_\Omega(\omega_0)|\\
		\leq&\frac{\pi N}{\beta}\sup_{\omega\in\mathbb{T}}\left|W(e^{i2\pi\omega})\right|+\|\mathbf{b}-\mathbf{b}_0\|_2\|\mathbf{E}_\Omega^H\mathbf{a}_\Omega(\omega_0)\|_2\\
		\leq&\frac{\pi N}{\beta}\sup_{\omega\in\mathbb{T}}\left|W(e^{i2\pi\omega})\right|+\varepsilon\|\mathbf{E}_\Omega^H\mathbf{a}_\Omega(\omega_0)\|_2.
	\end{aligned}
\end{equation}
Maximizing the both sides of \eqref{prop} yields that
\begin{equation}
	\label{bound}
	\begin{aligned}
		\|\mathbf{E}_\Omega\|_{\mathcal{A}_\Omega}^*\leq\frac{1}{1-\frac{\pi N}{\beta}-\varepsilon}\cdot\sup_{\mathbf{b}_0\in\mathcal{N}_1,\omega_0\in\mathcal{N}_2}|\mathbf{b}_0^H\mathbf{E}_\Omega^H\mathbf{a}_\Omega(\omega_0)|.
	\end{aligned}
\end{equation}

It is shown in \cite[Lemma 5]{AST} that if $\{x_t\}_{t=1,\dots,T}$ are complex Gaussian variables with unit variance, then
\begin{equation}
	\label{ref-chi}
	\mathbb{E}\left[\max_{t=1,\dots,T}|x_t|\right]\leq\sqrt{T+1}.
\end{equation}
Define 
\begin{equation}
	\label{Q-defn}
	\begin{aligned}
		Q(\mathbf{b}_0,\omega_0)=\frac{1}{\sqrt{\sigma M}}\mathbf{b}_0^H\mathbf{E}_\Omega^H\mathbf{a}_\Omega(\omega_0).
	\end{aligned}
\end{equation}
By the Gaussian assumption on $\mathbf{E}_\Omega$, $Q(\mathbf{b}_0,\omega_0)$ is a Gaussian variable with zero mean and variance
\begin{equation}
		\mathbb{E}\left[Q(\mathbf{b}_0,\omega_0)Q(\mathbf{b}_0,\omega_0)^H\right]=\frac{1}{\sigma M}\mathbb{E}\left[  |\langle\mathbf{E}_\Omega,\mathbf{a}_\Omega(\omega_0)\mathbf{b}^H_0\rangle|^2 \right]
		=1.
\end{equation}
Therefore, substituting $\{Q(\mathbf{b}_0,\omega_0)\}_{\mathbf{b}_0\in\mathcal{N}_1,\omega_0\in\mathcal{N}_2}$ into \eqref{ref-chi} gives
\begin{equation}
	\label{gaussian}
	\begin{aligned}
		&\mathbb{E}\left[ \sup_{\mathbf{b}_0\in\mathcal{N}_1,\omega_0\in\mathcal{N}_2}|\mathbf{b}_0^H\mathbf{E}_\Omega^H\mathbf{a}_\Omega(\omega_0)| \right]\\
		=&\sqrt{\sigma M}\cdot\mathbb{E}\left[ \sup_{\mathbf{b}_0\in\mathcal{N}_1,\omega_0\in\mathcal{N}_2}|Q(\mathbf{b}_0,\omega_0)|  \right]\\
		\leq&\sqrt{\sigma M}\sqrt{\log|\mathcal{N}_1\times\mathcal{N}_2|+1}\\
		\leq&\sqrt{\sigma M}\sqrt{\log\left[\left(\frac{2}{\varepsilon}+1\right)^{2L}\left(\beta+1\right)\right]+1}.
	\end{aligned}
\end{equation}

Consequently, combining \eqref{bound} and \eqref{gaussian}, we can conclude that
\begin{equation}
	\label{conclusion}
	\begin{aligned}
		\mathbb{E}\left[ \|\mathbf{E}_\Omega\|_{\mathcal{A}_\Omega}^*\right]&\leq\frac{1}{1-\frac{\pi N}{\beta}-\varepsilon}\\
		&\cdot\sqrt{\sigma M}\sqrt{\log\left[\left(\frac{2}{\varepsilon}+1\right)^{2L}\left(\beta+1\right)\right]+1}\\
		&\leq\frac{1}{1-\frac{\pi N}{\beta}-\varepsilon}\\
		&\cdot\sqrt{\sigma M}\sqrt{2L\log\left(\frac{2}{\varepsilon}+1\right)+\log\left(\beta+1\right)+1}.
	\end{aligned}
\end{equation}
Let $p={\beta}/{\pi N}$, and then we have
\begin{equation}
	\begin{aligned}
		\mathbb{E}\left[ \|\mathbf{E}_\Omega\|_{\mathcal{A}_\Omega}^*\right]
		&\leq\frac{\sqrt{\sigma M}}{1-p^{-1}-\varepsilon}\\
		&\cdot\sqrt{2L\log\left(\frac{2}{\varepsilon}+1\right)+\log\left(\pi Np+1\right)+1}.
	\end{aligned}
\end{equation}
Selecting $p=4L\log(6L+\log N)$ and $\varepsilon=1/8$ yields that
\begin{equation}
		\mathbb{E}\left[ \|\mathbf{E}_\Omega\|_{\mathcal{A}_\Omega}^*\right]
		\leq\frac{8\sqrt{\sigma M}}{7-8p^{-1}}
		\sqrt{2L\log17+\log\left(\pi Np+1\right)+1},
\end{equation}
which completes the proof.

}

\bibliographystyle{IEEEtran}
\bibliography{ref}

\end{document}